\documentclass[aps,prb,groupedaddress,twocolumn]{revtex4-1}

\usepackage{float}
\usepackage{color}
\usepackage{xcolor}
\usepackage{epsfig}
\usepackage{graphicx}
\usepackage{amsfonts}
\usepackage[figuresright]{rotating}
\usepackage{amssymb}
\usepackage{amsmath}
\usepackage{color}
\usepackage{mathrsfs}
\usepackage{graphicx}
\usepackage{placeins}
\usepackage{txfonts}
\usepackage{pxfonts}
\usepackage{graphicx,bm,units,yfonts}

\DeclareGraphicsExtensions{.eps, .PNG, .jpg}
\newcommand{\ket}[1]{| #1 \rangle}
\newcommand{\bra}[1]{\langle #1 |}

\begin{document}

\title{Signatures of metal-insulator and topological phase transitions in the entanglement of one-dimensional disordered fermions}
\author{Ian Mondragon-Shem and Taylor L. Hughes}
\affiliation{Department of Physics, University of Illinois, 1110 West Green St, Urbana IL 61801}
\date{\today}

\begin{abstract}
We study one-dimensional disordered fermions that either undergo metal-insulator transitions or topological phase transitions to become trivial Anderson insulators. We focus on using entanglement to elucidate how the spatial, momentum, and internal degrees of freedom of fermions are affected by the presence of disorder in such cases.  We develop entanglement tools that reveal the existence of metallic states in the presence of disorder and further show clear signatures of the corresponding localization transition even in the presence of interactions. In systems where the internal degrees of freedom are coupled with the motion of the electrons, topological phases develop. We subject a topological insulator model to different types of disorder and discuss how the topological aspects of the system can be captured through entanglement, even at strong disorder.
\end{abstract}

\maketitle

\section{Introduction}

The physics of Anderson localization transitions has been intensively studied for almost half a century, yet new aspects of this phenomenon continue to emerge \cite{Evers2008}. Among many recent developments, the theory of Anderson transitions has been used to classify topological insulators and analyze their robust properties \cite{Schnyder2008}, and to study thermalization behavior in many-body interacting systems \cite{Evangelou2004,Pal2010}. Furthermore, new systems have been proposed that exhibit Anderson localization physics, for example in photonic systems \cite{Lahini2008} and ultra-cold atoms in optical lattices \cite{Schulte2005, Kondov2011}.

One approach that has recently been applied to study Anderson transitions has been the use of quantum information and  entanglement measures to probe the correlations of many-body states. For example, the spatial entanglement of quantum states reveals the non-local properties of many-body wave functions \cite{Shi2004,Amico2008} which naturally suggests its usefulness for identifying localized and delocalized states. Additionally, entanglement is particularly useful in capturing universal properties that cannot be obtained through the expectation value of local operators \cite{Li2008,Fradkin2013}. In particular, it has  been used in condensed matter systems to gain insight into the nature of different phases of matter and the critical points that separate them \cite{Pollmann2010}.

Over the past few years there have been a number of articles that focus on applying entanglement methods to disordered fermion systems. In [\onlinecite{Varga2008}], for example, the scaling of the entanglement entropy was studied to understand a disordered system at criticality. In [\onlinecite{Jia2008}], the von Neumann entanglement entropy was analyzed using the multi-fractal spectrum of critical wave functions at localization transitions, and in [\onlinecite{Chen2012}], the multi-fractal spectrum of disordered systems was shown to have connections with the R\'enyi entropy of the single-particle states. In [\onlinecite{Prodan2010}] and [\onlinecite{Gilbert2012}], relations were found between the level spacing statistics of the Hamiltonians of disordered topological insulators and of their corresponding spatial entanglement spectrum from which critical delocalized states were identified. Recent work has also considered the entanglement properties of low-dimensional systems with correlated disorder\cite{MS2013,Pouranvari2013,Andrade2014}.

In this article, we complement these studies by analyzing the utility of several different entanglement methods for determining signatures of Anderson localization transitions. Spatial entanglement methods, such as those used in Refs. [\onlinecite{Jia2008,Prodan2010,Gilbert2012,Chen2012,Pouranvari2013}] have given insight into metal-insulator transitions and topological-to-Anderson-insulator transitions. Recently, in Ref. [\onlinecite{MS2013}], the momentum entanglement spectrum was proposed and used to elucidate the nature of metal-insulator transitions in one-dimensional disordered fermion systems. This was followed by a recent article which uses momentum-entanglement to study correlated disorder in low-dimensional models\cite{Andrade2014}. In our work we will focus solely on one-dimensional fermionic models. We study a particular set of models with Anderson transitions that can be precisely determined either analytically or numerically in order to calibrate our entanglement techniques. Although most of the focus is on free-fermion models, we also perform calculations on an interacting model, and discuss a generalization of the single-particle entanglement signatures to the many-body case. 

The outline of this work is as follows. We begin in Section \ref{sec:one} by reviewing the notion of bipartite quantum entanglement for generic systems and the simpler case of free-fermion models. We discuss the different types of entanglement partitions that we will apply in our analysis, and the dual nature of spatial and momentum entanglement for disordered systems. After developing this intuition we move on to some explicit calculations. In Section \ref{sec:entanglement1} we discuss two types of one-dimensional lattice models (i) the random n-mer models which have correlated disorder and (ii) the quasi-crystal Aubry-Andr\'e model. The localization transitions in these models have been previously studied and identified, and we evaluate several different types of entanglement measures to study the critical behavior. After calibrating with these two models we add interactions to the Aubry-Andr\'e model and study its entanglement properties. Finally in Section \ref{sec:entanglement2} we study a disordered topological insulator model and develop a hybrid momentum-orbital entanglement partition that clearly illustrates the localization transition.

\section{Quantifying the entanglement of fermionic systems}\label{sec:one}

\subsection{Bipartite entanglement}

Let us briefly review how bipartite entanglement of a given quantum state $\ket{\Omega}$ is quantified. We start by dividing the Hilbert space of many-body states into two subspaces, say $A$ and $B$, such that: (1) no states are shared between both subspaces; and (2) their union reproduces the original Hilbert space of the system. Given this partition, bipartite entanglement can be quantified using the so-called von-Neumann entropy, which is defined as
\begin{equation}
S_A=-\text{Tr}_A\left(\rho_A \log \rho_A\right),\label{SA}
\end{equation}
where the reduced density matrix $\rho_A$ is obtained by first constructing the density matrix $\rho=\ket{\Omega}\bra{\Omega}$ of the full system, and then tracing out the degrees of freedom that correspond to the subspace $B$.

To see why Eq. \ref{SA} is connected with the concept of entanglement, we can make use of the so-called Schmidt decomposition of $\ket{\Omega}$. Given $\ket{\Omega}$, we can always find its Schmidt decomposition, which reads
\begin{equation}
\ket{\Omega}=\sum_n \lambda_n \ket{\phi_n^A}\ket{\phi_n^B},\label{schmidt}
\end{equation}
\noindent where $\{\ket{\phi_n^A}\}$ and $\{\ket{\phi_n^B}\}$ are sets of many-body basis states that span the subspaces $A$ and $B$, respectively. From this expression, we see that the set of eigenvalues $\{\lambda_n\}$ controls the degree to which $\ket{\Omega}$ can be factorized into states from the subspaces $A$ and $B$. In other words, the set $\{\lambda_n\}$ controls the degree of quantum entanglement of the state $\ket{\Omega}.$ For example, we can use Eq. {\ref{schmidt}} to explicitly compute the reduced density matrix of subspace $A$. This leads to the expression
\begin{equation}
\rho_A=\text{Tr}_{B}\left[\ket{\Omega}\bra{\Omega}\right]=\sum_n  \lambda_n^2 \ket{\phi_n^A}\bra{\phi_n^A},
\end{equation}
from which we conclude, using Eq.\ref{SA},  that
\begin{equation}
S_A=-\sum_n \lambda_n ^2 \log  \lambda_n ^2.
\end{equation}
The von-Neumann entanglement entropy is a single number one can calculate from the $\lambda_n$ but the full set, which is the so-called ``entanglement-spectrum" represents the full information about the wavefunction under the $A$-$B$ partitioning\cite{Li2008}. 

\subsection{Single-particle entanglement spectrum}

The calculation of the entanglement spectrum and $S_A$ can be simplified for non-interacting fermions\cite{peschel2009,peschel2003}. In the absence of interactions, the quantum state of the full system $A\cup B$ can always be written as a Slater determinant. A consequence of this is that any correlation function can be expressed in terms of the correlation matrix 
\begin{equation}
\left[C\right]_{ij}=\bra{\Omega}c^{\dagger}_i c_j\ket{\Omega}=\text{Tr}\left(\rho c^{\dagger}_i c_j\right),
\end{equation}
using Wick's theorem; here $i,j$ represent some fermion degrees of freedom, e.g., lattice site, momentum, spin etc.  Since the reduced density matrix $\rho_{A}$ must also reproduce Wick's theorem when the indices are restricted such that $i,j\, \in \, A,$ then it can be shown\cite{peschel2003,peschel2009} that the reduced density matrix must take the exponential form
\begin{equation}
\rho_A=\mathcal{K}e^{- H_{ent}}, \quad \quad H_{ent}=\sum_{aa'} h_{aa'} c^{\dagger}_{a}c_{a'},
\end{equation}
\noindent where $\mathcal{K}$ is a normalization factor that ensures $\text{Tr}\rho_{A}=1,$ and the indices $a,a'$ label states within $A$. $H_{ent}$ is often referred to as the entanglement Hamiltonian, and the matrix $h$ is determined by the correlation matrix through the expression
\begin{equation}
[C]_{a a'}=\text{Tr}\left(\rho_Ac^{\dagger}_{a}c_{a'}\right)=\left[\frac{1}{e^{h}+1}\right]_{a a'}. \label{corr}
\end{equation}
From this construction we can define several sets of useful eigenvalue spectra. Since the correlation matrix $C$ is Hermitian, then $h$ is also Hermitian. In practice one typically diagonalizes $C$ to obtain the eigenvalues $\{\zeta_i\}$ (which lie between $0$ and $1$) commonly referred to as the single-particle entanglement spectra. From the $\{\zeta_i\}$ one can construct the eigenvalues of  $h$ to obtain the set  $\{\varepsilon_i\}$ (which lie between $-\infty$ and $+\infty$) which are commonly referred to as the single-particle entanglement energies. From the $\{\varepsilon_i\}$ one can derive the (many-body) entanglement spectrum  $\{\chi_i\}$ of $\rho_{A}$ through the relation $\chi_j=\sum_i \varepsilon_i n^{(j)}_i$, where $j$ indicates a many-body state of of $H_{ent}$ with occupation numbers $\{n^{(j)}_i\}$ for each single-particle entanglement energy state $i$ with energy $\varepsilon_i.$

The entanglement entropy can generically be written as
\begin{equation}
S_{A}=-\sum_{i}P_i \log P_i,  \quad  \quad P_i=\frac{e^{-\chi_i}}{\sum_{i}e^{-\chi_i}}.
\end{equation}
Using the expressions for $\chi_i$ in terms of $\varepsilon_i$ (which in turn can be written in terms of of the eigenvalues $\{\zeta_i\}$ of $C$) one finally obtains the simplified expression\cite{peschel2003,peschel2009}
\begin{equation}
S=\sum_i \left\{ -\zeta_i \ln \zeta_i-(1-\zeta_i)\ln(1-\zeta_i)\right\}.\label{Sspect}
\end{equation}
The entanglement entropy, and additionally, all entanglement quantities of  a free-fermion ground state $\ket{\Omega}$ can thus be understood by analyzing the $\zeta_i.$ Note that, from Eq. \ref{corr}, the eigenvalues $\{\zeta_i\}$ always lie between $0$ and $1$. Because of this, the entanglement of a system is determined by how close the $\zeta_i$ get to $1/2.$ We will make use of the single-particle entanglement spectrum to analyze the entanglement of the free-fermion models we study here in addition to the entanglement entropy itself. For  interacting models this simplification no longer holds and one must calculate the $\chi_j$ directly from the many-body reduced density matrix.  

\subsection{Possible entanglement cuts}

In this paper, we refer to the partition of the Hilbert space as an \emph{entanglement cut} that is performed on the system. Going back to the expression Eq. \ref{schmidt}, we note that the degree to which $\ket{\Omega}$ is close to being factorizable in the partitioned basis is sensitive to the type of partition that is chosen.  For example, if we were to choose a partition A and B for which all but one (say  n=1) of the $\lambda_n$'s in the Schmidt-decomposition are zero, then $\ket{\Omega}= \ket{\phi_1^A}\ket{\phi_1^B}$. In such a case, the state $\ket{\Omega}$ would not be entangled with respect to the partition that was chosen. But if we were to change the partition and choose different subspaces $A'$ and $B'$ we could find a partition such that all the $\lambda_n$'s are equal in magnitude, and then we would find $\ket{\Omega}\propto \sum_n  \ket{\phi_n^{A'}}\ket{\phi_n^{B'}} $, which is a highly entangled state. Because of this sensitivity of the entanglement to the choice of partition, we can obtain a more complete picture of the quantum properties of $\ket{\Omega}$ by making appropriate variations of the entanglement cut. Typically, physical motivation will guide this choice.  Some types of partitions, other than the conventional spatial cut, that have been studied in previous work include orbital \cite{Haque2007,Legner2013}, particle \cite{Zozulya2007},  spin \cite{Arnesen2001} and momentum \cite{Thomale2010, MS2013, Andrade2014}.

In the case of one-dimensional disordered fermions, there are a number of possible cuts that can be considered which are physically relevant. Spatial entanglement is the most conventional possibility. It is a basic measure of localization because it probes the degree to which extended states are able to correlate the two parts of the system. At the same time, localization is  related with the interference between forward and backward traveling states that ultimately is at the heart of disorder-induced localization. Hence, momentum entanglement will also serve as an insightful and complementary way to study the disordered ground state. Finally, interesting Anderson transitions occur in systems with internal degrees of freedom that are correlated with the electron motion, such as in topological insulators. Performing entanglement cuts with respect to such internal degrees of freedom will also exhibit interesting behavior as the system transits into the localized phase. We consider all of these cases below. 

\subsection{Comments on entanglement induced by disorder}

Before we begin our main discussion, let us briefly build up some simple intuition about how entanglement encodes information relevant to disordered fermions.  To begin, we consider how localization is exemplified in the spatial and momentum entanglement spectrum. 

The simplest situation we can consider is a quantum state of a single particle that is spatially delocalized. In other words, the particle moves freely throughout the system. The simplest example of this type of state is, of course, a plane-wave given by $\ket{k}=c^{\dagger}_k\ket{0}$, where $k$ is the momentum wave vector and $\langle x|k\rangle\sim e^{i k x}.$ The corresponding spatial correlation matrix reads
\begin{eqnarray}
C_{r_i r_j}&=&\bra{k} c^{\dagger}_{r_i} c_{r_j} \ket{k}=\frac{e^{-i k(r_i-r_j)}}{N}.
\end{eqnarray}
We want to cut the system in half, which is effected by restricting the indices to the range $\left[1,N/2\right].$  The $N/2$ entanglement modes that result from the eigenvalues of this correlation matrix, for any choice of $k$, are $1/2$ and $0$, with the latter being $(\frac{N}{2}-1)$-fold degenerate. It follows that a single particle in a plane-wave state has $S=\log(2)$ spatial entanglement entropy when the system is partitioned in half. This illustrates the intuition that generic delocalized states typically contribute order $\log(2)$ entropy to the spatial entanglement entropy, which indicates that the two halves of the system become correlated.

On the other hand, the corresponding momentum entanglement correlation matrix for the same state reads
\begin{equation}
C_{k_i k_j}=\bra{k} c^{\dagger}_{k_i} c_{k_j}\ket{k}=\delta_{k_i k} \delta_{k_j k}.
\end{equation}
As an example, we can choose to trace out momentum states in the range $k\in \left[0,\pi\right],$ one reason being that, in some of the models that we will consider, this range corresponds to left-moving states and the remaining modes are all right-moving. Hence, we are essentially probing correlations between right and left movers. Using this type of partition, the momentum entanglement modes are $1$ and $0$, with again the latter being $(\frac{N}{2}-1)$-degenerate. We thus obtain a momentum entanglement entropy $S=0$. This is, of course, a trivial result because the state $\ket{k}$ is trivially a product state in momentum space.

Now, upon the introduction of disorder, such plane wave states will scatter and eventually localize at sufficiently strong disorder.  We should then contrast the case of a delocalized state with that of a completely localized state at a position $r$ given by $\ket{r}=c_{r}^{\dagger}\ket{0}$. In this case, the spatial and momentum correlation matrices read
\begin{eqnarray}
C_{r_i r_i}&=&\bra{r} c^{\dagger}_{r_i} c_{r_j}\ket{r}=\delta_{r_i r} \delta_{r_j r}\\
C_{k_j k_j}&=&\bra{r} c^{\dagger}_{k_i} c_{k_j} \ket{r}=\frac{e^{i r(k_i-k_j)}}{N}.
\end{eqnarray}
In this completely localized case, it is the spatial entanglement entropy which vanishes. The momentum entanglement, on the other hand, has a $1/2$ entanglement mode that leads to a momentum entanglement entropy $S=\log(2)$ when using the same cut as above. Hence, the spatial and momentum entanglement behave in a dual fashion with respect to localized and delocalized states.

In the general situation of an occupied Fermi sea there will be more than one single-particle state filled. In this case, one might conjecture that a ground state with more than one delocalized particle will increase the spatial entanglement in an additive manner, for example, each additional occupied plane wave state could add an extra $\log(2)$ contribution when cut in half. It turns out, however, that the behavior for more than one particle can be affected significantly due to interference effects. The spatial correlation matrix for the case of $n$ particles occupying plane-wave states with wave vectors $\{k1, \ldots k_n\}$ is given by
\begin{eqnarray}
C_{r_i r_j}&=&\sum_{k \in \text{occ.}}\frac{e^{-i k(r_i-r_j)}}{N}. \label{interference}
\end{eqnarray}
Each matrix element represents a sum of phases that depend on the particular choice of the set $\{ k_1, \ldots k_n\}$. Because of this, the matrix elements of the correlation matrix can be either enhanced or suppressed depending on which states are occupied. An analogous case occurs for momentum entanglement when more than one spatially localized state is occupied. So, while the connection between localization/delocalization and entanglement immediately seems plausible, the details can be complicated. In this work we will study some interesting effects that this interference in the correlation matrix can have on the entanglement of the system.

Finally, we mention how the entanglement can carry signatures of the configuration of the internal degrees of freedom in the presence of disorder. In practice, this will be particularly relevant when the momentum vector and the internal degree of freedom are correlated, such as what occurs in systems with spin-orbit coupling. 
One simple family of states we can consider is a spin $1/2$ particle in a state parameterized by momentum $k$, given by $\ket{\psi_k}=\cos k \ket{k \uparrow}+\sin k \ket{k \downarrow}$. That is, this state is such that the spin rotates throughout the BZ and is explicitly entangled in the spin degree of freedom in a momentum dependent way. As we will discuss later, this type of rotation property in momentum space can have an important impact on the physics of the system. 

If we introduce disorder, one might suspect that scattering processes can potentially destroy the correlation between the spin and the momentum degree of freedom. For example, as localization is induced, there will be a corresponding broadening of the momentum space features, and the specific configuration of the internal degree of freedom in momentum space might become blurred. If this happens, we then expect that such an effect will be apparent in the structure of both the orbital and momentum entanglement. To what extent this happens is something that we wish to understand in this work and will discuss in Section \ref{sec:entanglement2}.

Now that we have introduced the notation, the concept of different entanglement cuts, and some simple examples, we will move on to discuss the entanglement properties of several types of 1D free-fermion models. In Sec. \ref{sec:entanglement1} we will discuss the random $n$-mer models and the Aubry-Andr\'e model, both of which exhibit metal-insulator transitions and have only a single on-site degree of freedom. We also discuss some effects of interactions on the  Aubry-Andr\'e model. In Sec. \ref{sec:entanglement2} we will discuss a chiral symmetric topological insulator model in the AIII class in the presence of disorder potentials that preserve the chiral symmetry. This type of model, in addition to having interesting momentum and spatial entanglement structure, will also yield orbital entanglement since there are two degrees of freedom per site.

\section{Effect of disorder on the motion of fermions}\label{sec:entanglement1}

In this section we focus on one-dimensional disordered models with delocalized states that can exhibit metal-insulator transitions. One dimensional disordered systems generically exhibit localization for any disorder strength as long as the system does not obey any particular symmetry \cite{Andreson1958}.   This would seem to imply that it is not possible to study delocalization transitions at a finite disorder strength in 1D. It has been shown, however, that there are special models with correlated disorder that preempt the occurrence of Anderson localization, and thus possess metallic ground states.  An example is the so-called Random Dimer Model (RDM) \cite{Wu1992,Phillips1991} and its generalizations, and the quasi-periodic system referred to as the Aubry-Andr\'e model (AAM) \cite{AA1980}. Although the AAM is not actually a disordered system, it does have a localization transition for reasons similar to that of the RDM. Both models exhibit metallic behavior at finite disorder potential strengths, and eventually become Anderson insulators. 

We will thus explore the localization properties of these models using entanglement techniques. For such systems, there are two natural ways in which the Hilbert space can be partitioned, namely the spatial and momentum entanglement cuts we mentioned previously. Because of the nature of the delocalized state, it will turn out that for these classes of models the momentum entanglement cut will prove particularly insightful.

\subsection{Correlated disorder potential}

\subsubsection{The random dimer model and its generalizations}

The random n-mer models describe some of the simplest one-dimensional systems with disorder that allow for a metallic phase in one dimension\cite{Wu1992,Phillips1991}. The Hamiltonian of the general model is
\begin{equation}
H= -\sum_{m=1}^{N} t \left( c^{\dagger}_{m+1}c_{m}+ c^{\dagger}_{m}c_{m+1}\right)+\sum_{m=1}^{N} V_m c_m^{\dagger} c_m, \label{RD}
\end{equation}
\noindent where $m$ labels the lattice sites, $N$ is the lattice size, $t$ is the tunneling matrix element between nearest neighbors, and $V_m$ takes on two values $\epsilon_a$ and $\epsilon_b$. The important property of $V_m$ is that the value $\epsilon_a$ always has to be placed on $n$ consecutive sites so that ``n-mers'' are distributed throughout the lattice. The key to explain why this type of disorder leads to delocalized states in the energy spectrum is the fact that there is a complete suppression of scattering between certain momenta. To see this, note that the Fourier elements of the disorder potential $V_m$ are given by
\begin{equation}
V_{\Delta k} =(\epsilon_b-\epsilon_a)\left[\delta_{k,k'}+ \, f\left({\Delta k},\{r_i\}\right) S(\Delta k) \right],
\end{equation}
where $\Delta k=k'-k$, the $r_i$ denote the random positions of the n-mers, $f(\Delta k, \{r_i\})= \frac{1}{\sqrt{N}}\sum_{i} e^{i r_i (k-k')}$ depends on the particular disorder realization, and $S_n(\Delta k)=\frac{1}{\sqrt{N}}\sum_{m=0}^{n-1} e^{im(k-k')}$ is independent of any randomness. The key observation is that the structure factor $S(\Delta k)$ has zeroes at $\mathcal{Q}_n(m)=2\pi m/n,$  $ m=1,\cdots,n-1$. Scattering is consequently suppressed for all $\Delta k=\mathcal{Q}_n(m)$.  If we calculate the energies of the points in the translationally invariant bandstructure that are nested by the $Q_{n}(m)$ of these suppressed scattering events we obtain
\begin{eqnarray}
 E_n(m)=\epsilon_a-2t \cos\left(\frac{\pi m}{n}\right).
\end{eqnarray}
In other words, near each of these special energies $E_n(m)$ there will be a suppression of left and right-mover scattering. This suggests that localization is inhibited for states near the energies $E_n(m)$, which is indeed the case. Furthermore, it can be shown that in the neighborhood of these so-called resonant energies, a finite fraction of the eigenstates have a localization length of the order of the lattice size. Thus, the $n-1$ zeroes of $V_{\Delta k}$ lead to metallic phases in one-dimensional disordered fermions when $E_f$ is tuned to the neighborhood of $E_n(m).$

\begin{figure*}
\includegraphics[trim = 0mm 0cm 50cm 0mm, clip, scale=0.06]{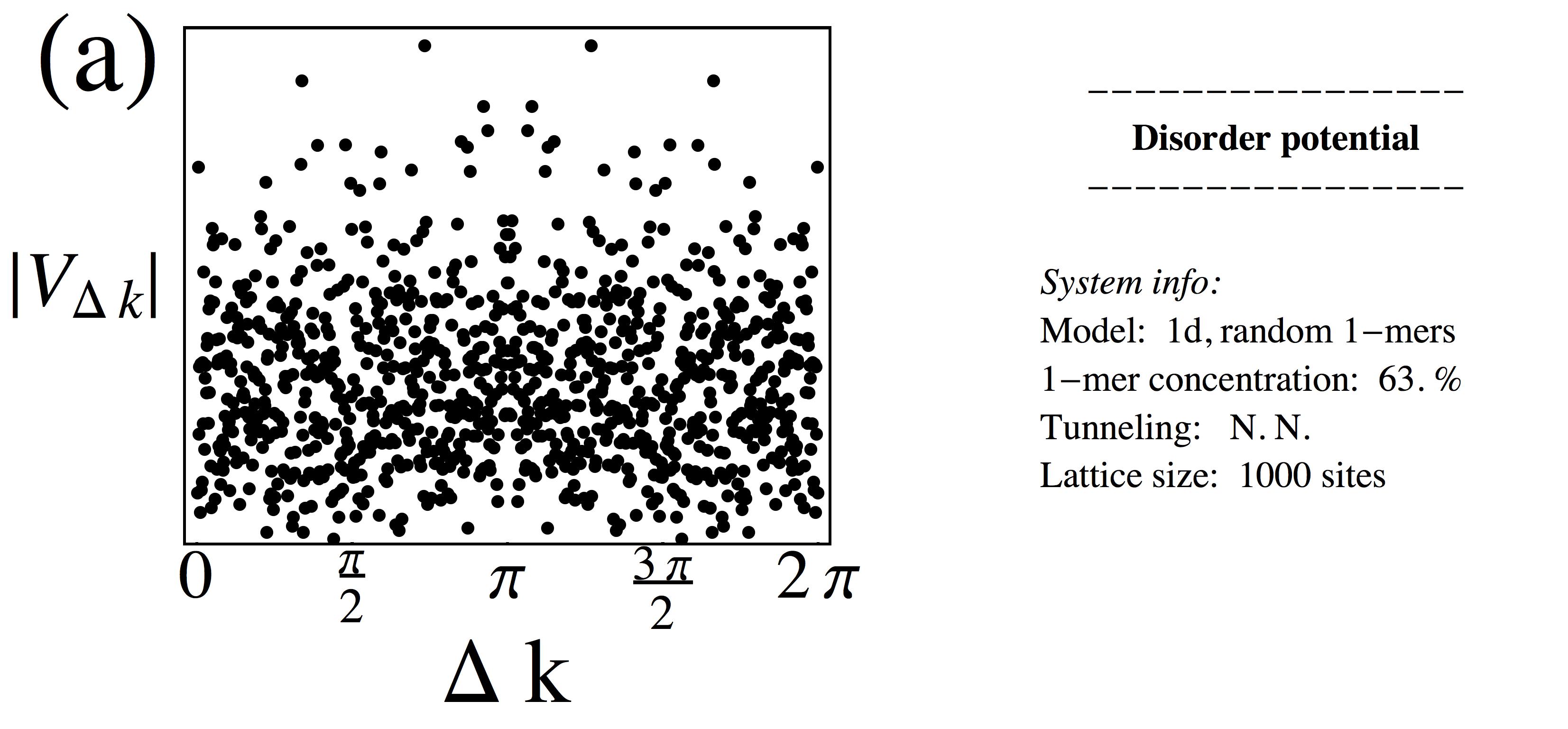}
\includegraphics[trim = 0mm 0cm 37cm 0mm, clip, scale=0.042]{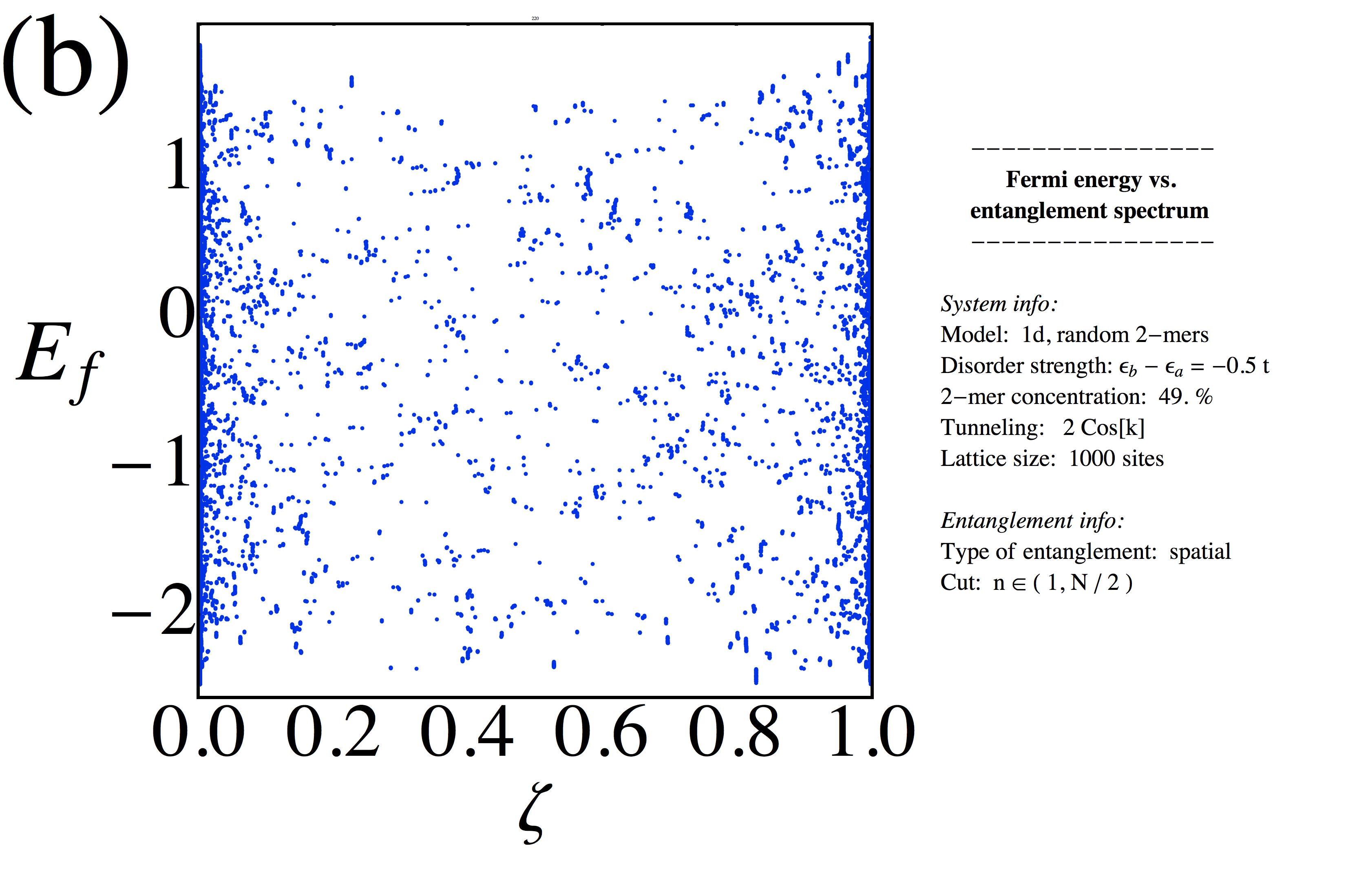}
\includegraphics[trim = 0mm 0cm 38cm 0mm, clip, scale=0.042]{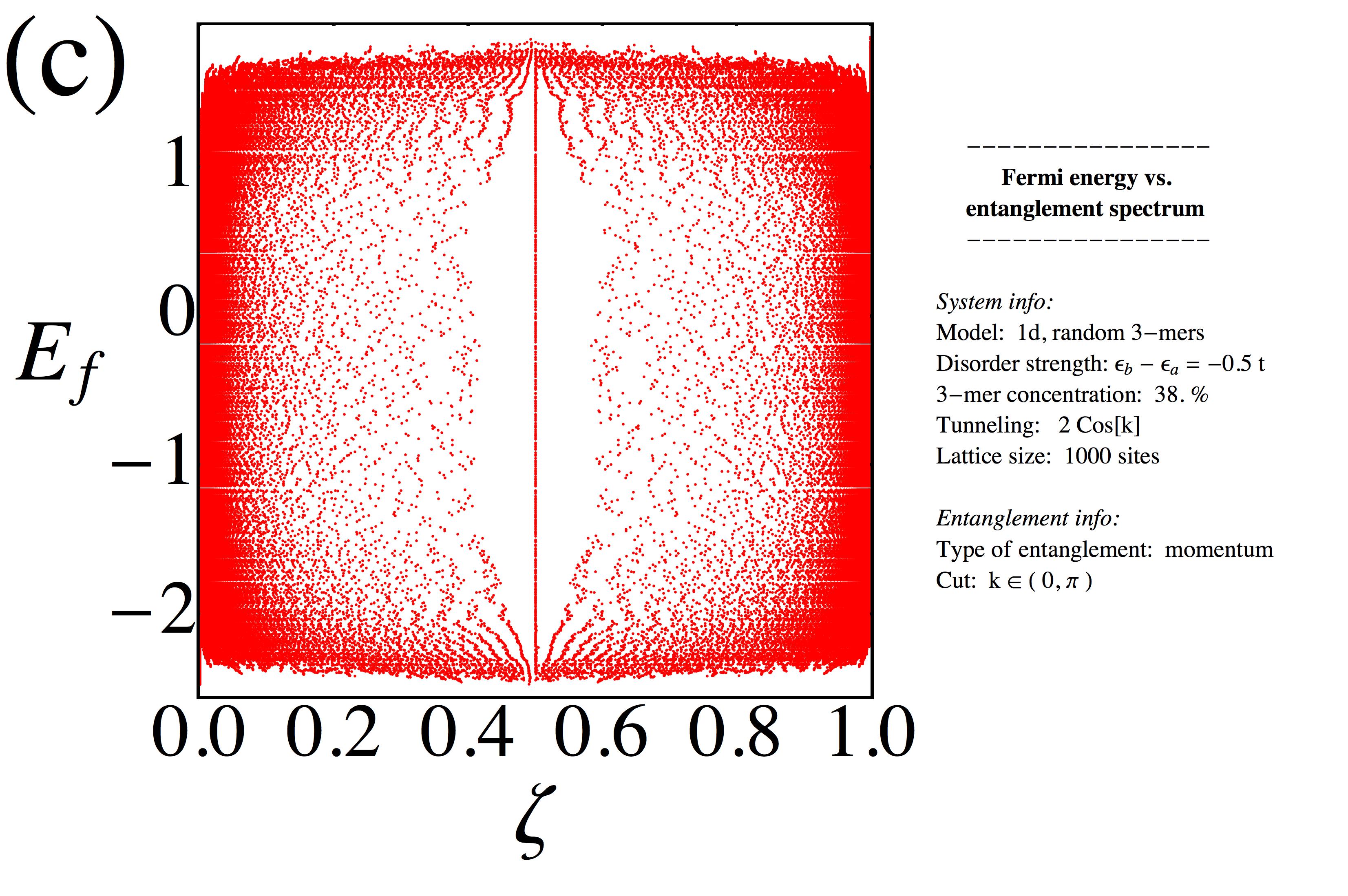}
\includegraphics[trim = 0mm 0cm 43cm 0mm, clip, scale=0.059]{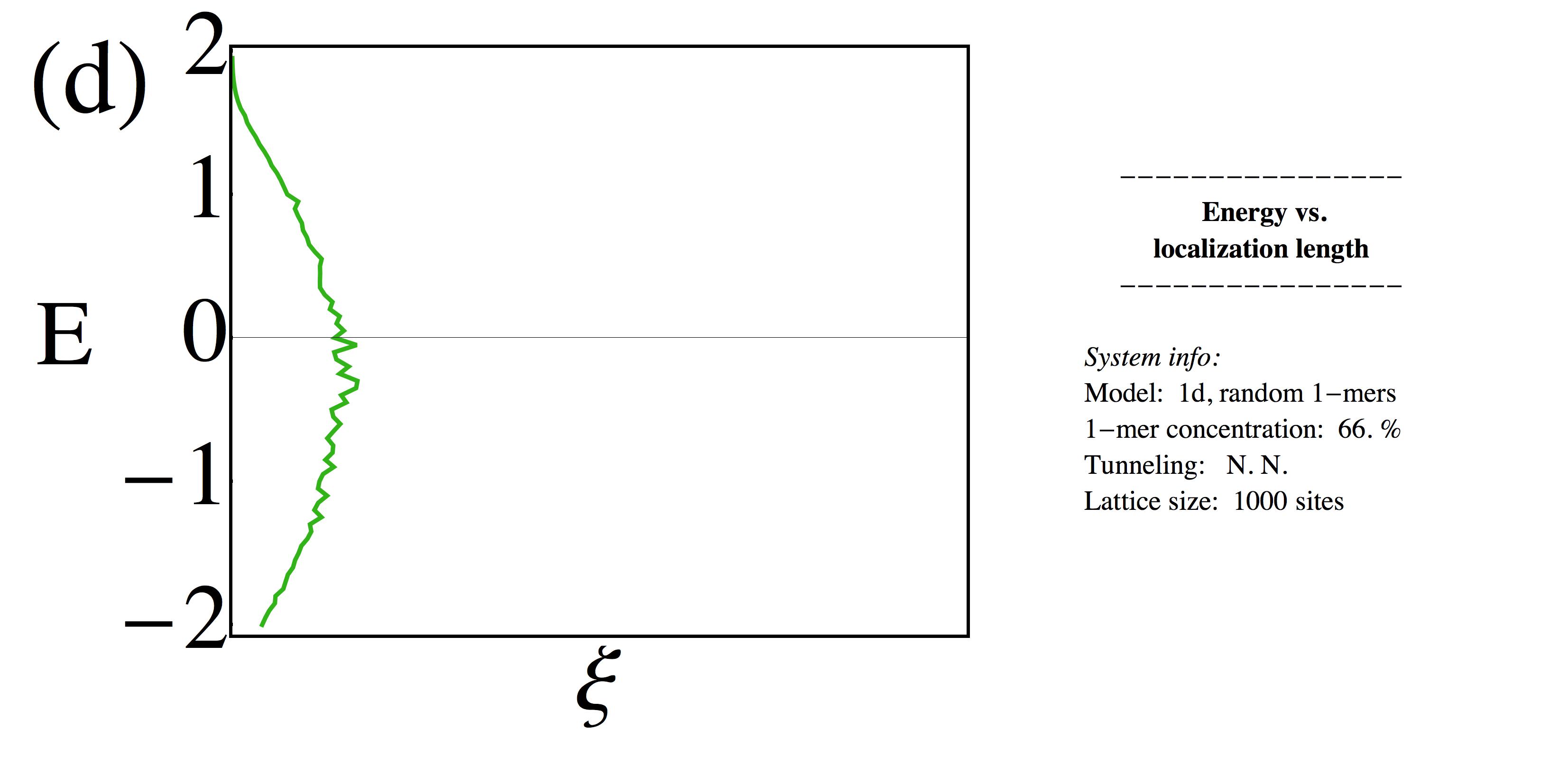}

\includegraphics[trim = 0mm 0cm 50cm 0mm, clip, scale=0.06]{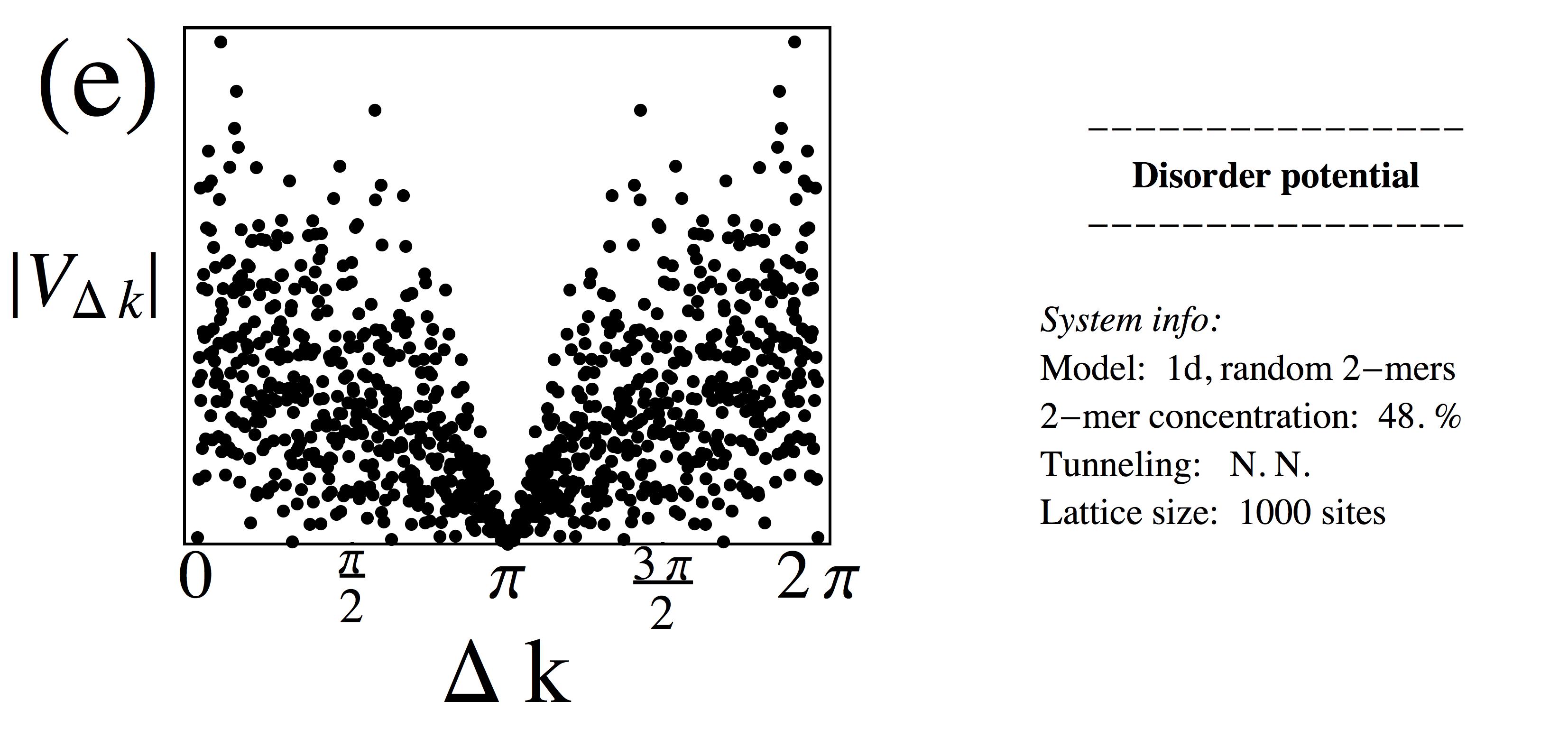}
\includegraphics[trim = 0mm 0cm 37cm 0mm, clip, scale=0.042]{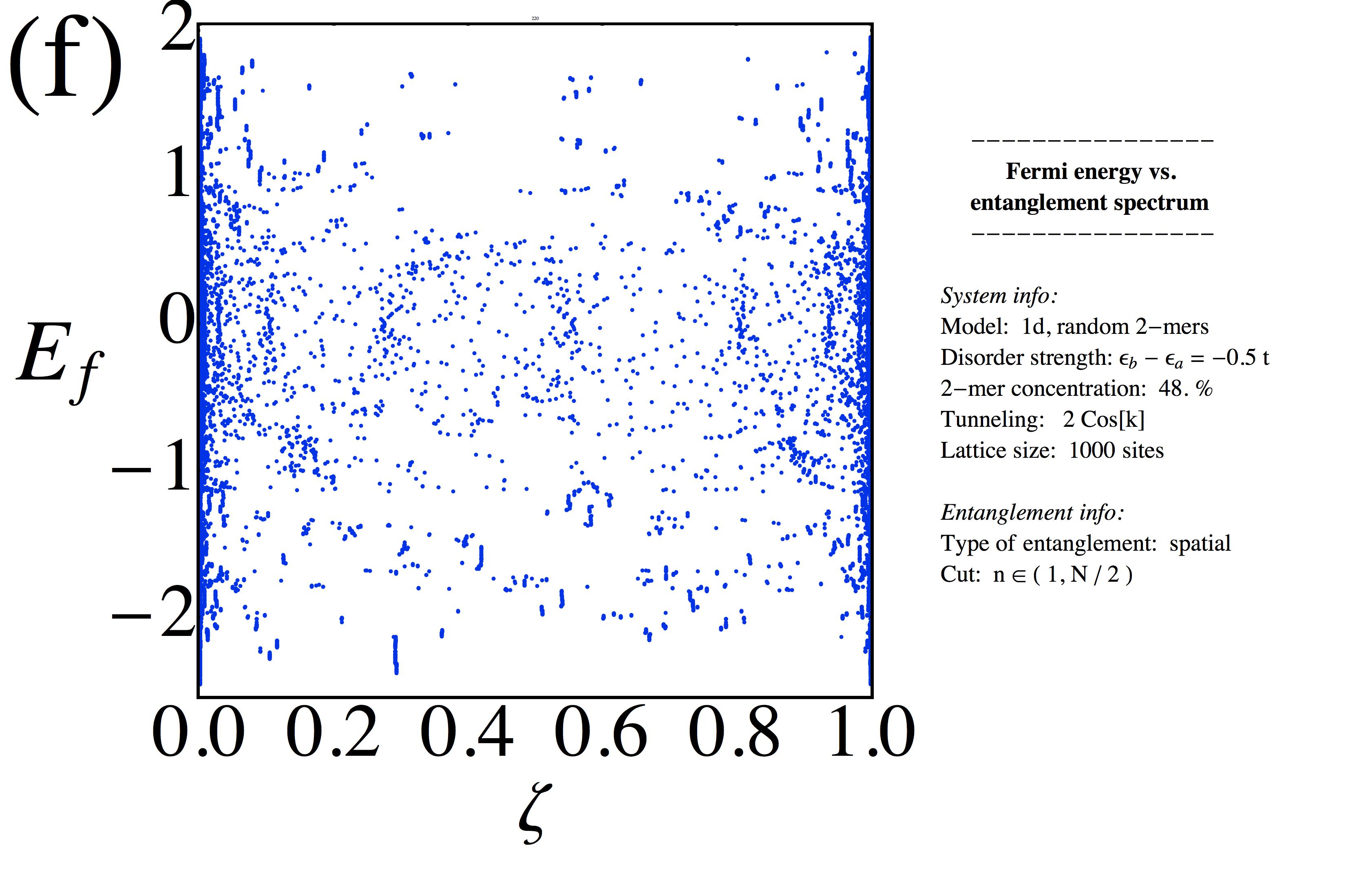}
\includegraphics[trim = 0mm 0cm 38cm 0mm, clip, scale=0.042]{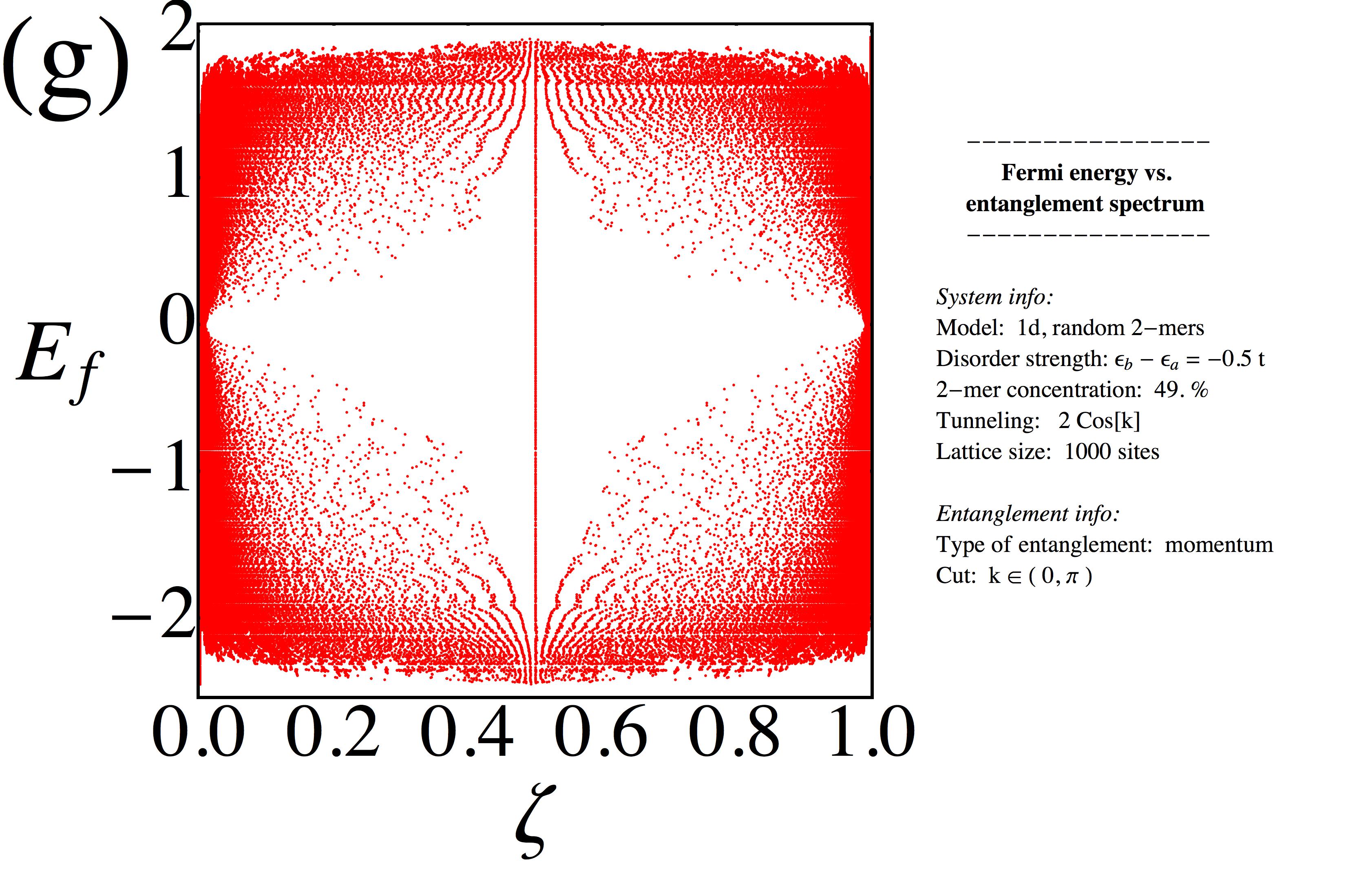}
\includegraphics[trim = 0mm 0cm 43cm 0mm, clip, scale=0.059]{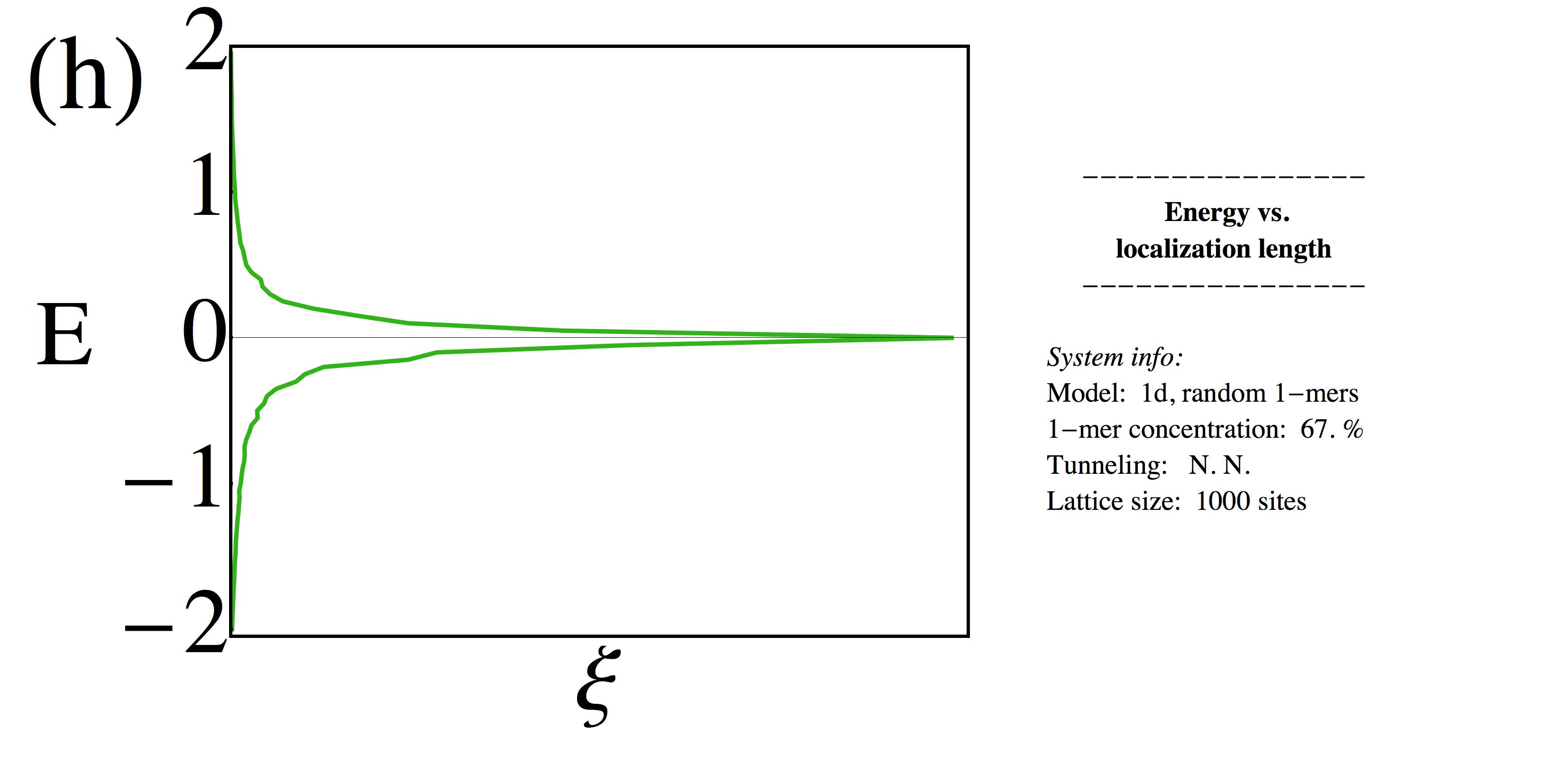}

\includegraphics[trim = 0mm 0cm 50cm 0mm, clip, scale=0.06]{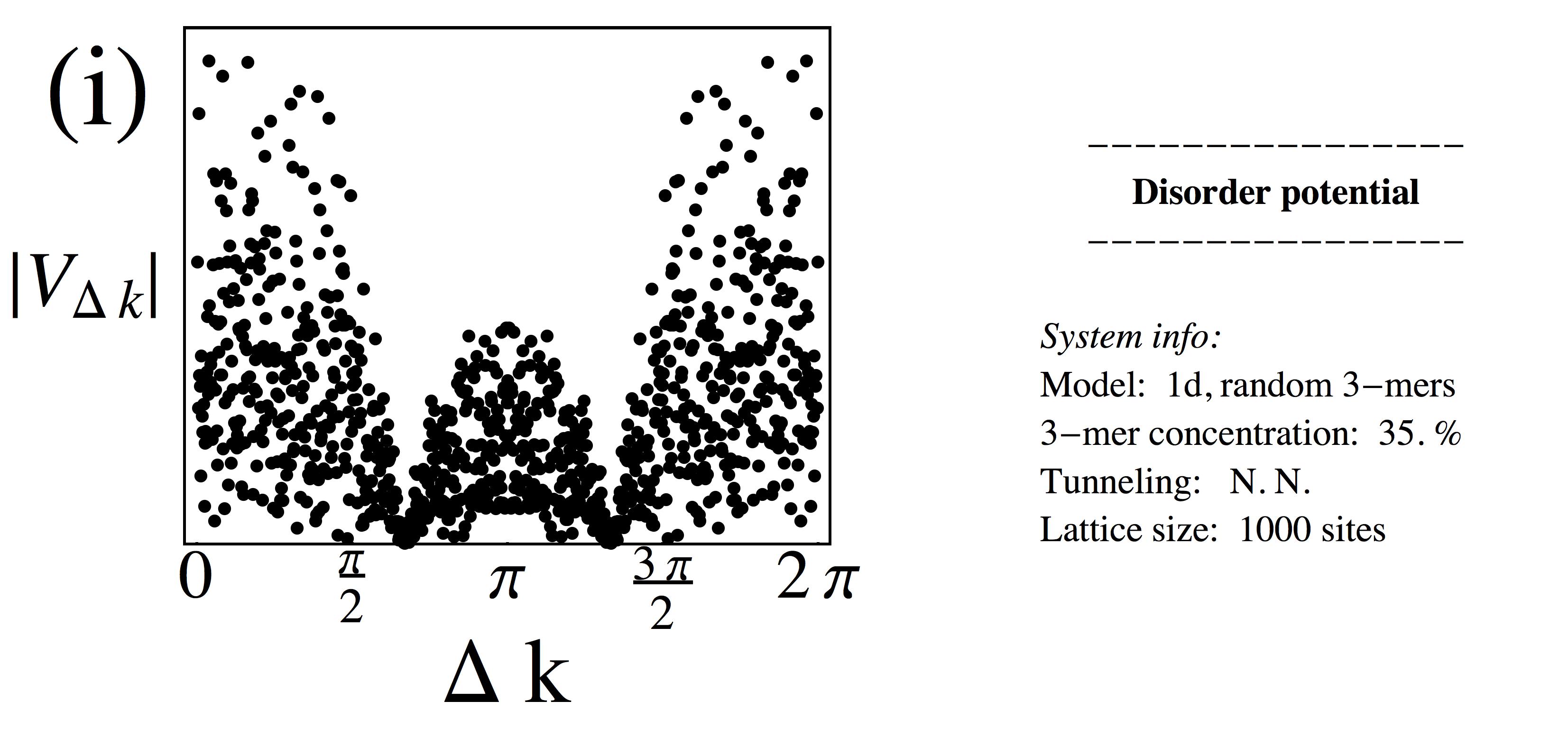}
\includegraphics[trim = 0mm 0cm 37cm 0mm, clip, scale=0.042]{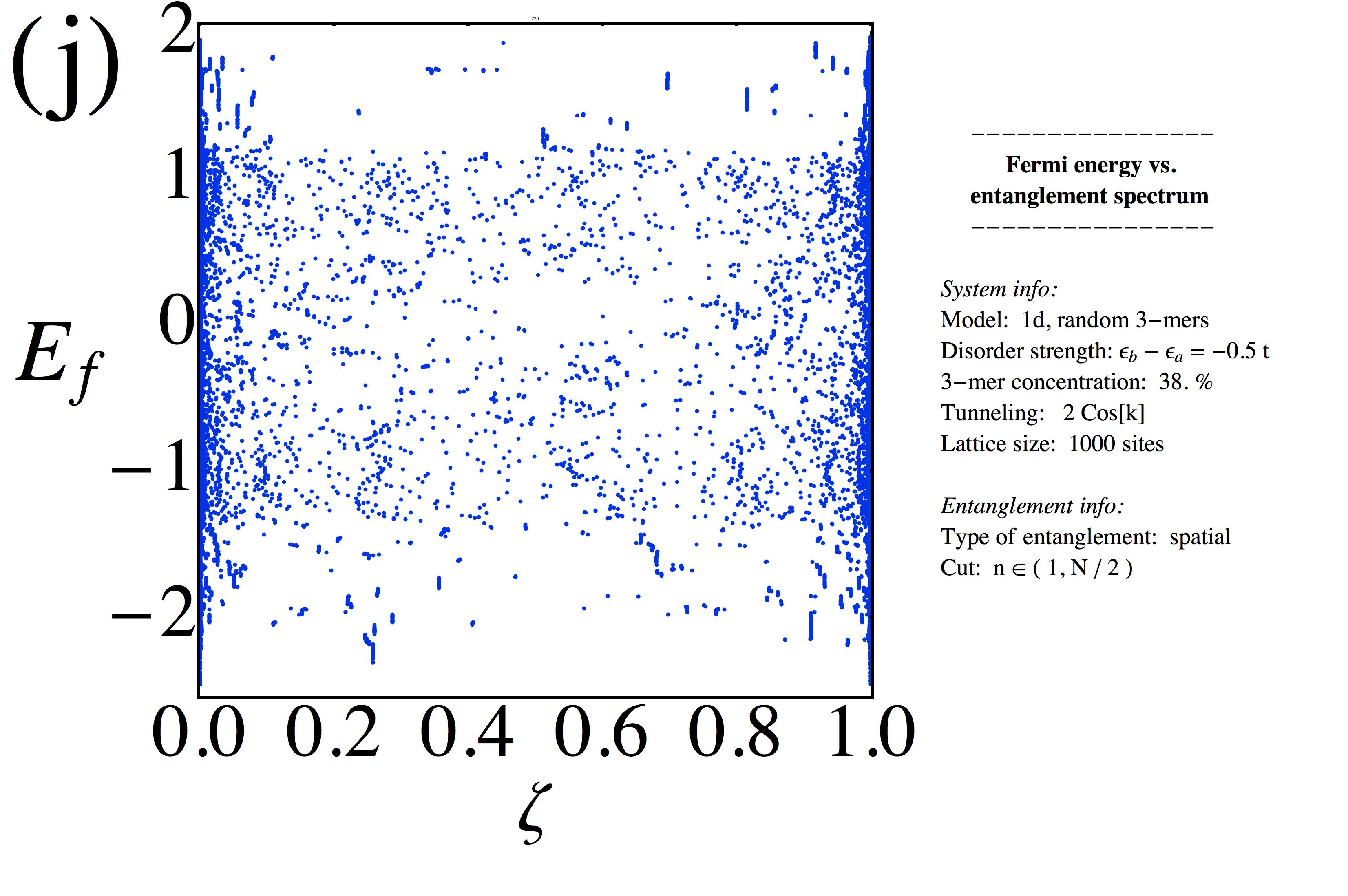}
\includegraphics[trim = 0mm 0cm 38cm 0mm, clip, scale=0.042]{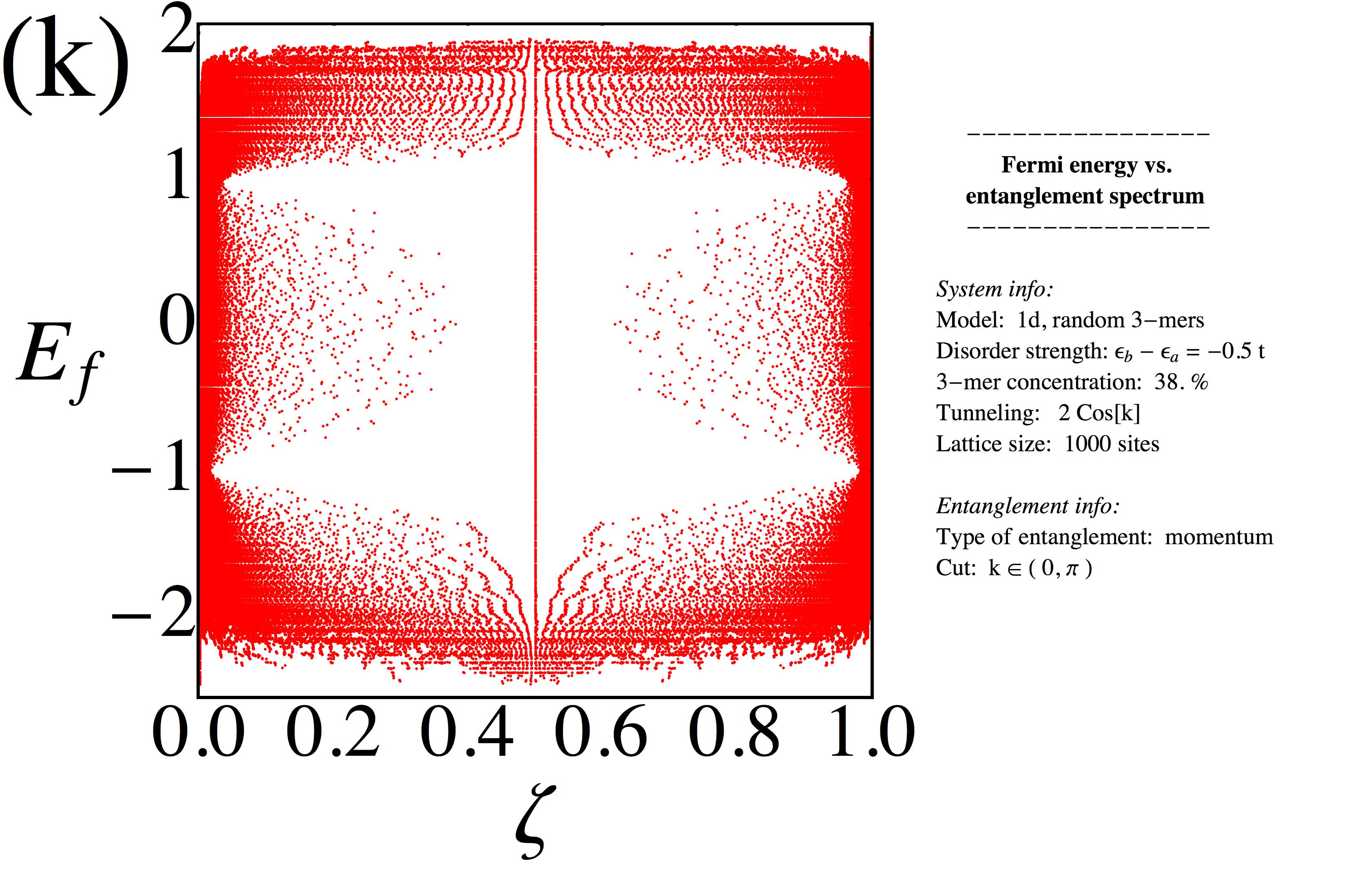}
\includegraphics[trim = 0mm 0cm 43cm 0mm, clip, scale=0.059]{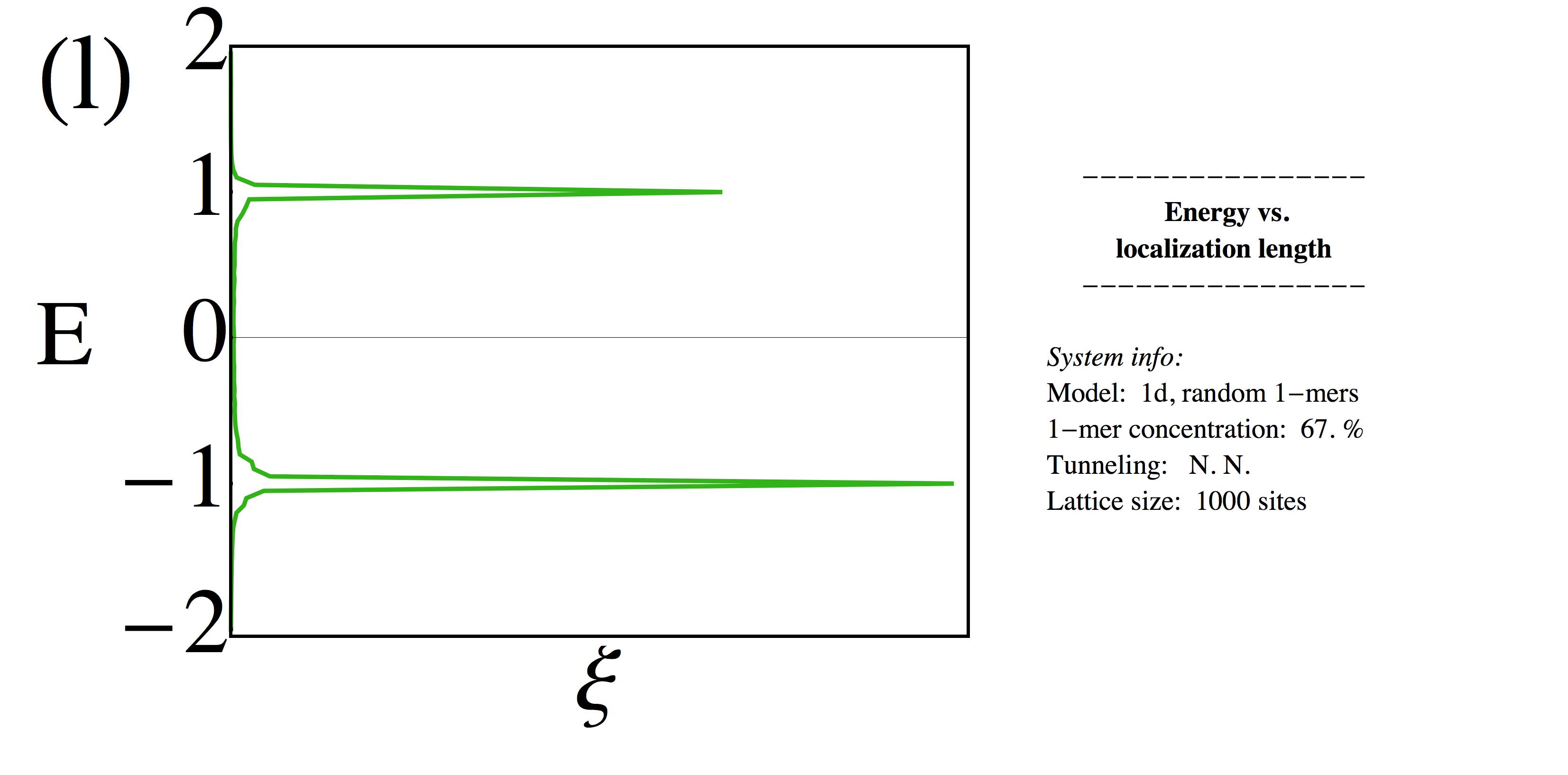}
\caption{Fourier components of the scattering potential (black, column 1), the spatial (blue, column 2) and momentum (red, column 3) entanglement spectrum, and the localization length (green, column 4) for three models: uncorrelated disorder (a),(b),(c),(d); RDM (e),(f),(g),(h); and RTM (i),(j),(k),(l). For all of these plots, the disorder strength was fixed at $\epsilon_b=-0.5t$ and the lattice size is $N=1000$.}\label{Fig_set17}
\end{figure*}

If the disorder becomes sufficiently strong, all single-particle states will localize even at the resonant energies. For the n-mer models, the delocalized states exist up to critical values of the disorder strength $\epsilon_b-\epsilon_a$ given by \cite{Wu1992}
\begin{equation}
-2t\left \{1+\cos\left(\frac{\ell \pi}{n}\right)\right\}\le \epsilon_b -\epsilon_a\le 2t \left\{1-\cos\left(\frac{\ell \pi}{n}\right)\right\}, \label{res}
\end{equation}
\noindent where $\ell=1,\cdots,n-1$. In what follows, we will set $t=1$ and $\epsilon_a=0$, so that the disorder strength will be measured by $\epsilon_b$.  We will focus on the n=1 (uncorrelated disorder), n=2 (random dimer model), and n=3 (random trimer model) cases. The RDM has an extended state at $E=0$ which survives up to the critical disorder $\epsilon_b=\pm2$. For the random trimer model (RTM) there are two extended-state regions, but we will focus on the extended state that exists at $E=+ 1$ which survives for $\epsilon_b\in[-3, 1]$.

\begin{figure}[b]
\includegraphics[trim = 0mm 15cm 0cm 0mm, clip, scale=0.3]{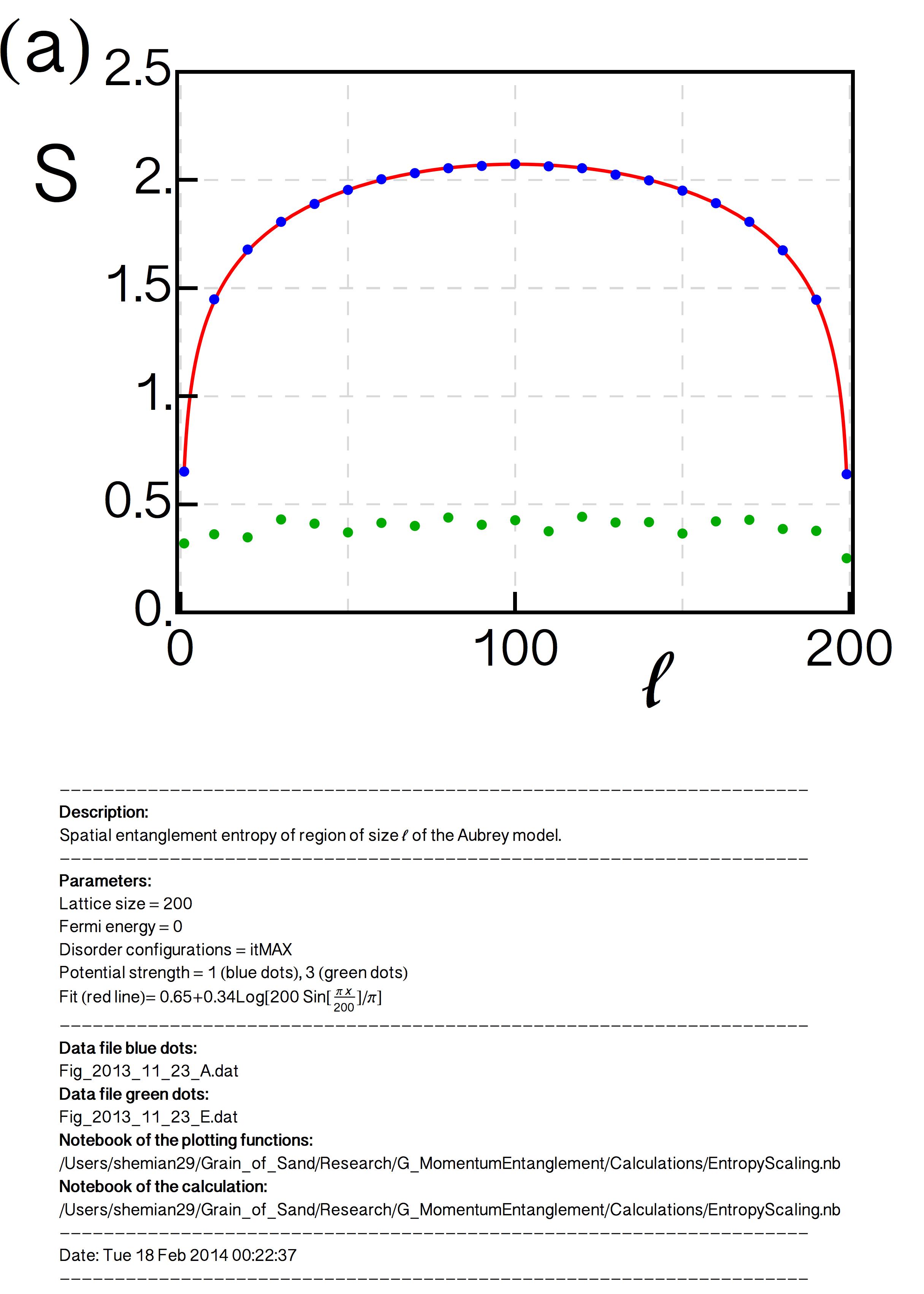}
\includegraphics[trim = 0mm 15cm 0cm 0mm, clip, scale=0.3]{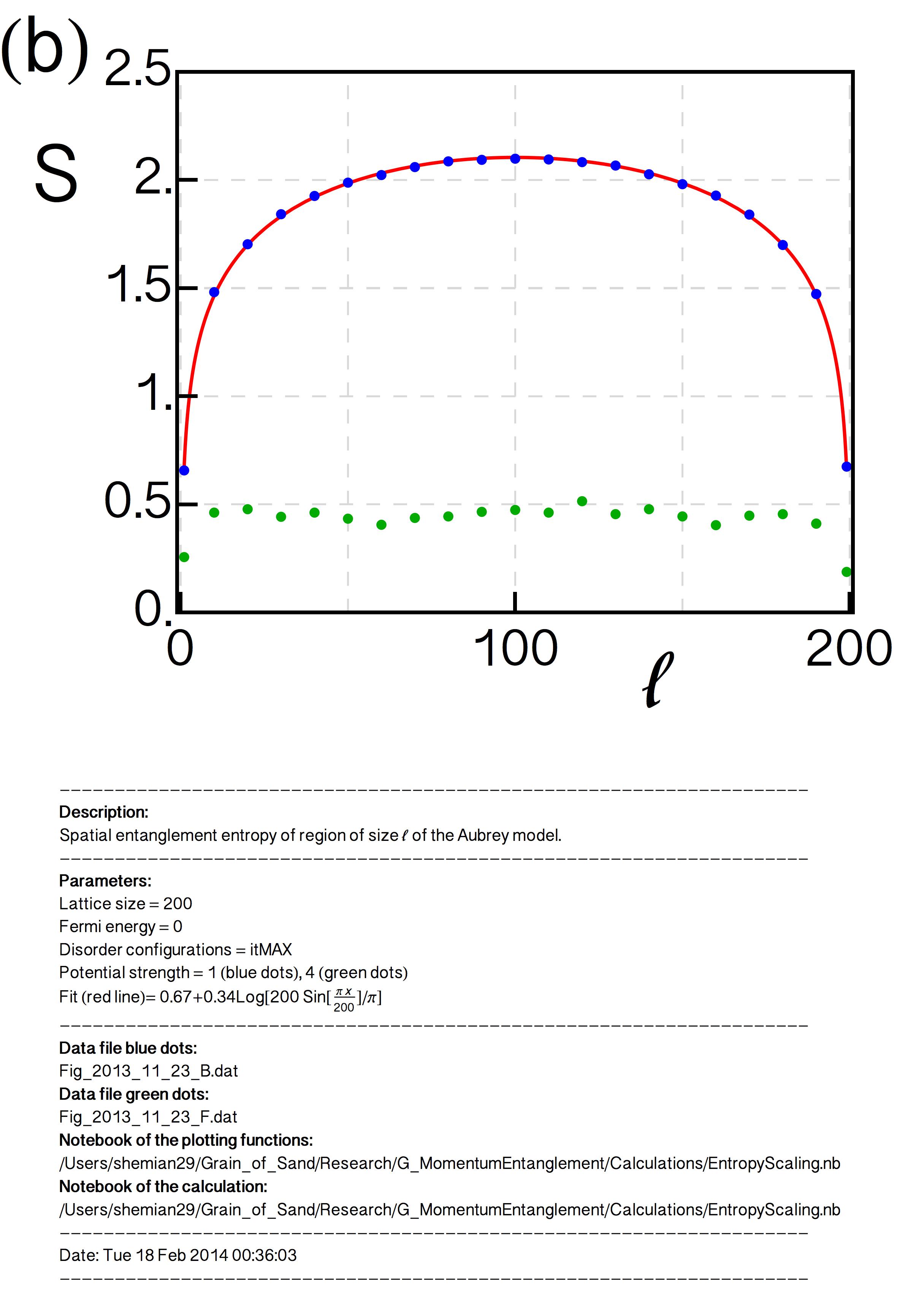}
\caption{Disorder-averaged scaling of the spatial entanglement entropy for a sub-system region of size $\ell$ for the (a) RDM and the (b) RTM . The blue dots denote the numerical calculation when there are extended states in the system and the Fermi-level is tuned to the delocalized resonance state (with $\epsilon_b=-0.5$ for the RDM and $\epsilon_b=1.0$ for the RTM), the red line denotes the corresponding analytical expression and the green dots denote the result when the disorder strength is strong enough to localize all states. The number of sites for this calculation is $N=200,$ and the number of disorder realizations used is 100.}\label{Fig_criticality}
\end{figure}

\subsubsection{Entanglement signatures of the localization transition}

Having discussed the basic aspects of the random $n$-mer models, we now analyze how the metal-insulator transition manifests itself in the entanglement properties. The momentum entanglement and some spatial entanglement for these models was studied earlier in Ref. [\onlinecite{MS2013}], and additionally the spatial entanglement was considered in Ref. [\onlinecite{Pouranvari2013}]. To study these models, there are two parameters at our disposal that allow us to tune the localization transitions, namely the Fermi energy $E_f$ and the disorder strength $\epsilon_b.$ By varying both of these parameters, we can drive the system between the metallic and insulating states. 

We will start with the Fermi energy $E_f$. In Fig. \ref{Fig_set17}, we show four columns: (i) the Fourier components of the scattering potential showing the zeroes of scattering, (ii, iii) the spatial and momentum entanglement spectrum as the Fermi level is varied, and (iv) the localization length as a function of energy. In Appendix \ref{AppA} we explain how the localization length is calculated. The rows correspond to the system with uncorrelated disorder (first row), the RDM (second row), and the RTM (third row). For all of these plots, the disorder strength is $\epsilon_b=-0.5t$, which means that both the RDM and the RTM have resonant states in their energy spectrum. The localization length is shown here to confirm the energies for which delocalization occurs. 

From the spatial entanglement spectra in Fig. \ref{Fig_set17}, one can see by eye that the entanglement modes exhibit level repulsion when the Fermi-level is in the delocalized region, i.e., when the ground state is metallic. This is consistent with the results of Ref. [\onlinecite{Prodan2010}] which made similar observations when the Fermi-level was tuned to a delocalized state in a Chern insulator band. The level repulsion causes the entire region between $[0,1]$ to be filled with eigenvalues since the levels are more rigidly spaced. In regions near the edges of the bandwidth, where localized states should dominate, the entanglement eigenvalues do not appear to have any particular structure. The filling of the region between $[0,1]$ is much more sparse as the levels can cluster together since they are not repelling. For the case of uncorrelated disorder ($n=1$) there are no obvious features in the entanglement nor in the localization length. 

Although the signatures in the spatial entanglement spectrum do not appear to be particularly clear, the disorder-averaged entanglement entropy does reveal the delocalized nature of the extended state. In particular, it is well known that the entanglement entropy in a critical state of a subsystem of size $\ell$  (with total system size $N$) varies as 
\begin{equation}
S=\frac{c}{3} \log\left[\frac{N}{\pi} \sin\left(\frac{\pi \ell}{N}\right)\right]+s_0, \label{scal}
\end{equation}
where $s_0$ is a non-universal constant, and $c$ is a universal coefficient that depends on the critical properties of the ground state (and is the central charge when considering conformal field theories). We show the disorder-averaged spatial entanglement entropy for the RDM and RTM in Fig. \ref{Fig_criticality}a,b. The scaling function is shown as the red curve, and the blue dots are the numerical calculation when $E_f$ is at aligned to the delocalized resonant state. The green dots correspond to when the disorder strength is strong enough to localize everything. The match between the scaling function and the numerical calculation is clear. For the clean system, i.e., the 1D tight-binding chain,  $c=1$ for all Fermi-energies away from the band-edges. In certain disordered systems, such as the disordered XX spin-$1/2$ chain, this coefficient acquires an additional $\log 2$ factor\cite{Refael2004}. For the RDM and RTM, however, we obtain $c=1$ even with disorder. This result is also obtained in Ref. \onlinecite{Pouranvari2013}. The simple reason that we find this result is due to the fact that the RDM and RTM potentials do not couple the states at the resonance energy and thus these fermions remain free. The low-energy theory will remain a $c=1$ Dirac fermion.

\begin{figure*}
\includegraphics[trim = 0mm 0cm 37cm 0mm, clip, scale=0.08]{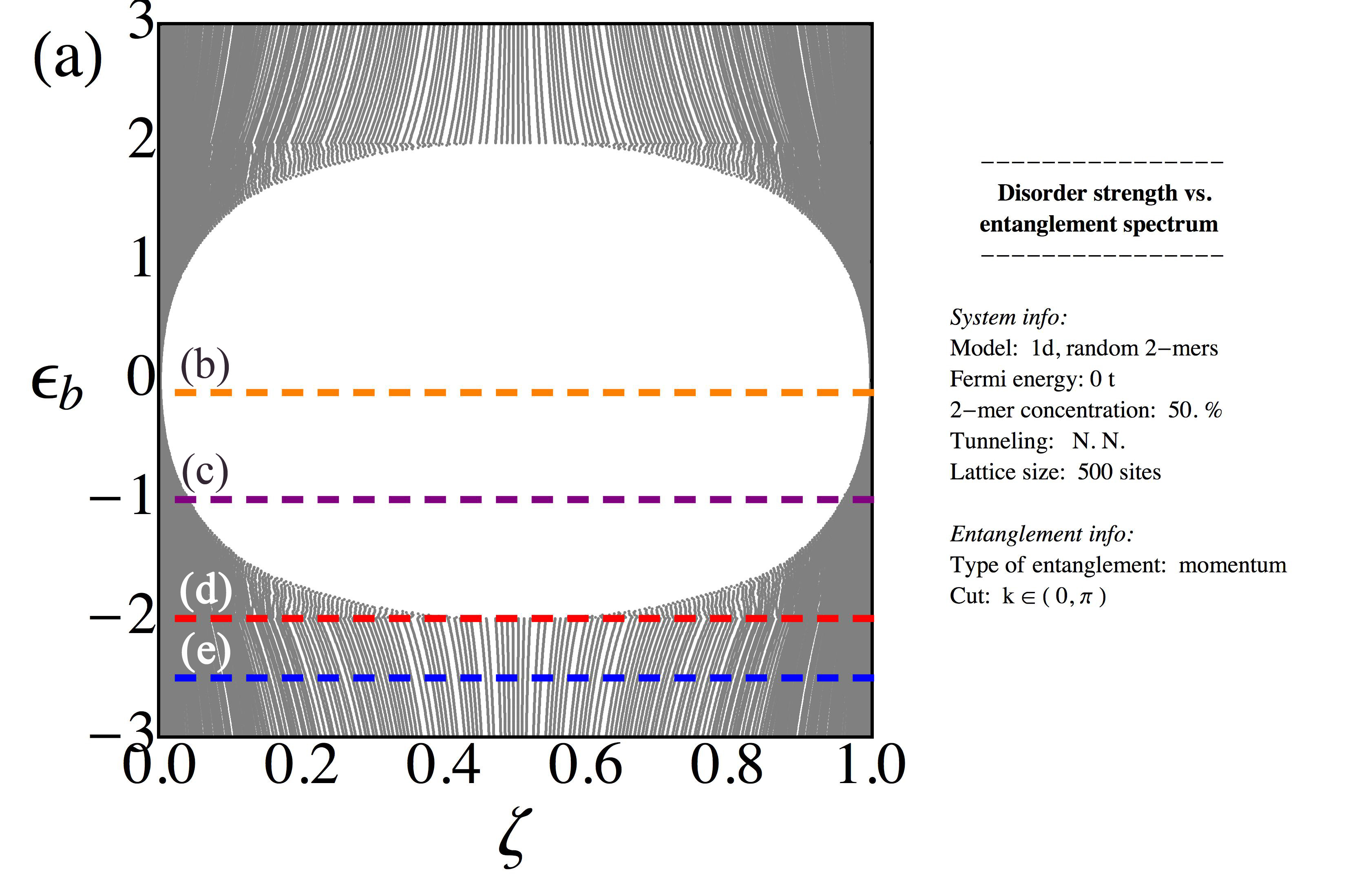}
\hspace{1cm}\includegraphics[trim = 0cm 0cm 34.5cm 0mm, clip, scale=0.08]{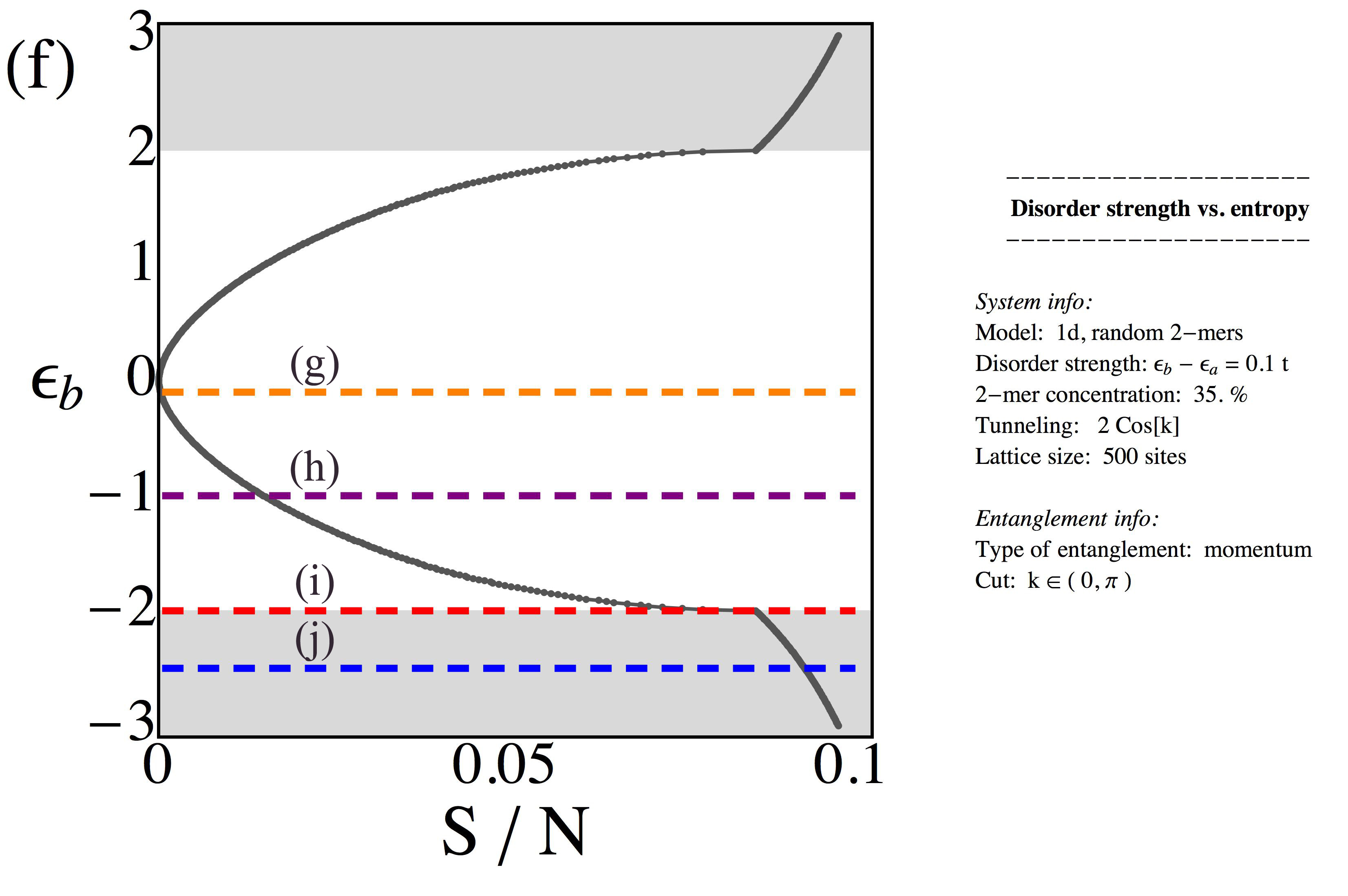}

\includegraphics[trim = 0mm 0cm 37cm 0mm, clip, scale=0.04]{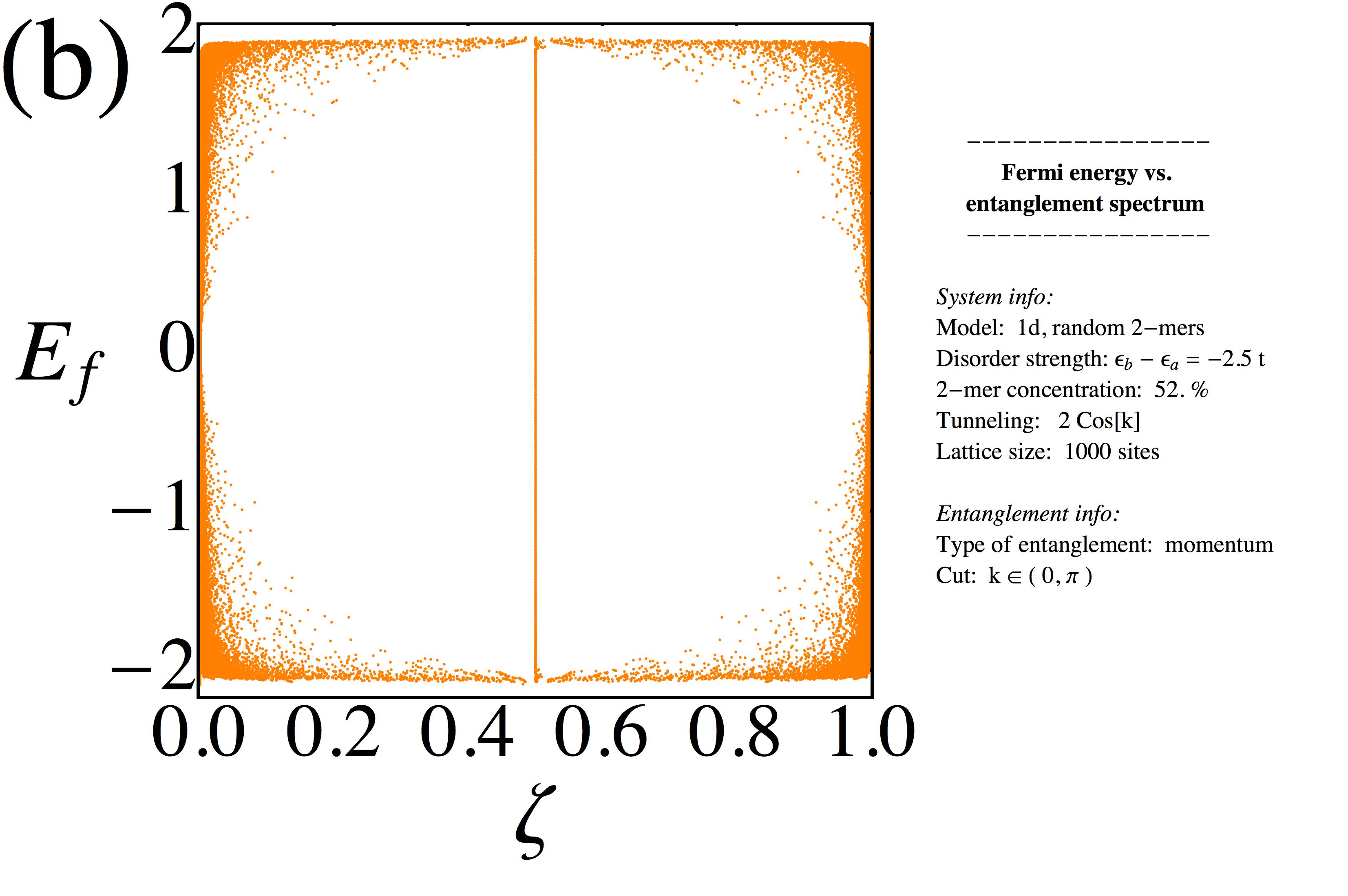}
\includegraphics[trim = 0mm 0cm 37cm 0mm, clip, scale=0.04]{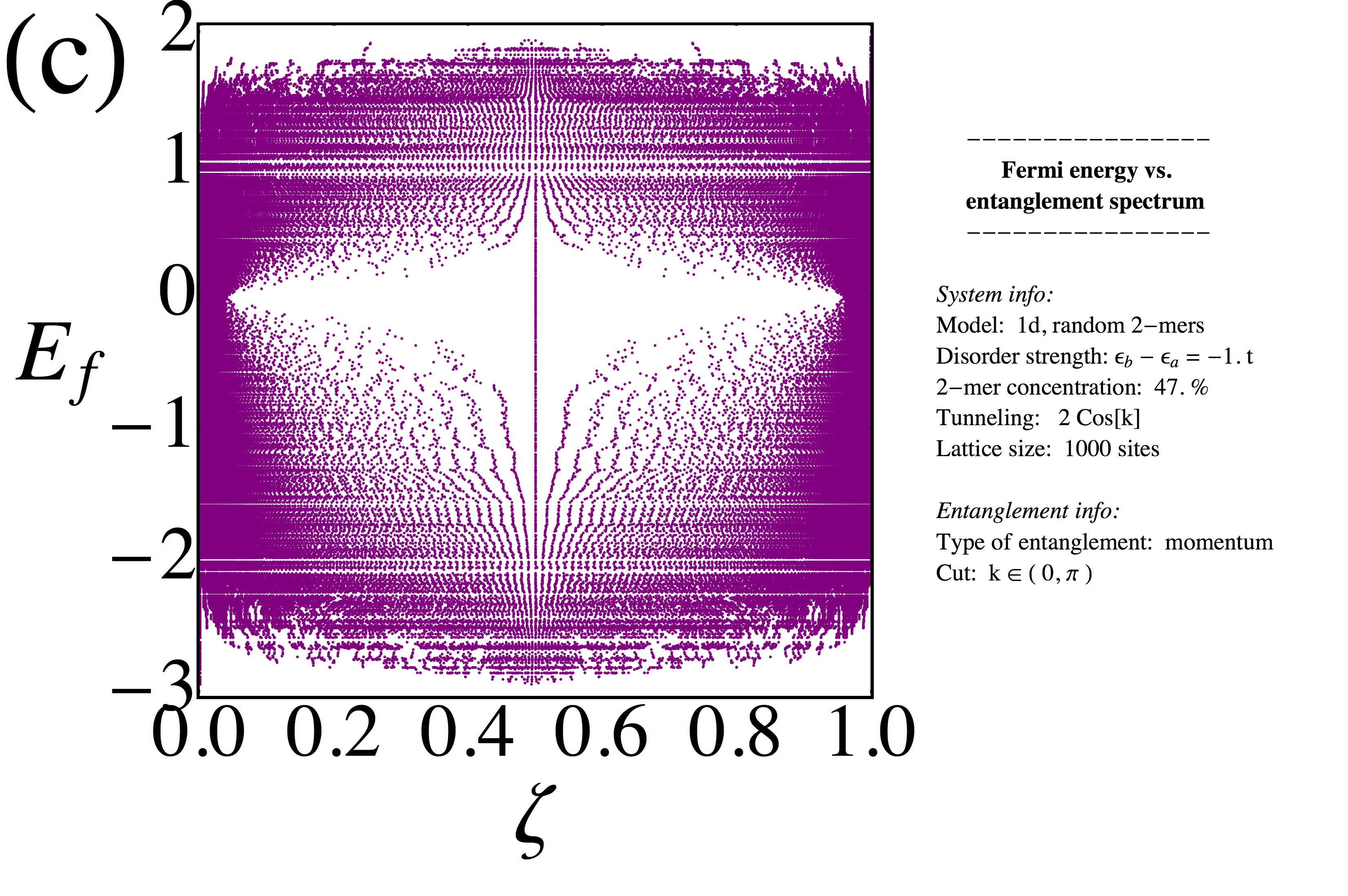}\hspace{1cm}
\includegraphics[trim = 0mm 0cm 37cm 0cm, clip, scale=0.04]{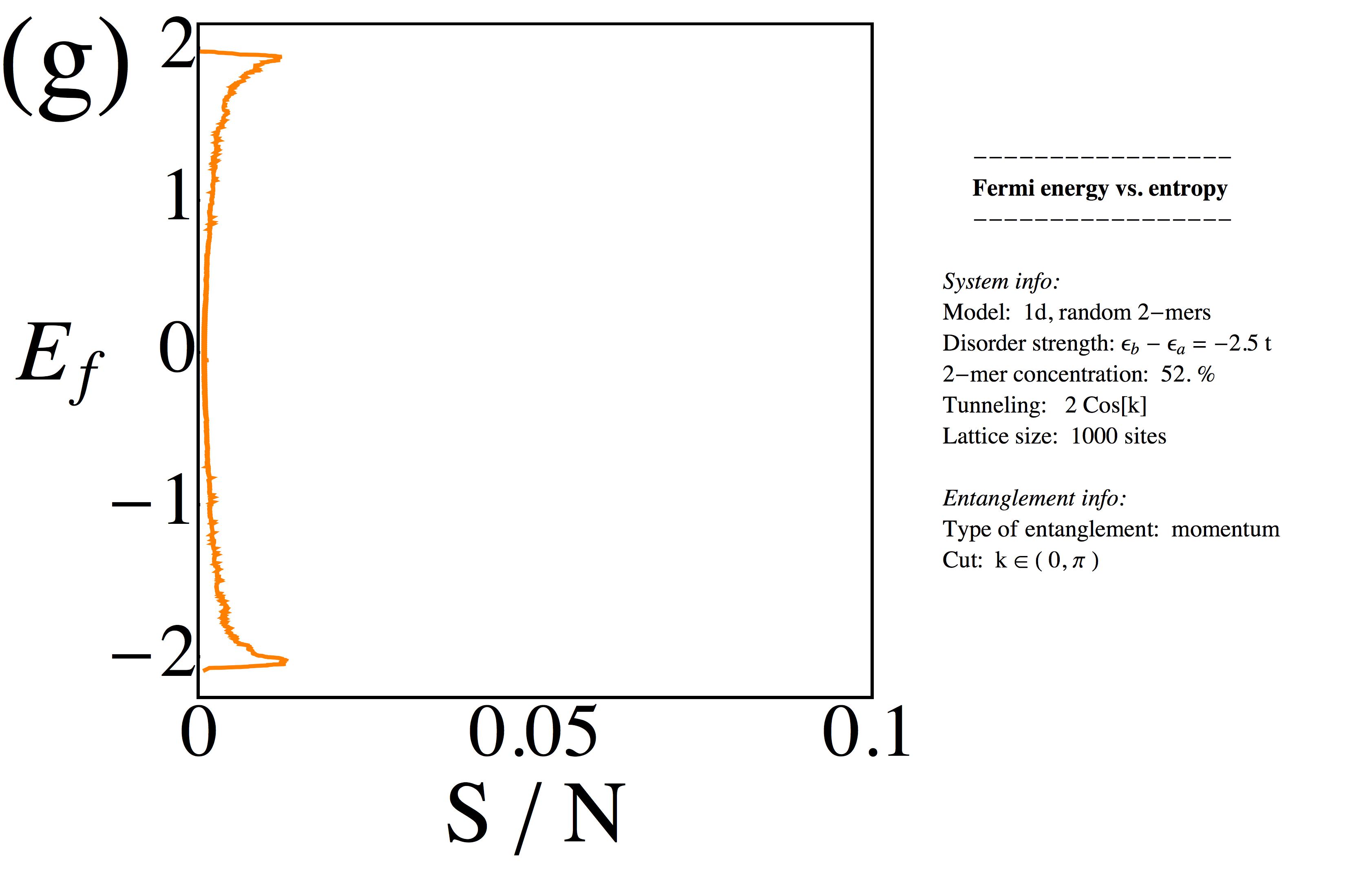}
\includegraphics[trim = 0mm 0cm 37cm 0cm, clip, scale=0.04]{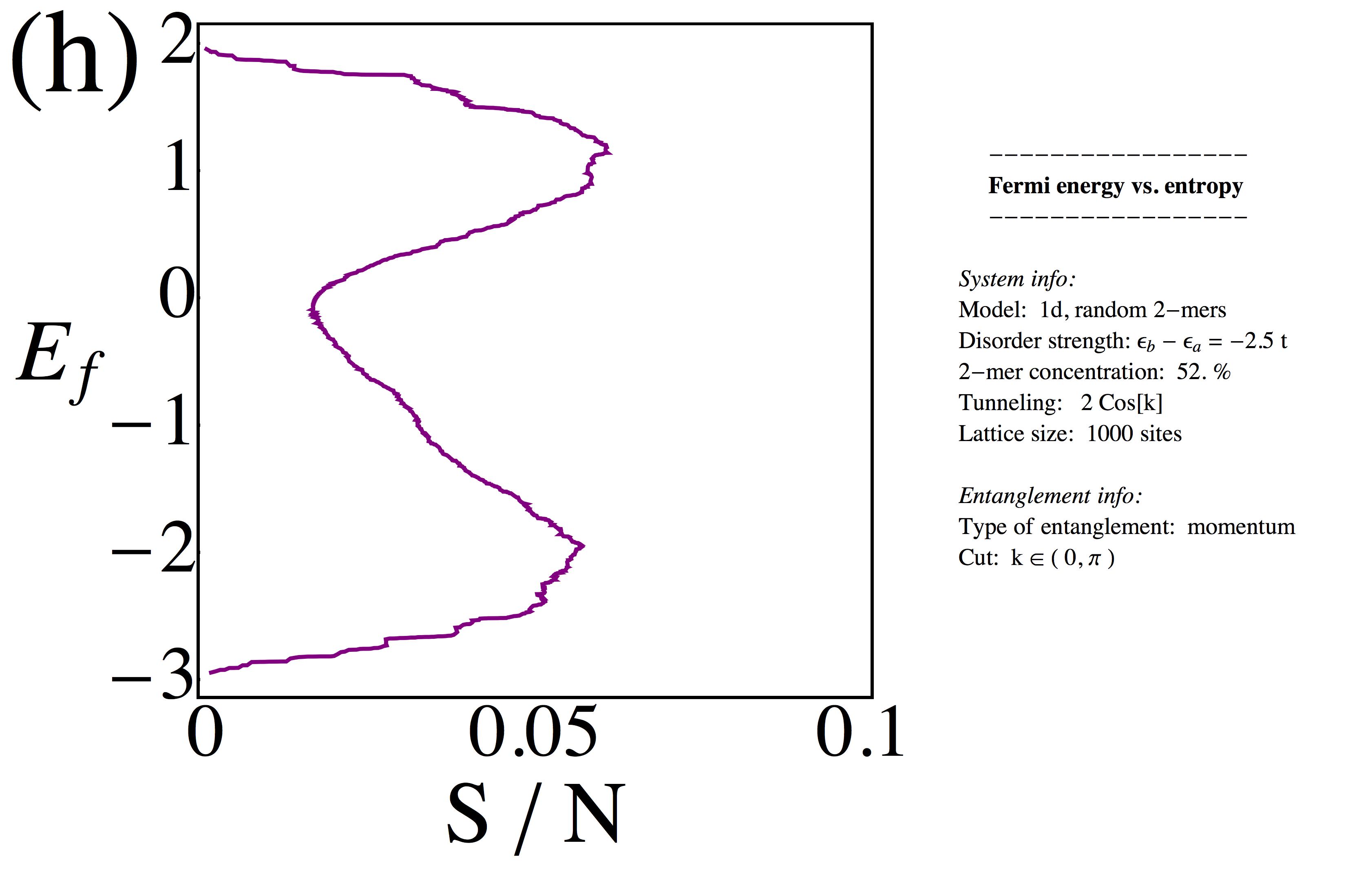}

\includegraphics[trim = 0mm 0cm 37cm 0cm, clip, scale=0.04]{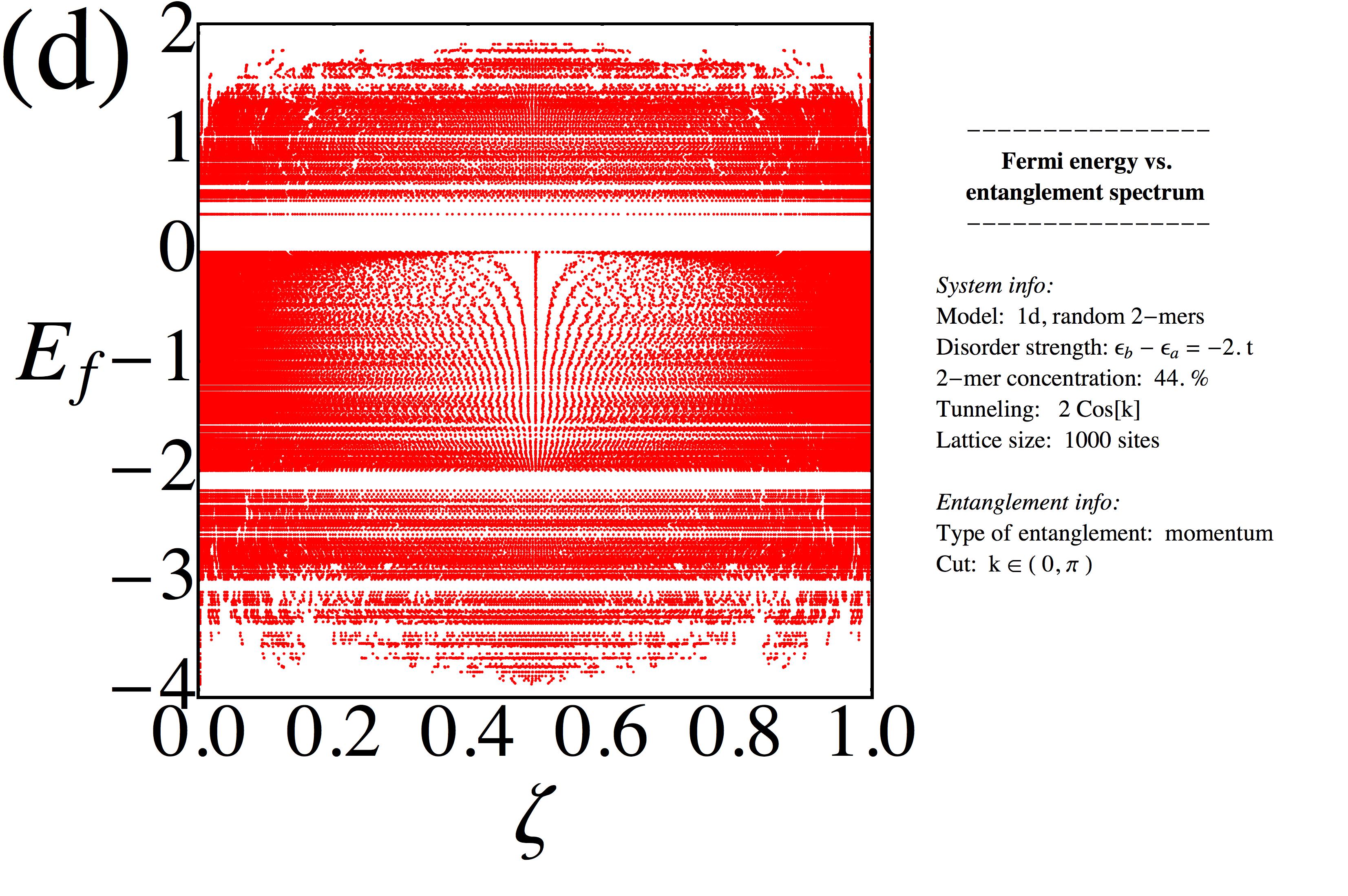}
\includegraphics[trim = 0mm 0cm 37cm 0mm, clip, scale=0.04]{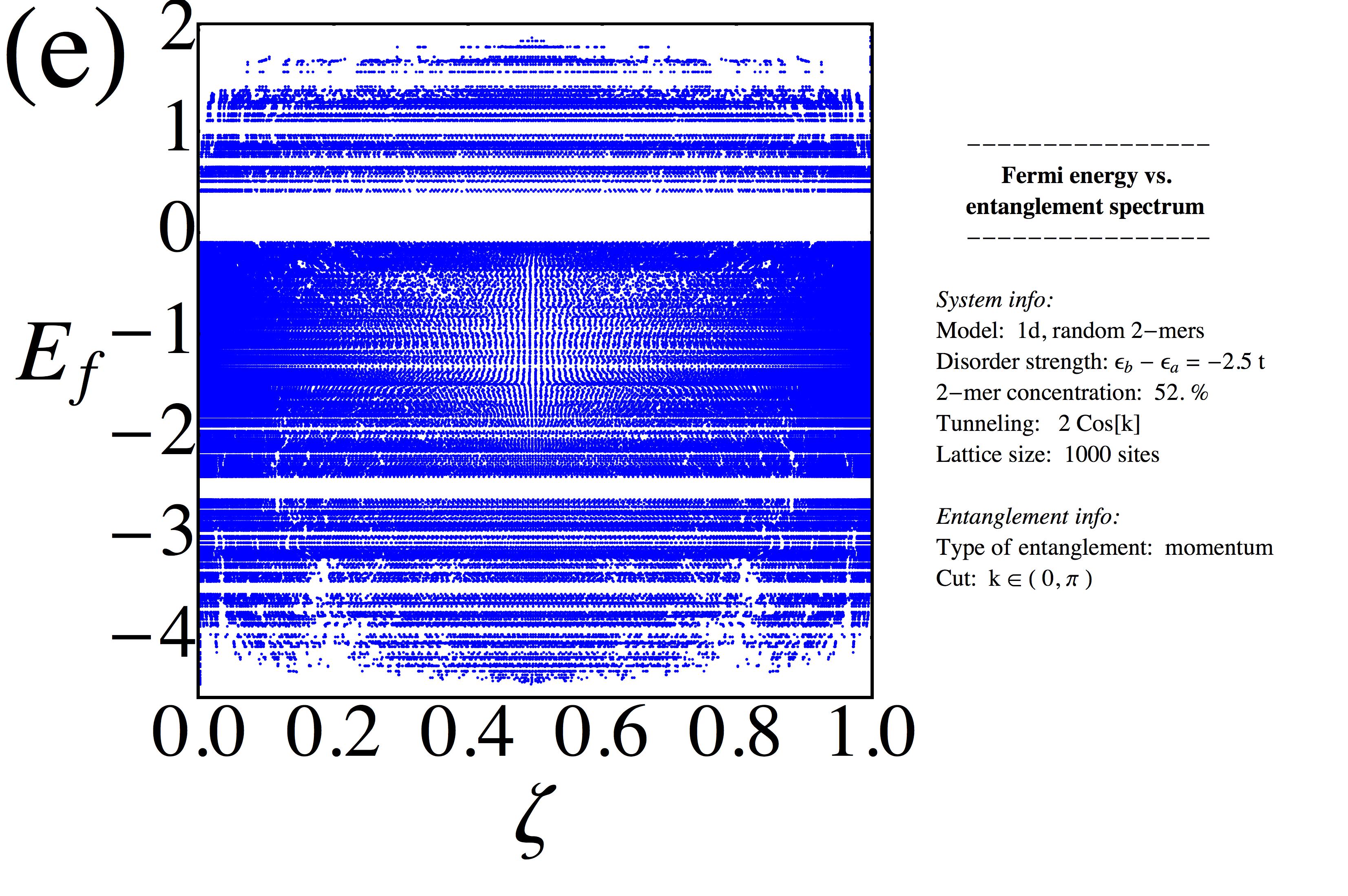}\hspace{1cm}
\includegraphics[trim = 0mm 0cm 37cm 0cm, clip, scale=0.04]{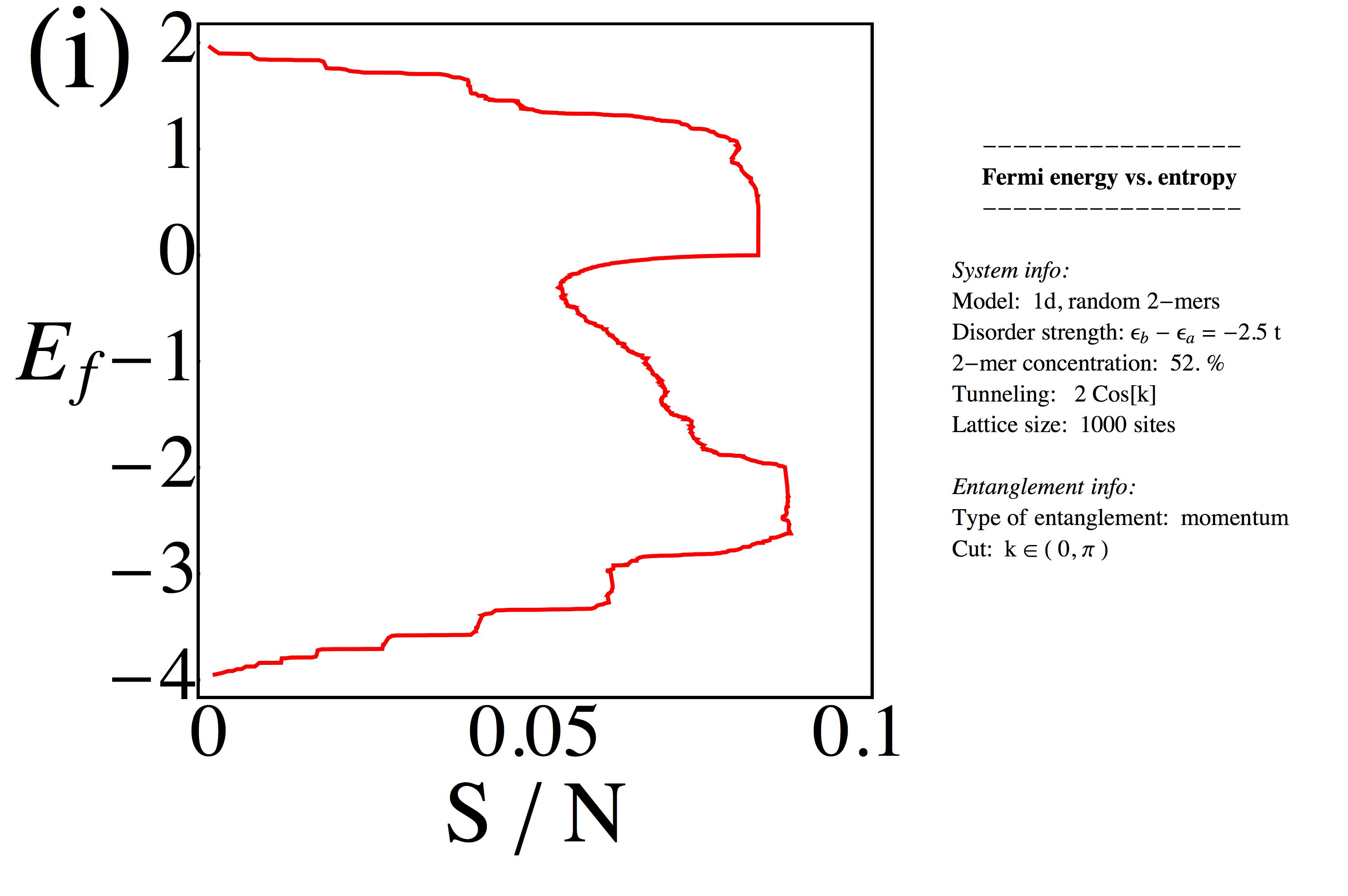}
\includegraphics[trim = 0mm 0cm 37cm 0cm, clip, scale=0.04]{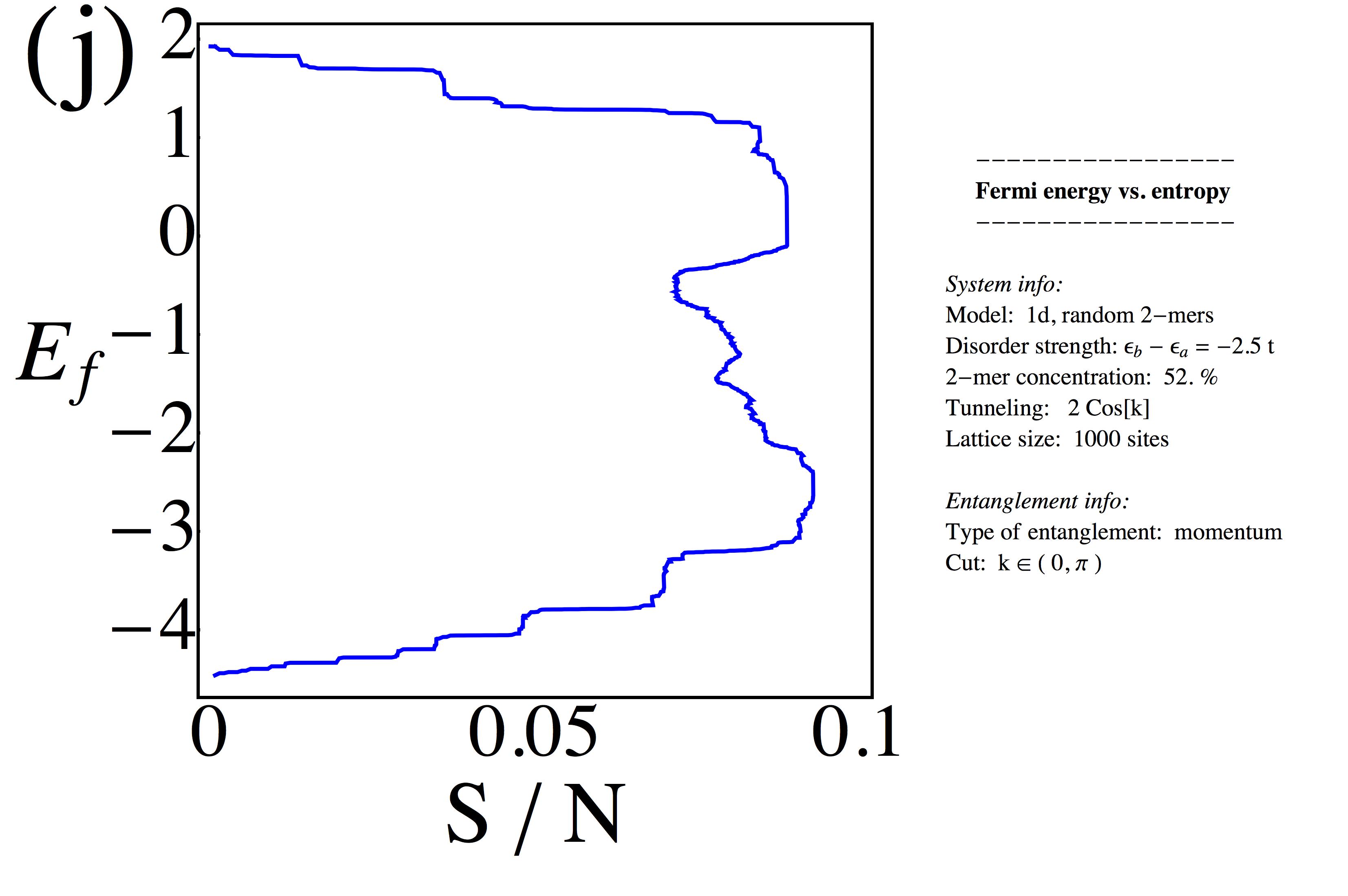}

\caption{Momentum entanglement spectrum and entropy of the random dimer model: The momentum entanglement spectrum is shown in (a) as a function of the disorder strength and (b) (c) (d) (e) as a function of the Fermi energy for different disorder strengths. The momentum entanglement entropy is shown in (f) as a function of the disorder strength and (g) (h) (i) (j) as a function of the Fermi energy for different disorder strengths. The different disorder strengths are correlated with the locations of the horizontal lines in (a), (f) and the colors of the lower plots correspond to the colored dashed lines in the top plots. For all of these plots, we used $N=1000$.}\label{Fig_set18}
\end{figure*}

Let us now proceed to discuss the momentum entanglement in these models. The momentum entanglement spectrum shows much sharper behavior than the spatial entanglement spectrum: in the metallic state the entanglement modes are largely clustered near $0$ and $1$, whereas they fill the region in between $0$ and $1$ as the Fermi level is moved away from resonance and into localized states. As we mentioned before, the scattering profiles of the RDM and RTM potentials exhibit exact zeroes in the Fourier components which are responsible for stabilizing the delocalized states of such models.  In particular, there is a clear connection between the existence of the metallic phase at finite disorder, and the behavior of its corresponding momentum entanglement. Since momentum entanglement yields information about the build up of correlations between left and right movers, the fact that momentum entanglement is suppressed in the metallic state can be interpreted as a decrease of correlations between left and right movers, which thus preempts the Anderson insulator at these energies. This connection is natural, since the return probability is an important ingredient in the localization transition \cite{Phillips2012}.

Note that, in all of these examples, the momentum entanglement turns out to be sharply suppressed even though the occupied Fermi sea of the ground state contains single-particle states below the resonant energies that are actually localized. As we mentioned in a previous section, a single localized state by itself leads to an entanglement of order $\log 2$ for this type of cut. One would naturally expect that all of the entanglement from the entire set of occupied localized states would add up. However, the fact that there is not a large accumulation of entanglement when $E_f$ is at resonance suggests that there is significant destructive interference in the correlation matrix similar to what we discussed for Eq. \ref{interference}.

To understand this observation, one can think of the zeroes of the scattering potential as a mechanism that prevents mixing between states in the energy spectrum that are separated by the resonant energies. For example, for the case of the RDM, the suppressed scattering in the neighborhood of $E=0$ is very effective in keeping states below and above this resonant energy from mixing with the opposite group at weak disorder. As a consequence, the set of states below zero energy (call this set $\Lambda$) hybridize primarily among themselves. Effectively, the absolute suppression at $E=0$ divides the spectrum into two pieces $E<0$ and $E>0$ that are not mixed for weak-disorder. This means that, upon constructing a Slater determinant with $E_f=0,$ we can approximate the single-particle states by $\gamma^{\dagger}_n \approx \sum_{k\in \Lambda}\alpha_{n}(k)c^{\dagger}_{k} $. As a result, there is a significant simplification in the form of the Slater determinant ground state
\begin{equation}
\ket{\Omega}\approx\prod_{n\in \text{occ.}}\gamma^{\dagger}_{n}\ket{0}= \prod_{n\in\text{occ.}}\left(\sum_{k\in \Lambda}\alpha_{n}(k)c^{\dagger}_{k} \right)\ket{0}=\prod_{k\in \Lambda}c^{\dagger}_k\ket{0}
\end{equation}\noindent where the last equality is true up to a global phase (which does not enter the calculation of the density matrix).
In other words, the Fermi sea will approximately reproduce the ground state of delocalized states of the clean system. In fact, while the energies of the states below $E_f$ will be affected by disorder, the same orbitals that were occupied in the clean-limit will be occupied in the weak-disorder limit.  In terms of the correlation matrix, when $E_f$ is set at resonance, this is equivalent to saying that
contributions from localized wave functions in the momentum correlation matrix interfere destructively, because for every $k$, $-k$ is included, and this leads to a matrix that is approximately diagonal in momentum space.

Parallel to this, we note that there are single momentum entanglement modes at $1/2$ in all of the examples we present here (see e.g., Fig. \ref{Fig_set18}b). Although it is hard to see by eye in the figures we present here, closer inspection reveals that these $1/2$ modes depend on whether the number of occupied states is even or odd. To understand this behavior, consider the weak disorder limit. In this case, the degenerate plane waves at $k$ and $-k$ will hybridize easily and form the states $\gamma^{\dagger}_{\pm}=\frac{1}{\sqrt{2}}\left(c^{\dagger}_{k}\pm c^{\dagger}_{-k}\right)$ which are split in energy by an amount that is linear in the disorder strength (though the effective disorder strength vanishes if $2k=\mathcal{Q}_n(m)$). Now, if the ground state only has one of these excitations occupied, say the state $\ket{\psi}=\gamma_{-}^{\dagger}\ket{0}$, then the momentum entanglement spectrum will automatically have a $1/2$ mode. However, if the ground state has both states filled, then we obtain the factorized state $\ket{\psi}=\gamma_{+}^{\dagger}\gamma_{-}^{\dagger}\ket{0}=c^{\dagger}_{k}c^{\dagger}_{-k}\ket{0}$, which has zero momentum entanglement. By successively filling the hybridized $\gamma^{\dagger}_{\pm}$ states, we will thus obtain a sequence of $1/2$ modes for an odd number of occupied states regardless of the scattering properties of the disorder potential. Since the hybridization is immediate for the degenerate modes, even for extremely small disorder,  they will appear as soon as the disorder is increased from zero. Eventually, when the disorder is strong enough to make scattering between $k$-states at different energies play a more important role, such entanglement modes are obscured because of new entanglement modes entering from the edges near $0$ and $1$ to fill the region near $1/2$.

After our discussion of the effects of tuning $E_f$ we now turn to the description of the metal-insulator transition as the disorder strength $\epsilon_b$ is varied to tune the transition. We take the RDM as an example and set the Fermi energy at its resonant value $E_f=0$.  Fig. \ref{Fig_set18}a,f shows the momentum entanglement spectrum and entropy, respectively as a function of $\epsilon_b.$  Fig. \ref{Fig_set18}b,c,d,e shows the entanglement spectrum as a function of Fermi energy for a few selected disorder strengths referenced by dashed lines in figures Fig. \ref{Fig_set18}a. We show a similar array of figures for the momentum entanglement entropy calculated from these same entanglement spectra in Fig. \ref{Fig_set18}f as a function of disorder strength and Fig. \ref{Fig_set18}g,h,i,j as a function of Fermi energy for different disorder strengths.

\begin{figure}[b]
\begin{center}
\includegraphics[trim = 0mm 0cm 4cm 0mm, clip, scale=0.2]{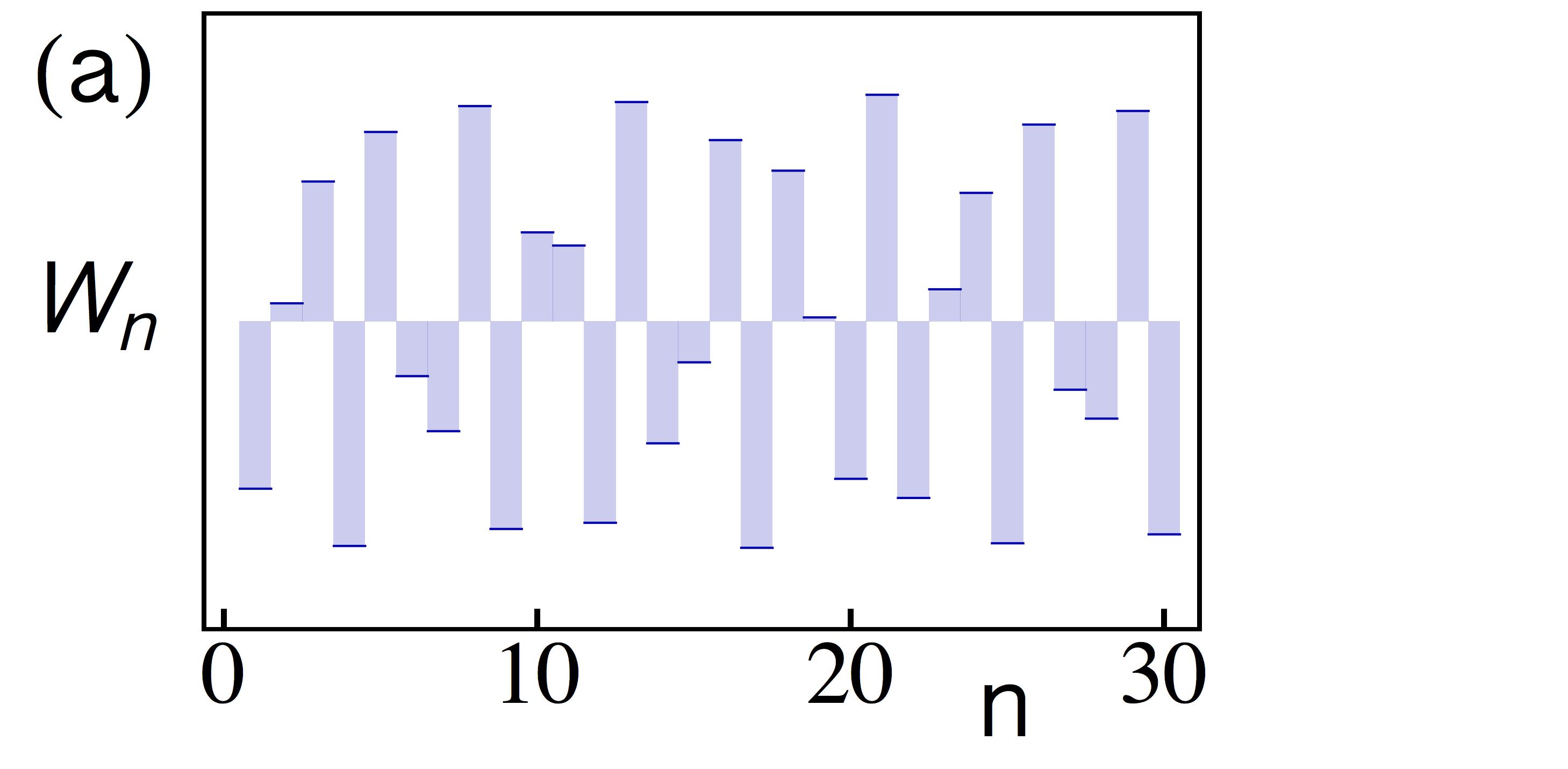}
\includegraphics[trim = 0mm 0cm 5cm 0cm, clip, scale=0.2]{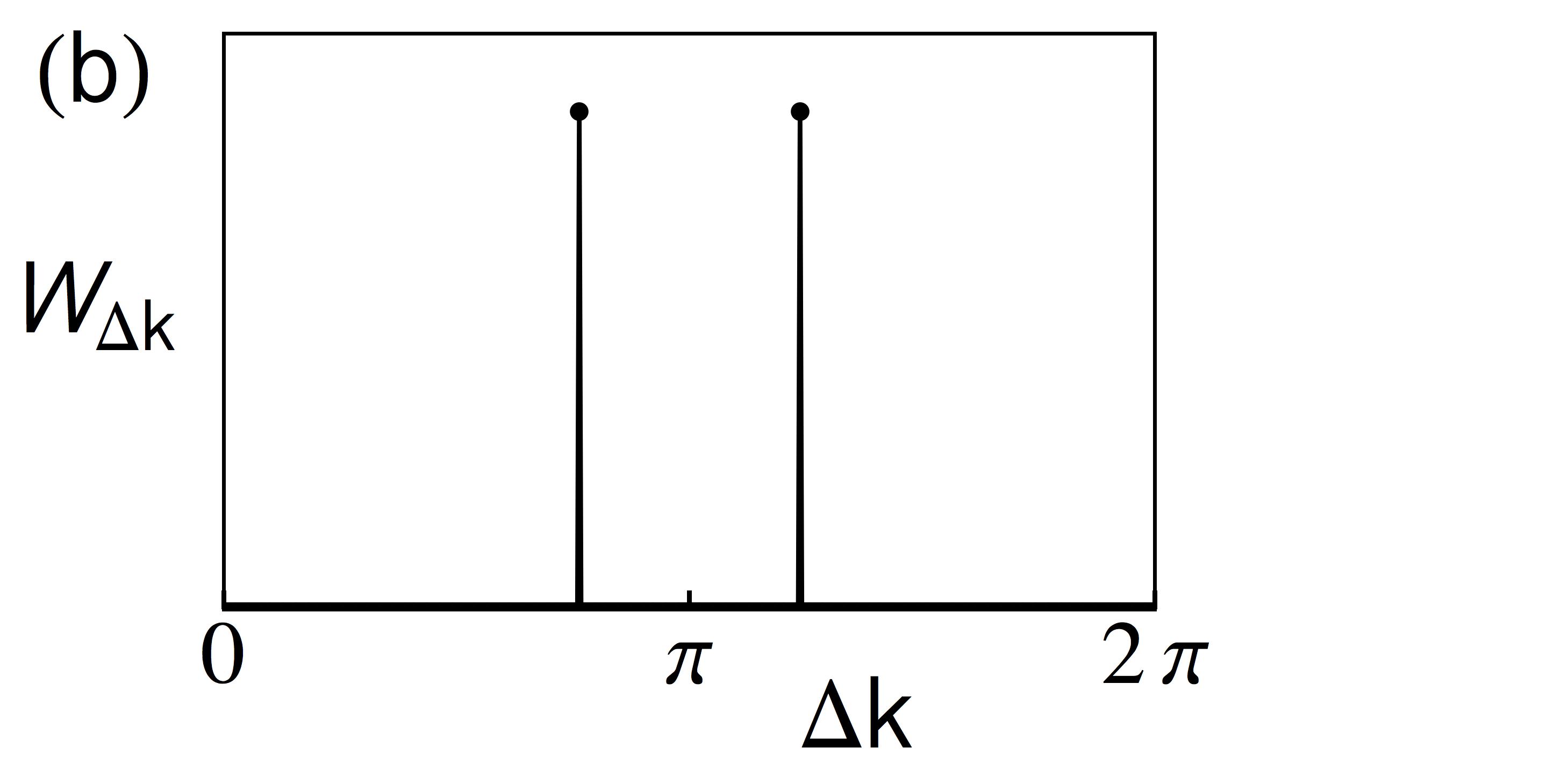}

\includegraphics[trim = 0cm 0cm 0.5cm 0mm, clip, scale=0.18]{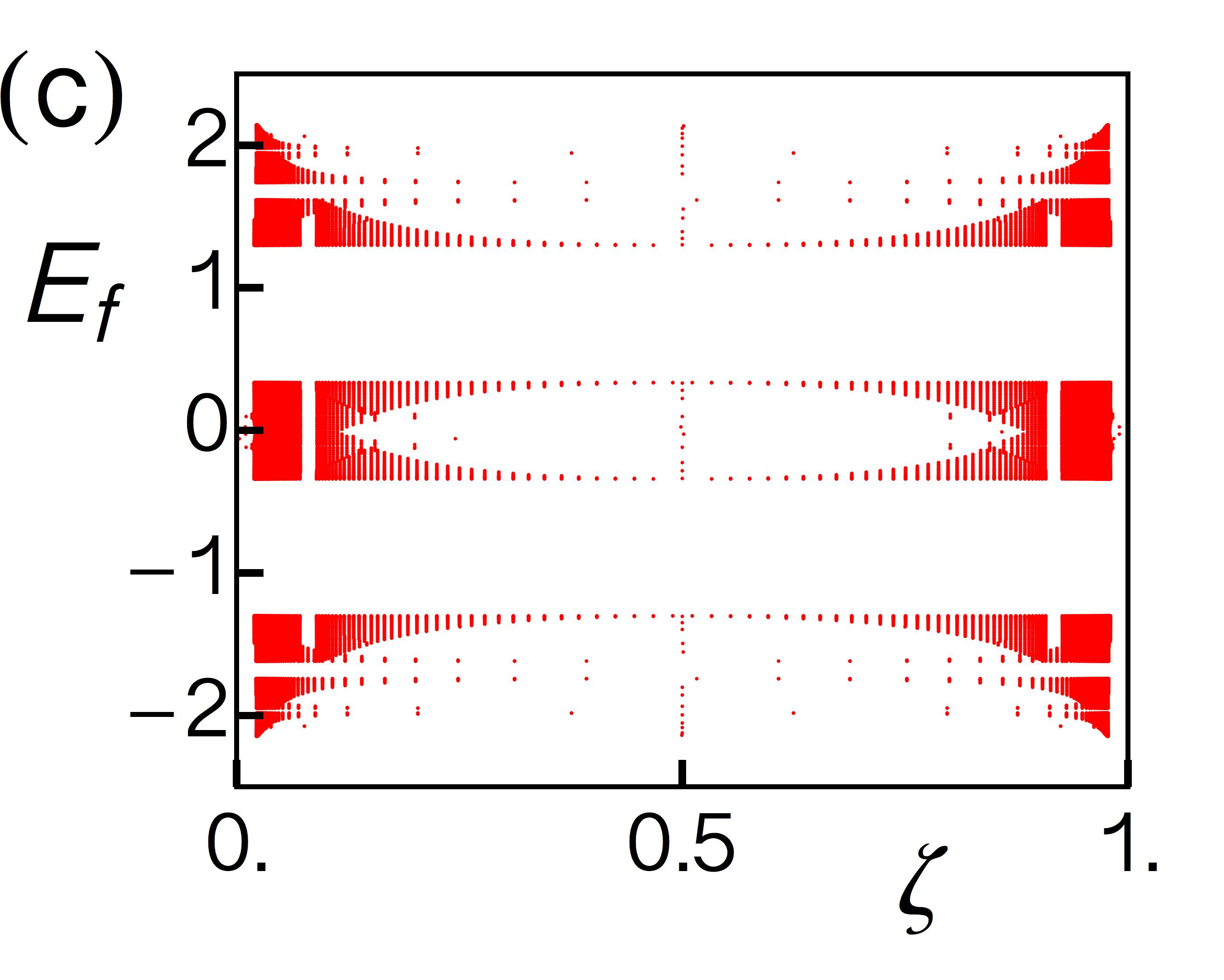}
\includegraphics[trim = 0cm 0.8cm 0cm 0mm, clip, scale=0.225]{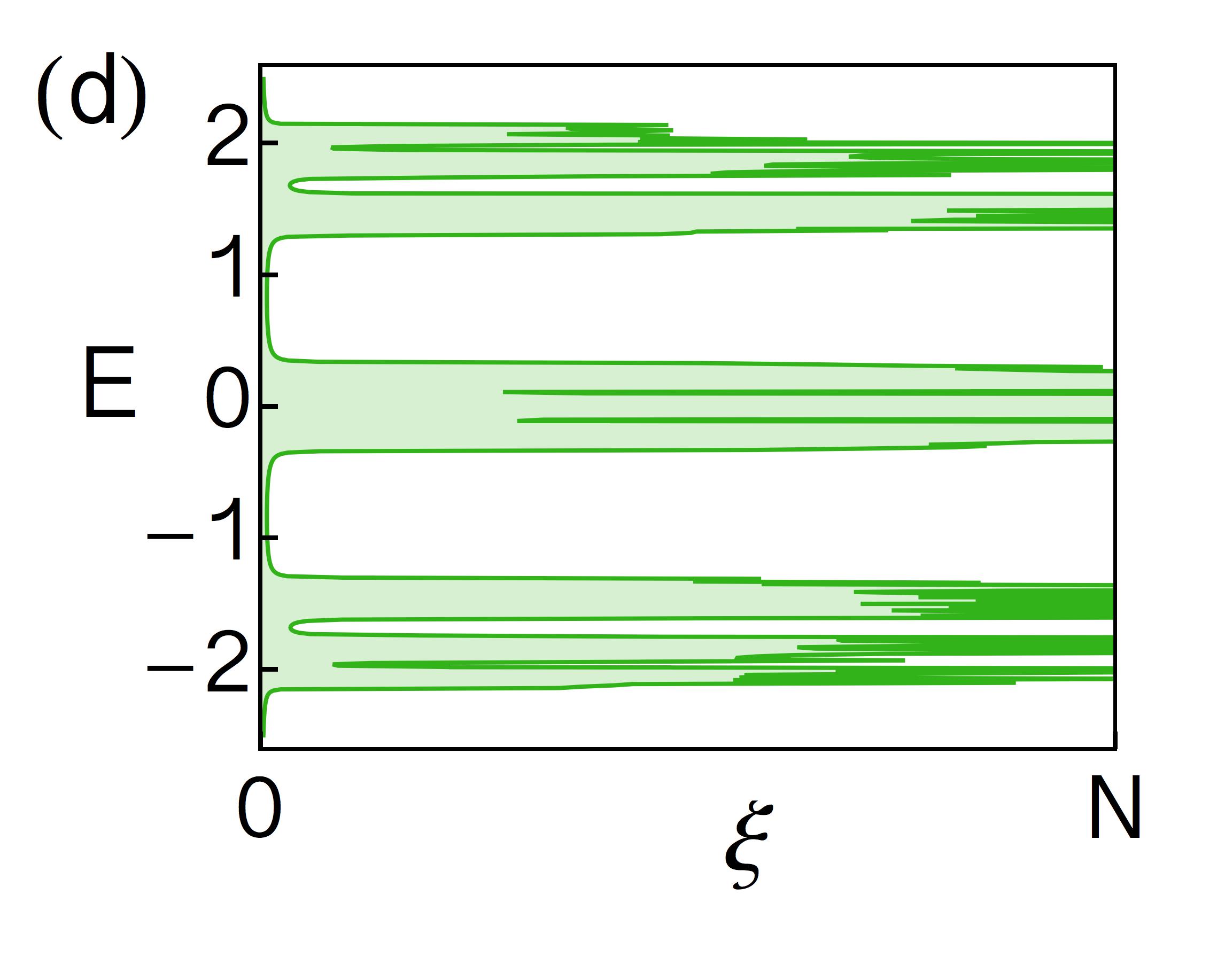}
\caption{(a) Aubry-Andr\'e potential in position space. (b) The corresponding Fourier components. (c)  Momentum entanglement spectrum as a function of the Fermi energy. (d) Localization length as a function of energy. Subfigures (c), (d) are for $W=1$.}\label{AApotential}
\end{center}
\end{figure}

From Fig. \ref{Fig_set18}a we see that, as the disorder strength $\epsilon_b$ is varied, the momentum entanglement modes can change quite drastically. Near the clean limit $(\epsilon_b=0)$ there is minimal entanglement and the eigenvalues are accumulated near $1$ and $0$, as expected. As $\epsilon_b$ increases, the entanglement modes begin to spread out smoothly from $0$ and $1$ toward $1/2$, which indicates an increase in momentum entanglement. Correspondingly, the entanglement entropy grows smoothly until the point where the delocalized states disappear $(\epsilon_b=2).$ At this critical disorder strength, a number of entanglement modes reach $1/2$ and there is a discontinuous change in the slope of the entanglement entropy as a function of $\epsilon_b$. This seems to indicate that the Anderson insulator transition occurs when there is a saturation of momentum entanglement modes near $1/2$ followed by a slowdown in the rate of increase of momentum entanglement. This qualitative change suggests itself as a way to characterize the Anderson localization transition. We note that all of this structure is apparent without having to perform disorder averages, as all of these figures are for a single disorder realization. The qualitative change in the entanglement entropy could be useful, and efficient, to determine the phase diagram of systems that undergo Anderson transitions. We will explore this possibility further in later sections.

\subsection{Quasiperiodic potential}

\subsubsection{The Aubry-Andr\'e model}

We will now move on to discuss another special 1D free-fermion model, namely the Aubry-Andr\'e model (AAM). The Hamiltonian for this system consists of one-dimensional fermions in a quasicrystal potential and has the form
\begin{equation}
H= -\sum_{m=1}^{N} t \left( c^{\dagger}_{m+1}c_{m}+ c^{\dagger}_{m}c_{m+1}\right)+\sum_{m=1}^{N} W_m c_m^{\dagger} c_m, \label{RD}
\end{equation}
where the Aubry-Andr\'e potential is given by
\begin{equation}
W_n=W\cos\left(2\pi \alpha n\right).
\end{equation}
Here, $\alpha$ is conventionally chosen to be an irrational number, usually the golden ratio. The main property of such a quasicrystal is that it leads to a localization transition when the potential strength is tuned past a known value. This can be shown by comparing the spatial and momentum space versions of the Hamiltonian and noting that both models are dual to each other: low values of $W$ lead to localization in momentum space whereas high enough values of $W$ lead to localization in position space. The self-dual point is consequently $W_c=2t$, which is the critical point (we will usually set $t=1$). With the potential turned off the system is simply a single-band tight-binding Hamiltonian with the conventional dispersion $E(k)=-2t\cos k.$ For this dispersion the states for $k\in (-\pi,0)$ are left-movers and those from $k \in(0,\pi)$ are right-movers. Thus, when we perform a momentum space partition we will pick region A to be one of these subspaces, and region B to be the other so that we separate the left-movers from the right-movers.
Additionally, there is one important subtlety to note: since in a finite lattice with periodic boundary conditions we must have $W_{n+N}=W_{n}$, we have to approximate $\alpha$ using a fraction of relative prime numbers. For this work, we will use a lattice size $N=102$ for the non-interacting calculations, which means that the best approximation to the golden ratio using periodic boundary conditions is $\alpha=167/102\approx	1.637$. For the interacting case we use $N=14$ at half-filling, and choose $\alpha=24/14\approx1.64$.

To compare with our previous analysis for the random $n$-mer models, the potential profile and the corresponding Fourier components of the Aubry-Andr\'e  potential are shown in Figs. \ref{AApotential} a,b.  Furthermore, in Figs. \ref{AApotential}c,d we show the momentum entanglement spectrum as a function of Fermi energy as well as the localization length as a function of energy, both for the case $W=1$. The energy gaps that we see in  Fig. \ref{AApotential}c correspond to actual gaps in the single-particle energy spectrum of the system. These gaps are created by the nonzero scattering elements of the disorder potential shown in Fig. \ref{AApotential}b. Within each of the resulting energy bands the scattering is suppressed, which leads to delocalized states in each band. This is evidenced by the divergence of the localization length in each band. In this sense, the AAM shares some similarities with the n-mer models. There is, however, an important difference: it is a well-known property of the AAM that \emph{all} single-particle states remain delocalized when $W<W_c$, which means that there is no mobility gap and thus no mobility edge in the spectrum. It is only when $W$ crosses the critical point that \emph{all} single-particle states in the energy spectrum become localized. 

\begin{figure}
\includegraphics[trim = 0mm 0cm 0cm 0mm, clip, scale=0.18]{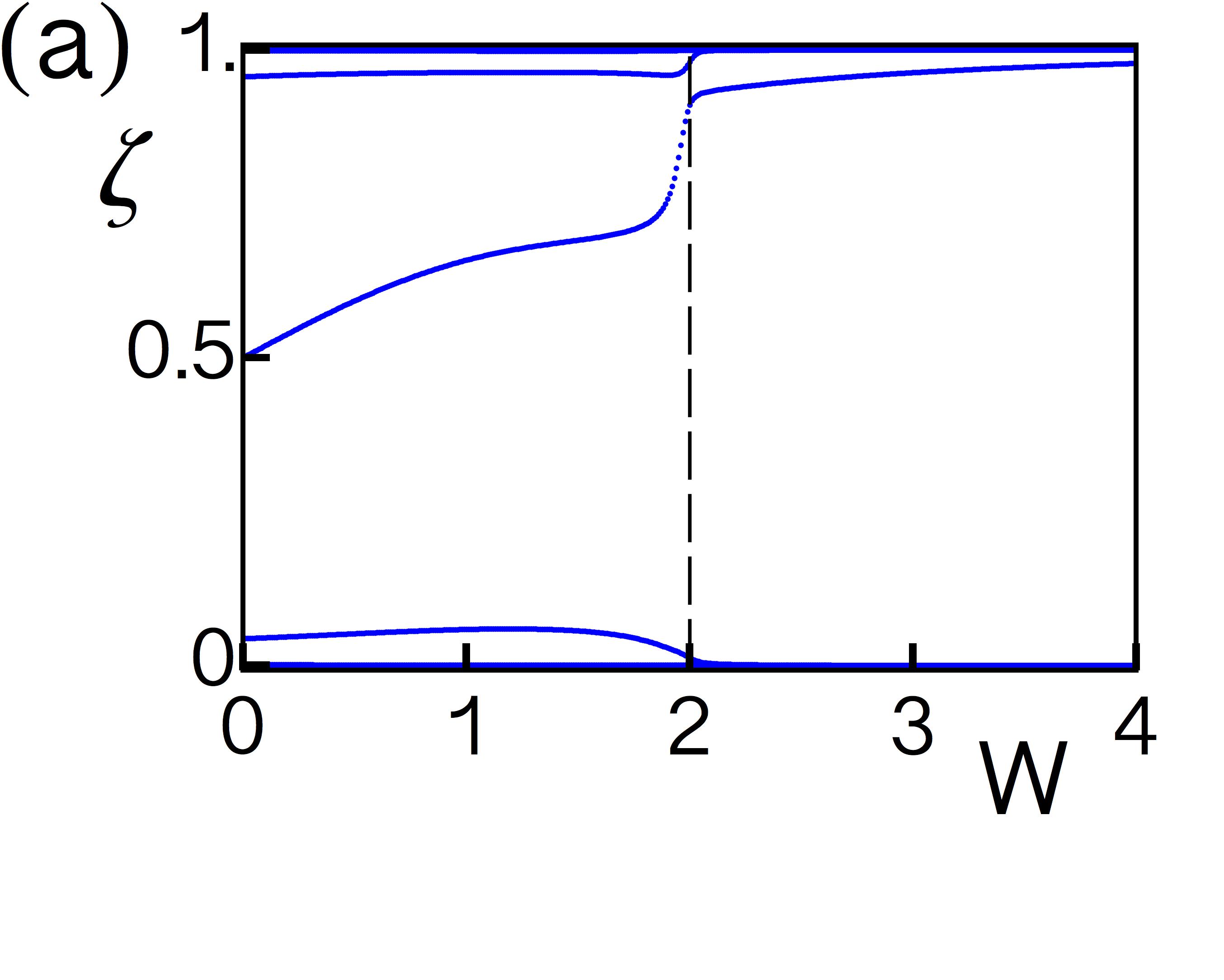}
\includegraphics[trim = 0mm 0cm 0cm 0mm, clip, scale=0.18]{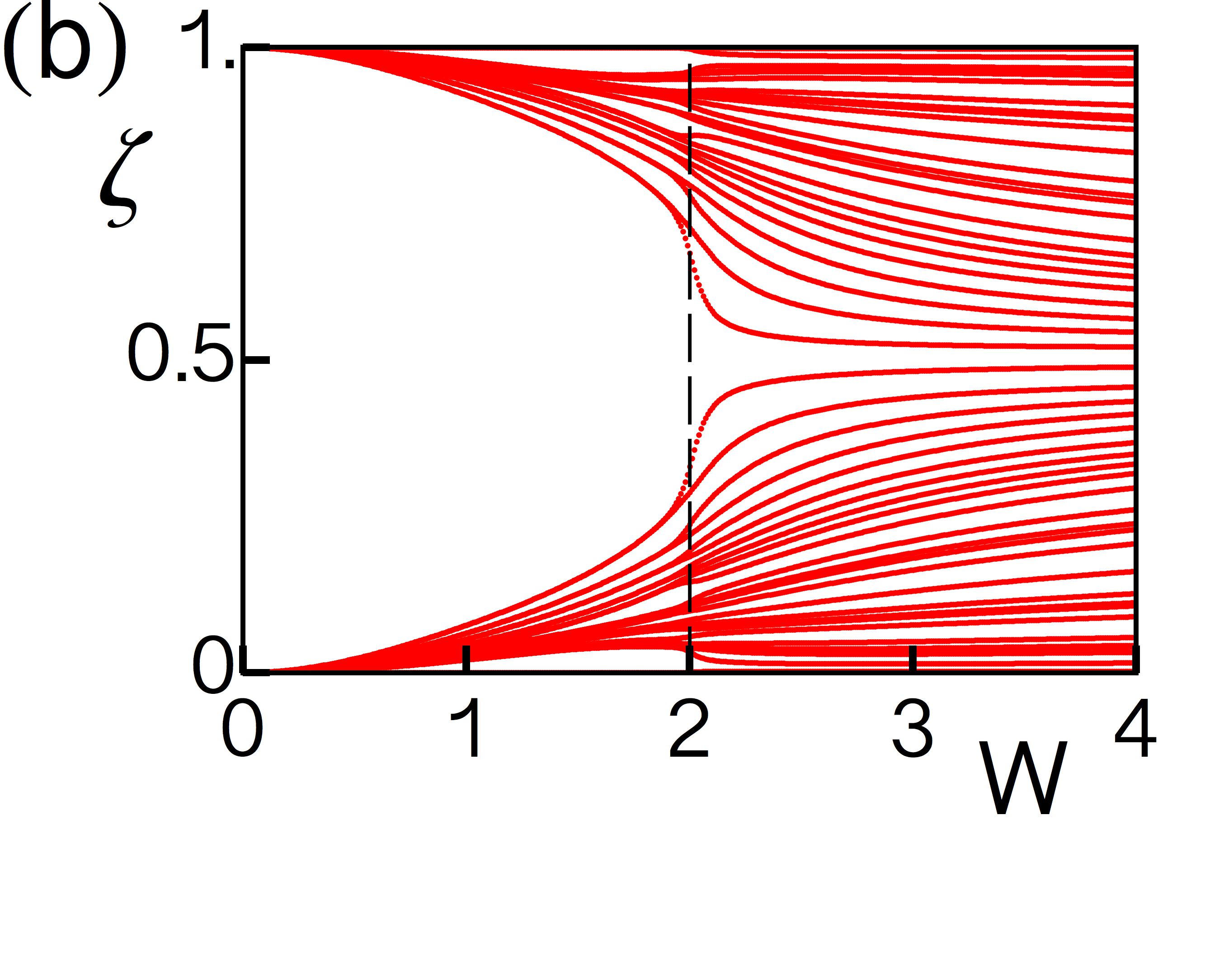}

\includegraphics[trim = 0mm 0cm -0.5cm 0mm, clip, scale=0.18]{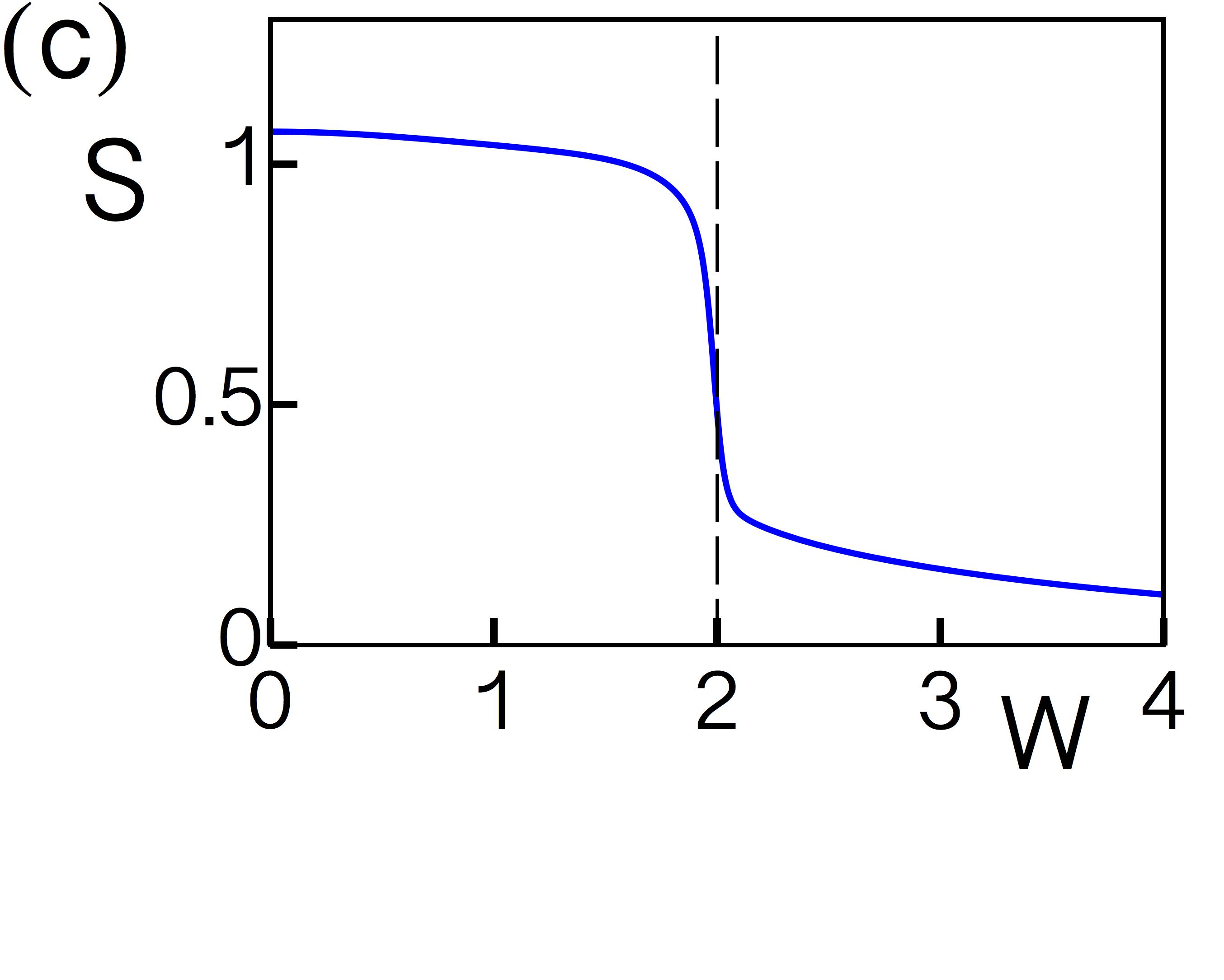}
\includegraphics[trim = 0mm 0cm -0.5cm 0mm, clip, scale=0.18]{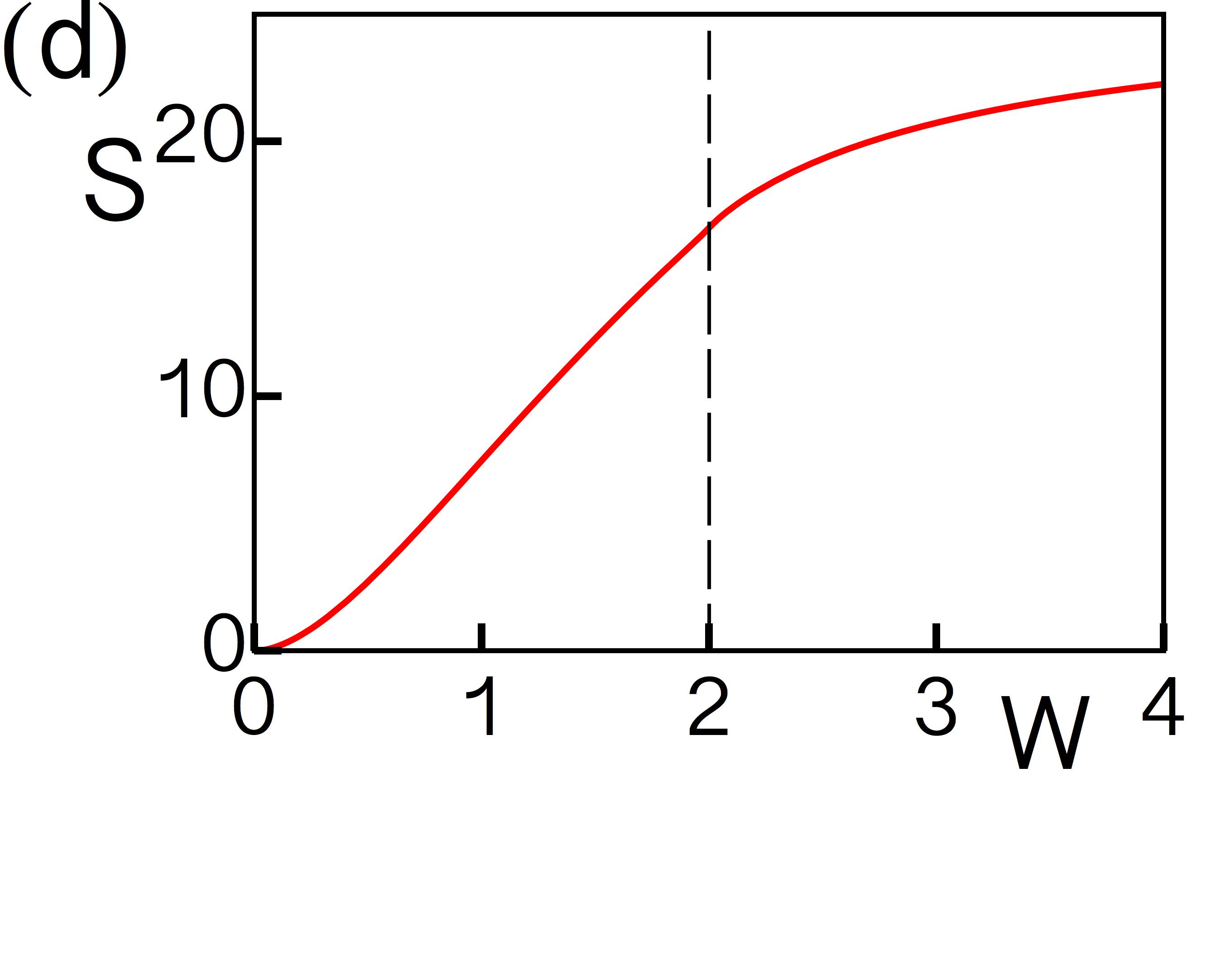}

\includegraphics[trim = 0mm 0cm 0cm 0mm, clip, scale=0.186]{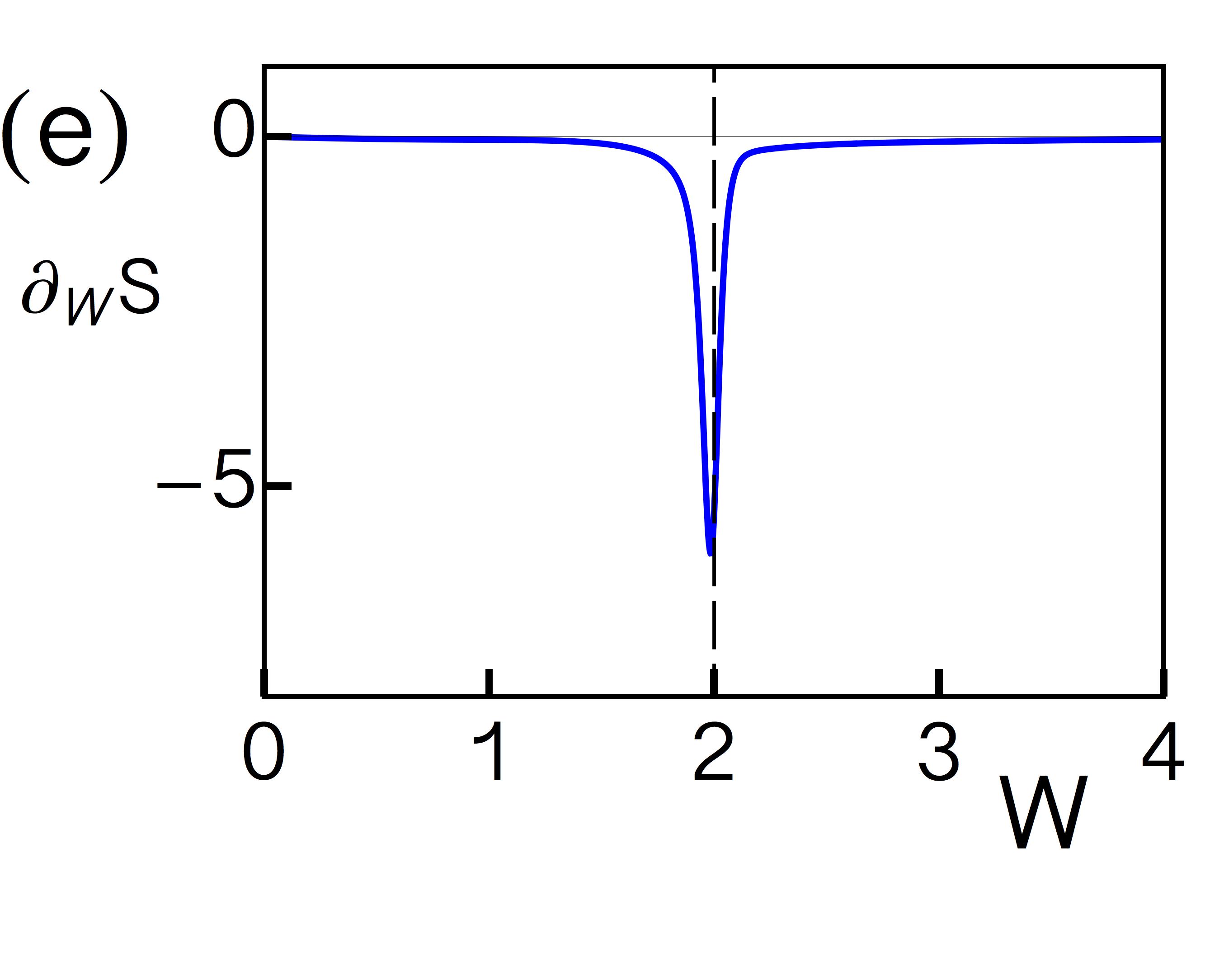}
\includegraphics[trim = 0mm 0cm 0cm 0mm, clip, scale=0.186]{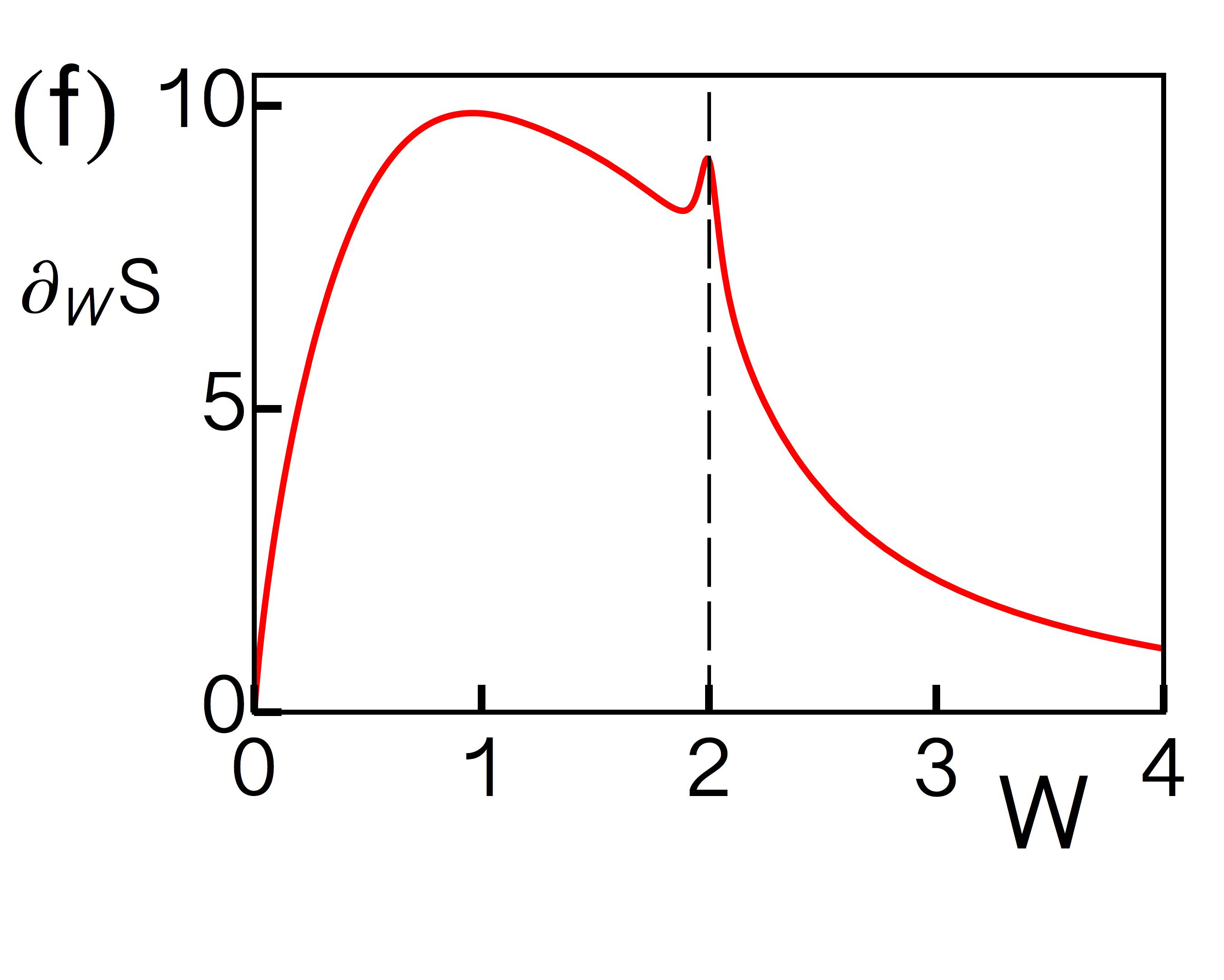}
	
\caption{Entanglement of the Aubry-Andr\'e model: (a) and (b) show the momentum and spatial entanglement spectrum as  a function of the AAM potential height; (c) and (d)  show the corresponding momentum and spatial entanglement entropy; (e) and (f)  show the derivative of the  momentum and spatial entanglement entropies with respect to $W$, both showing peaks at the phase transition when $W_c=2.$}\label{Fig_set12}
\end{figure}

An additional goal of this section is to use the AAM to explore the effect of interactions on the momentum entanglement at the localization transition. Knowing the location of the critical point ($W_c$) in the noninteracting system will simplify the identification of the signatures in the entanglement which are relevant to the localization transition; and it is known that the AAM exhibits a localization transition even in the interacting case  \cite{Shankar2013}. Interestingly, we will find that both spatial and momentum entanglement are still viable tools for identifying this transition, even though it is not at all obvious that the useful momentum-entanglement characteristics will survive when interactions are turned on. However, before we move on to the interacting case, let us narrow down some qualitative features of the entanglement of the non-interacting AAM that will be useful for an interacting generalization.

\subsubsection{Signatures of the phase transition in the non-interacting Aubry-Andr\'e model}

We now consider signatures of the disorder-induced phase transition in the entanglement of the system. In Fig. \ref{Fig_set12}, we show both the spatial and momentum entanglement spectra (Fig. \ref{Fig_set12}a,b) and entanglement entropies (Fig. \ref{Fig_set12}c,d) as a function of disorder ($W$) for $E_f=0$. For this particular model, there is a clear signature of the localization transition in the entanglement spectrum for both types of entanglement cuts. 

The spatial entanglement shows a pattern of modes very similar to the case of a translationally invariant system for weak potential strength. That is, in the translationally invariant limit, or for weak disorder, most of the modes are clustered near 0 and 1 except a few that have peeled off from the unentangled ``band-edges" and additionally O(1) modes that lie in the mid-gap region. Toward the critical potential value, the entanglement modes start to deviate sharply toward one and zero, indicating that the eigenstates localize in the system in a sharp manner at the transition point. The spatial entanglement entropy appears to be discontinuous at the transition, and the first derivative of the entanglement entropy clearly shows a sharp peak precisely at $W_c$ as evidenced in Fig. \ref{Fig_set12}e.
 
The momentum entanglement spectrum, for weak potential strength, is largely clustered toward one and zero in the delocalized state with no mid-gap modes, and has very little entropy. As the potential strength is increased toward the transition point, the entanglement modes gradually start spreading toward a value of $1/2$. At the transition point, there is a sharp increase of the density  of mid-gap  entanglement modes near 1/2, very reminiscent of what we observed in the random n-mer models. The first derivative of the momentum entanglement entropy correspondingly shows a maximum at the transition point as evidenced in Fig. \ref{Fig_set12}f.

Since we know the exact location of the critical point, we can try to use the non-interacting signatures near $W=W_c$ (e.g., the trends of the spatial and momentum entanglement entropy) to help identify a localization transition when interactions are included. We turn to this case now. 

\begin{figure}
\includegraphics[trim = 0cm 3cm 3cm 0cm, clip, scale=0.205]{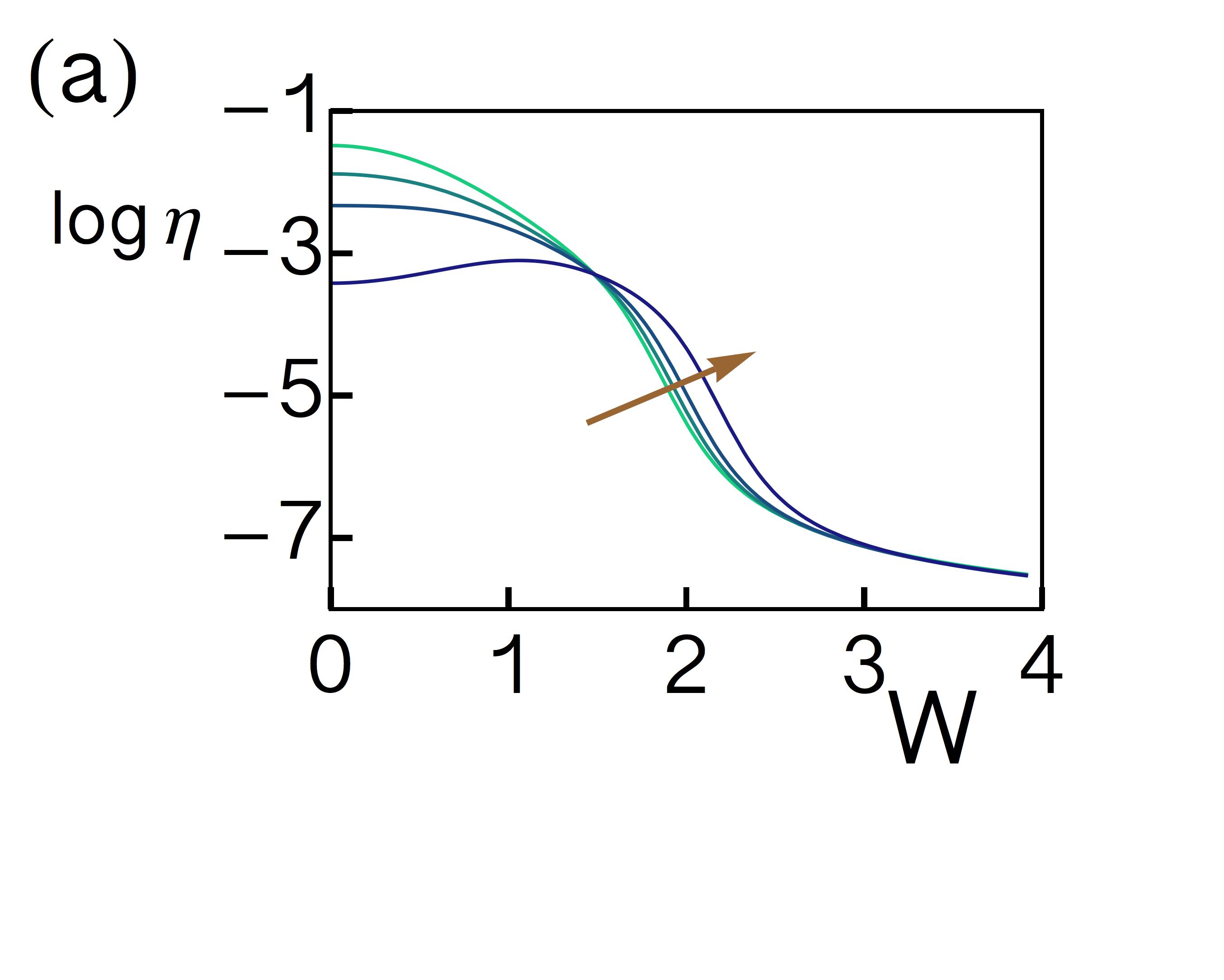}
\includegraphics[trim = 0mm 4cm 3cm 0cm, clip, scale=0.215]{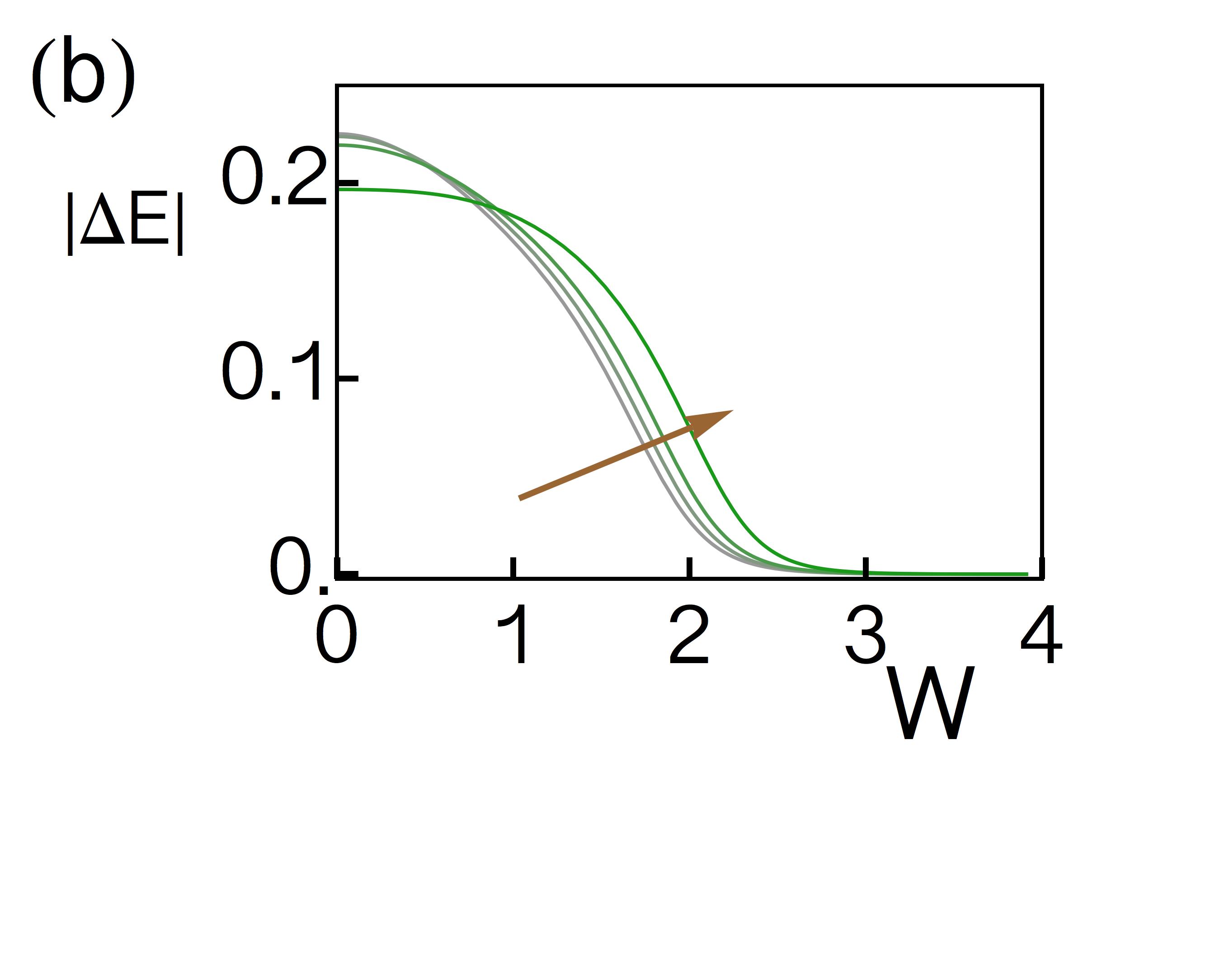}
\includegraphics[trim = 0mm 4cm 3cm 0cm, clip, scale=0.205]{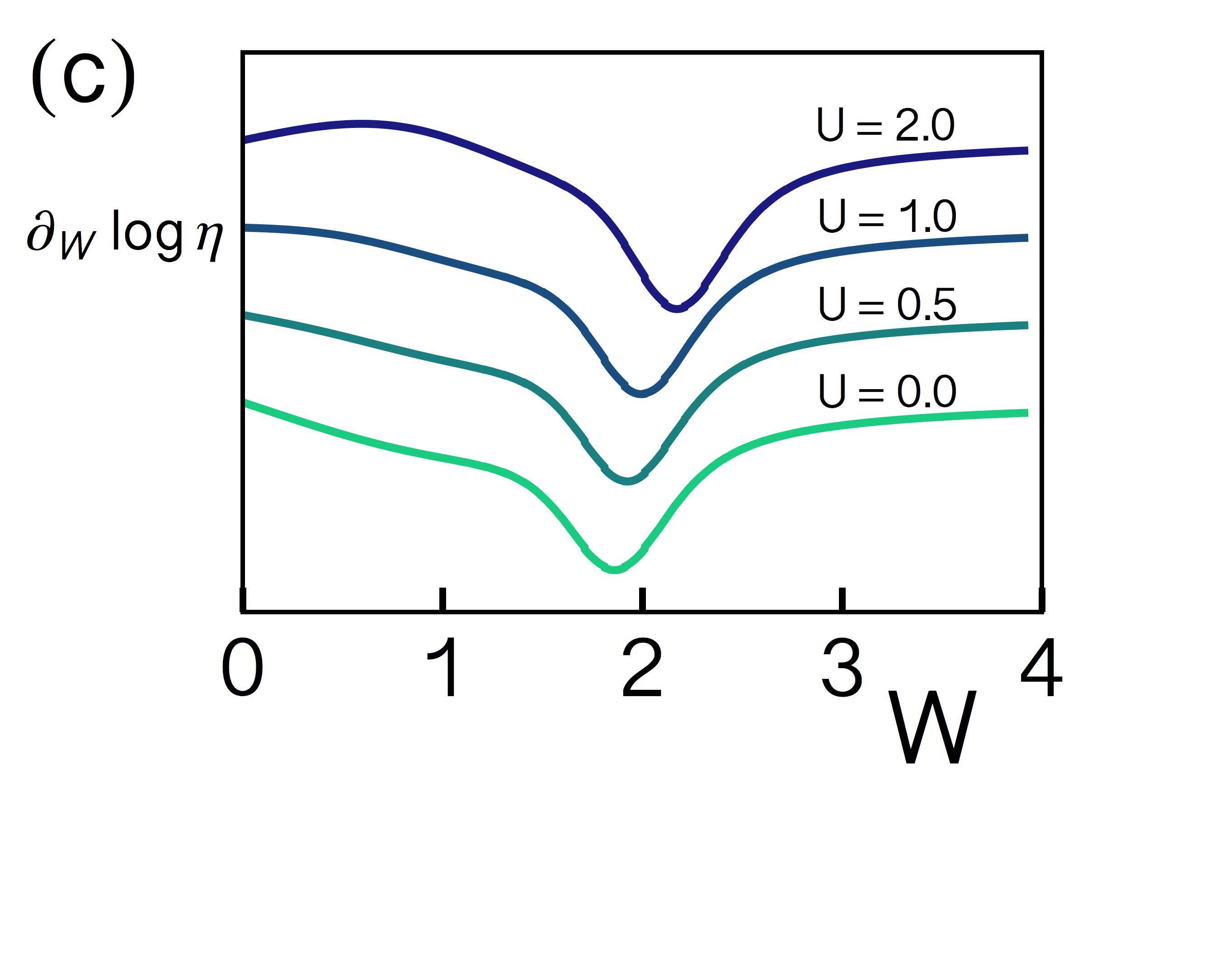}
\includegraphics[trim = 0mm 4cm 3cm 0cm, clip, scale=0.205]{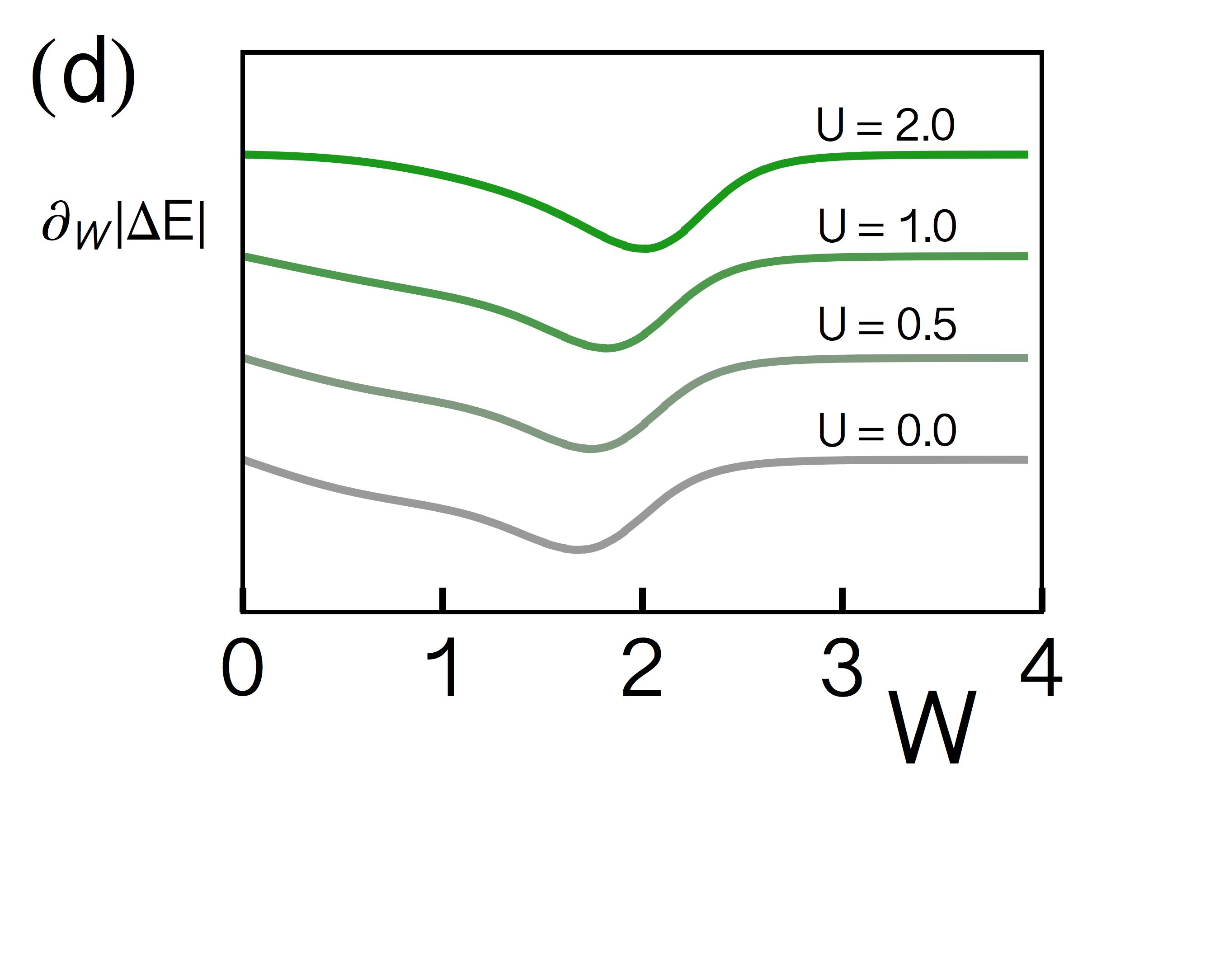}
\caption{The top two figures show (a) the log of the configuration-basis inverse participation ratio $\eta$ and (b) the change in ground state energy after twisting the boundary conditions $\vert \Delta E \vert$, both as a function of disorder strength. Each curve corresponds to a Hubbard interaction $U=0.0,0.5,1.0,2.0$, with the arrow denoting the sense in which the interaction increases. The two lower plots show the corresponding derivative of each curve so as to highlight the point where the transition occurs and how it shifts as the interaction is increased.} \label{Fig_set28FDEC}
\end{figure}

\begin{figure*}
\includegraphics[trim = 0mm 2cm 0cm 0cm, clip, scale=0.19]{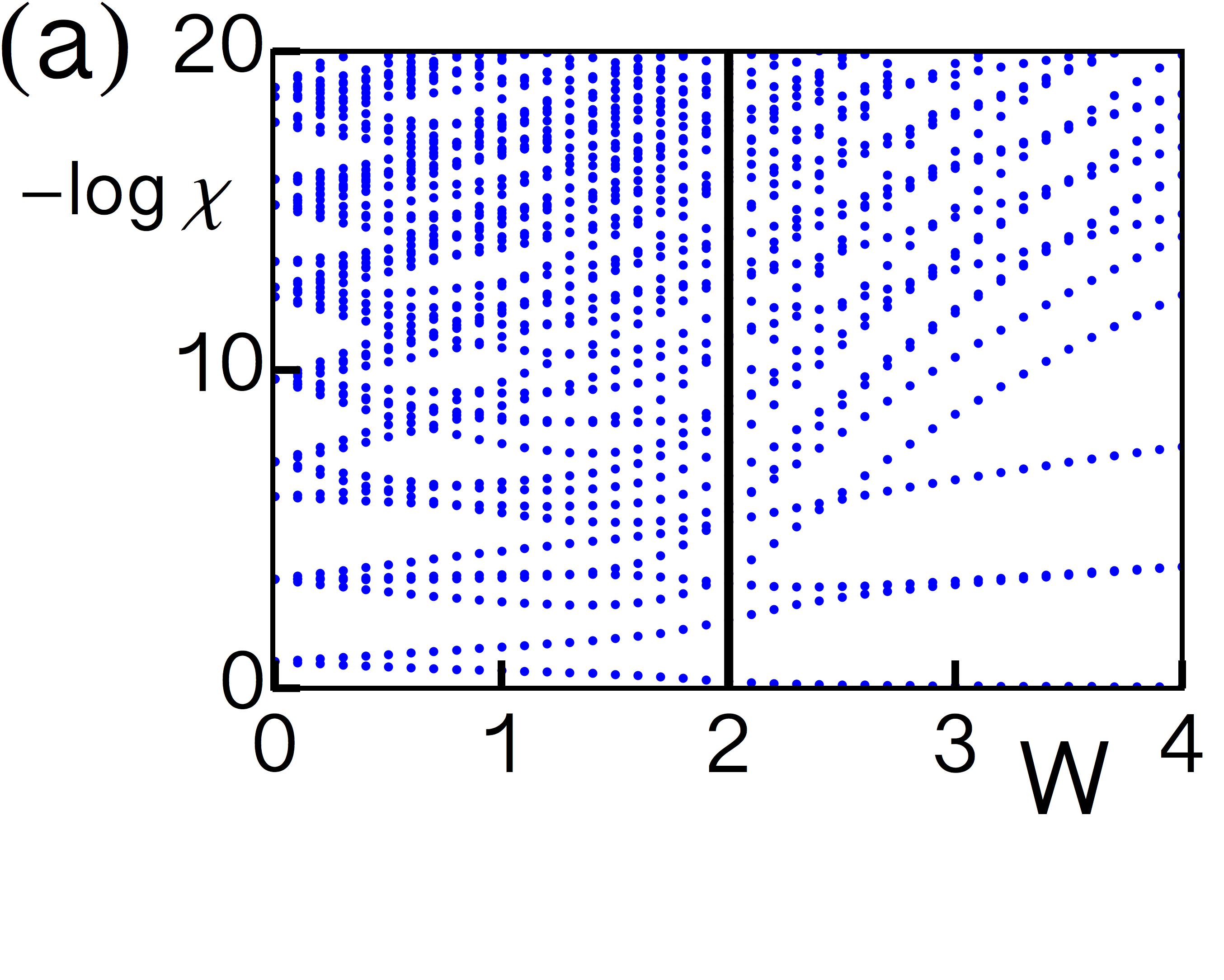}
\includegraphics[trim = 0mm 2cm 0cm 0cm, clip, scale=0.19]{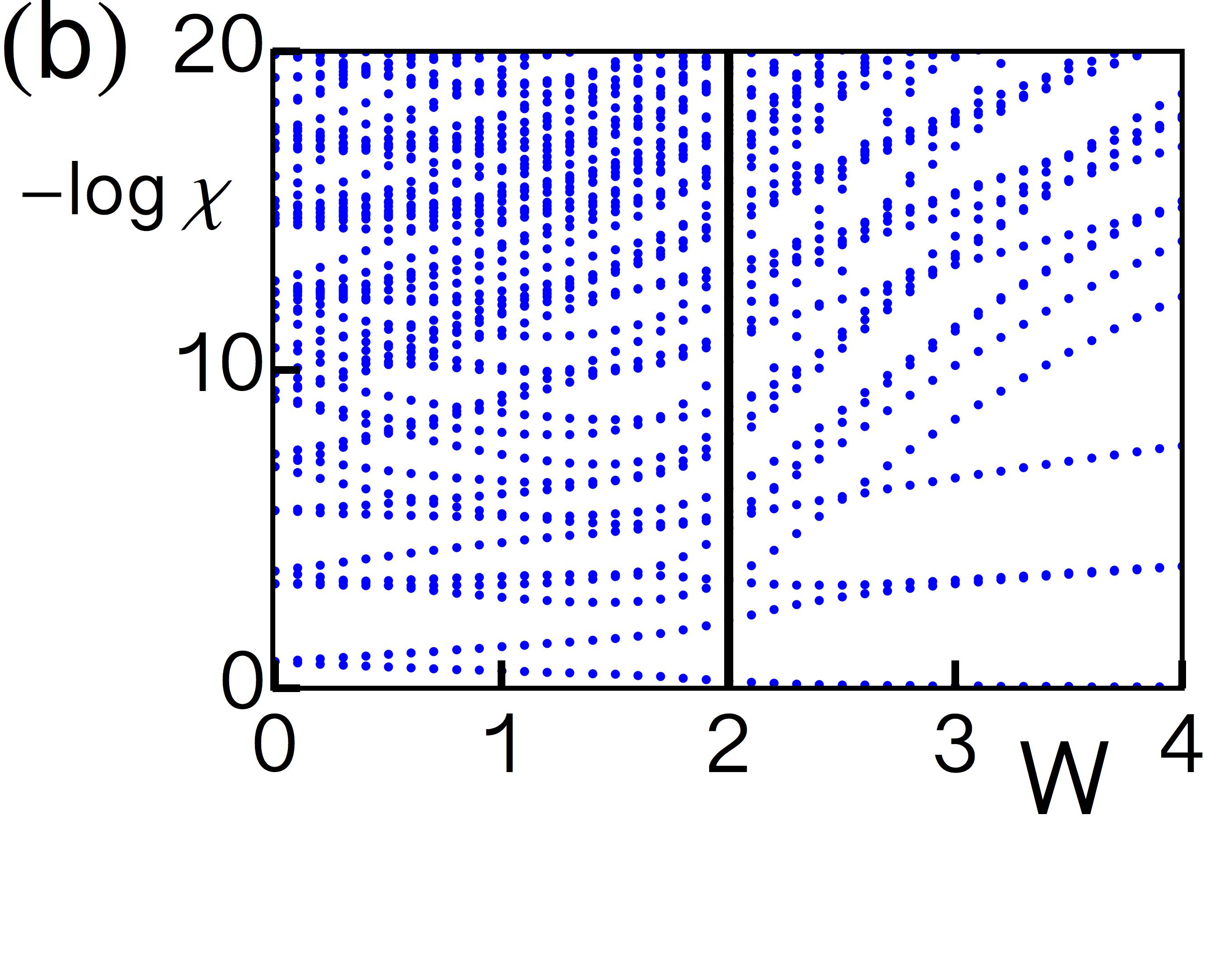}
\includegraphics[trim = 0mm 2cm 0cm 0cm, clip, scale=0.19]{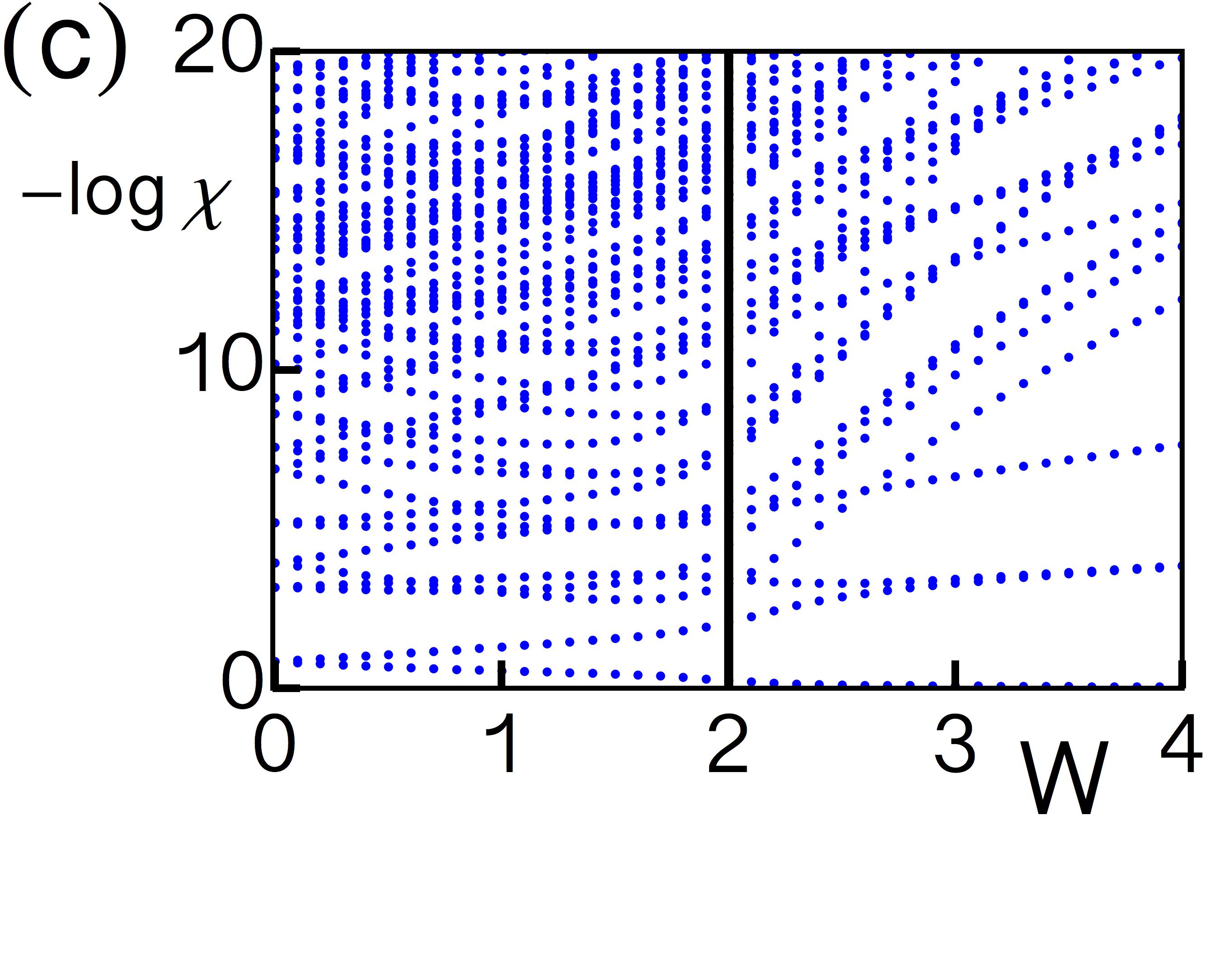}
\includegraphics[trim = 0mm 2cm 0cm 0cm, clip, scale=0.19]{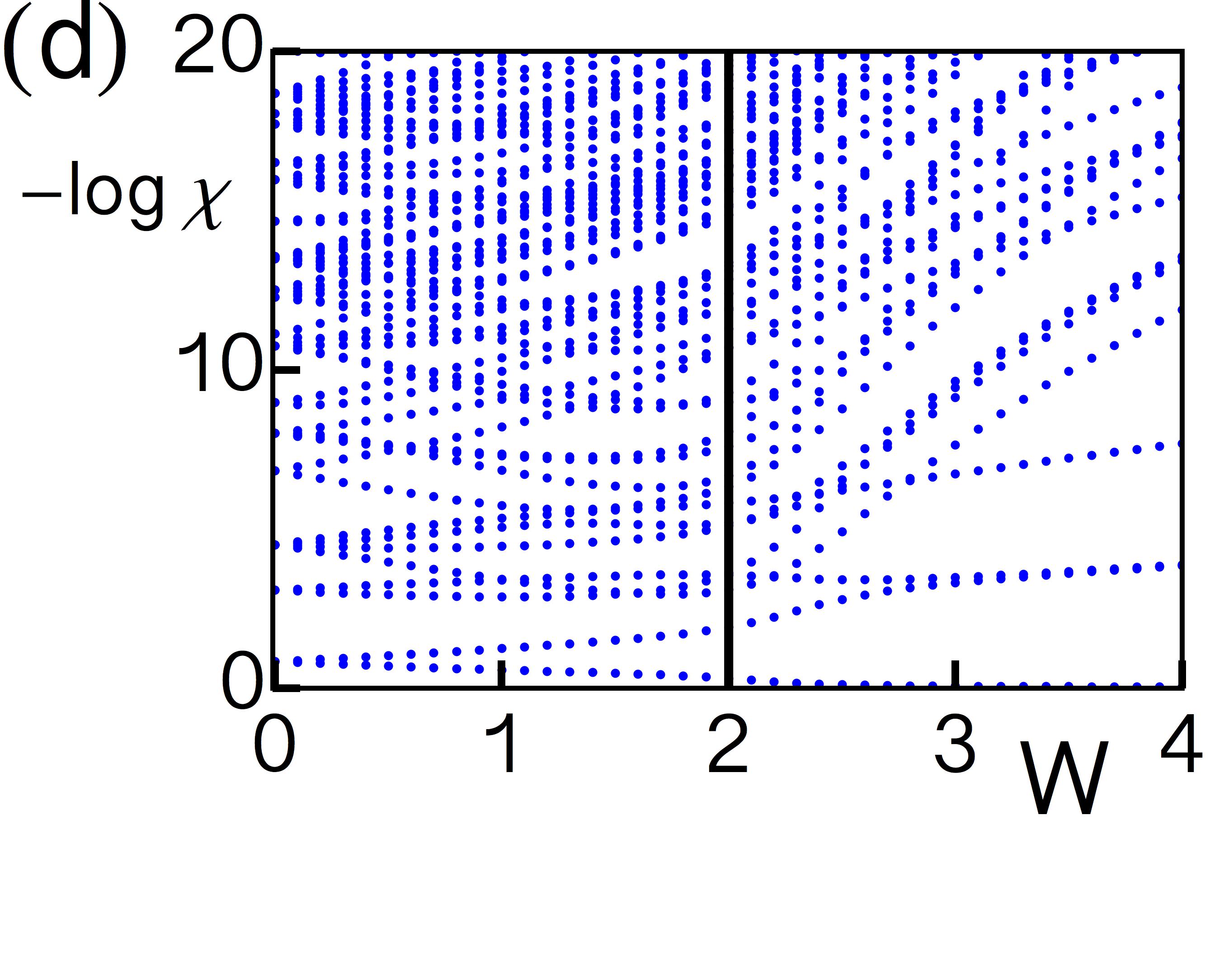}

\includegraphics[trim = 0mm 2cm 0cm 0cm, clip, scale=0.19]{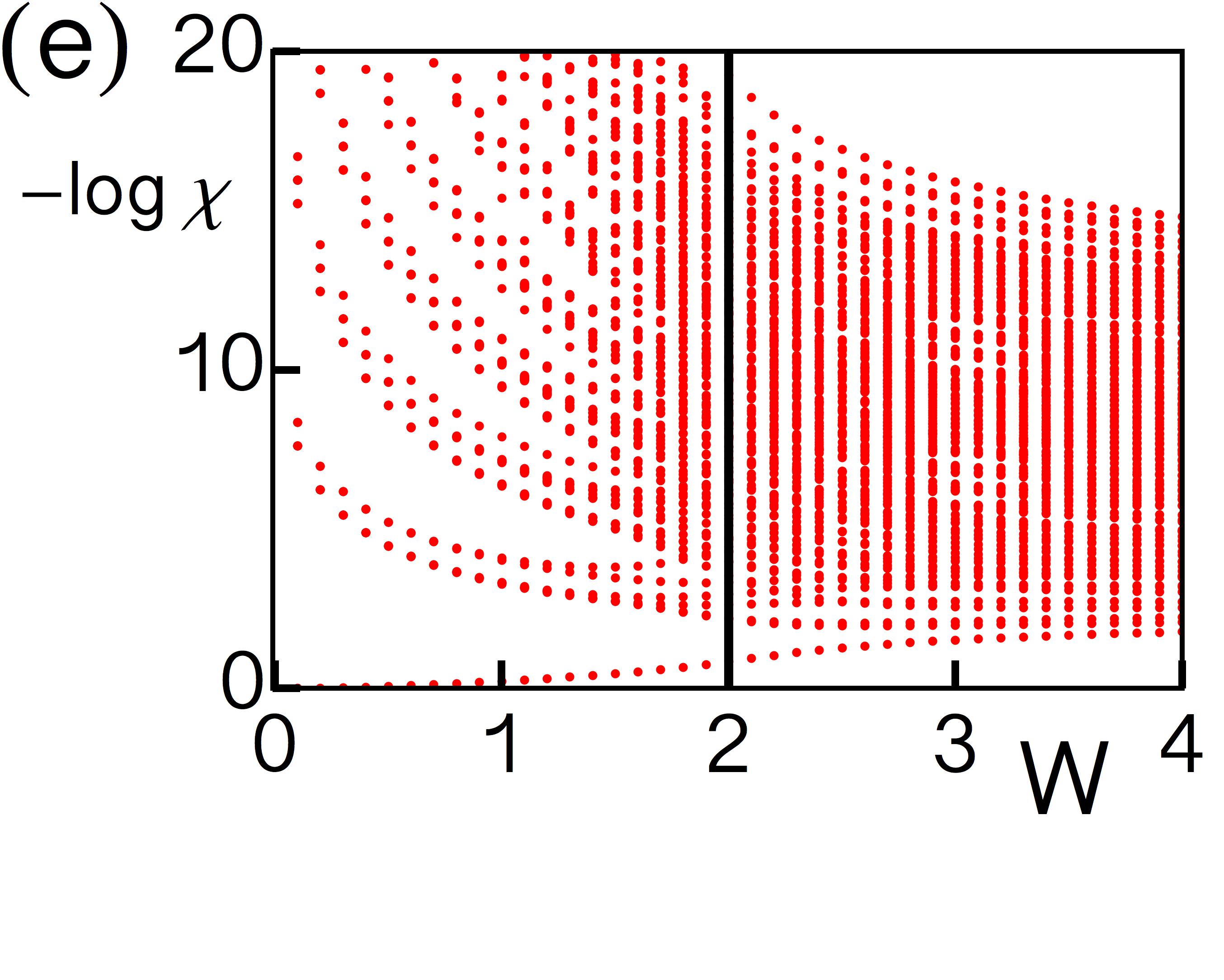}
\includegraphics[trim = 0mm 2cm 0cm 0cm, clip, scale=0.19]{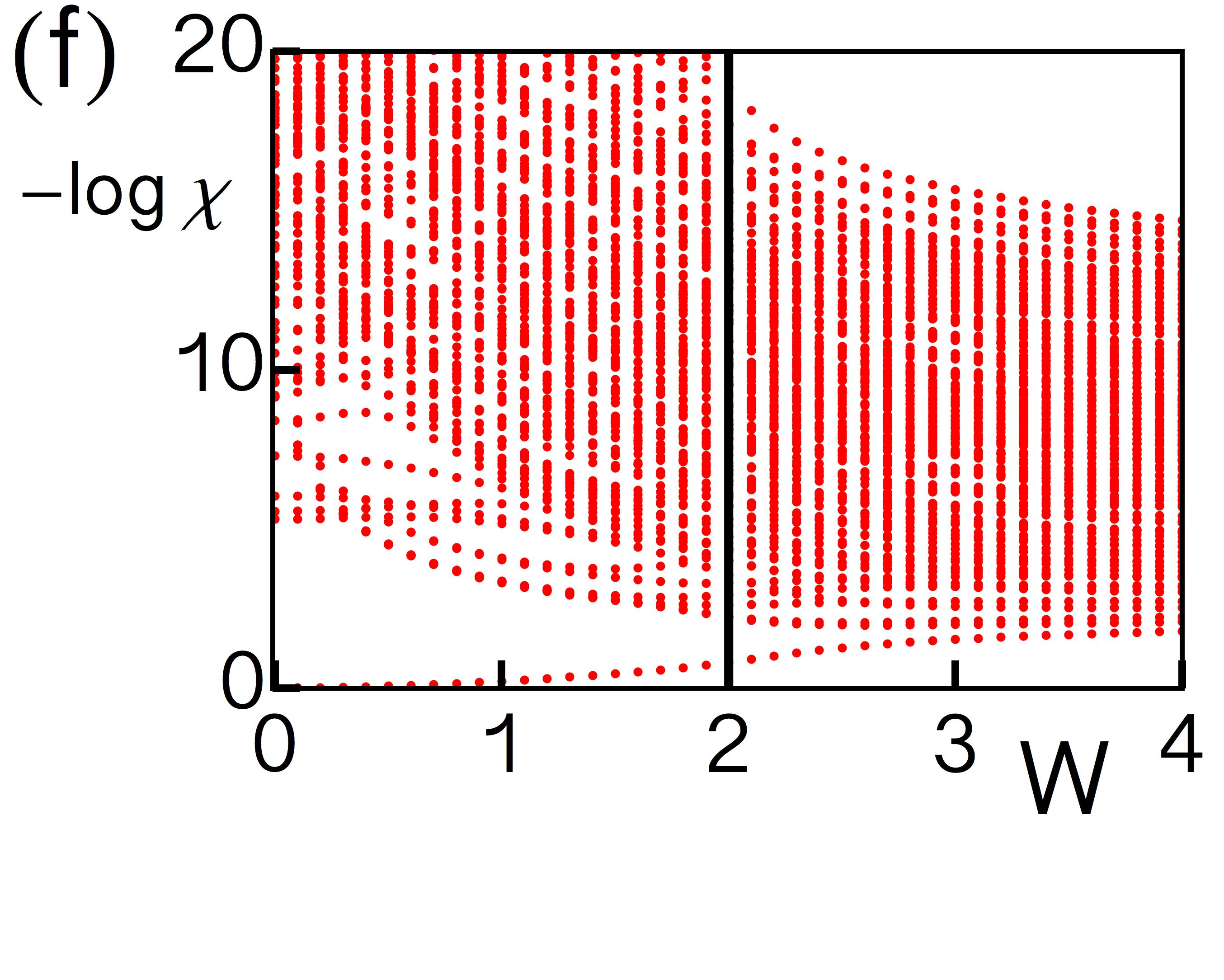}
\includegraphics[trim = 0mm 2cm 0cm 0cm, clip, scale=0.19]{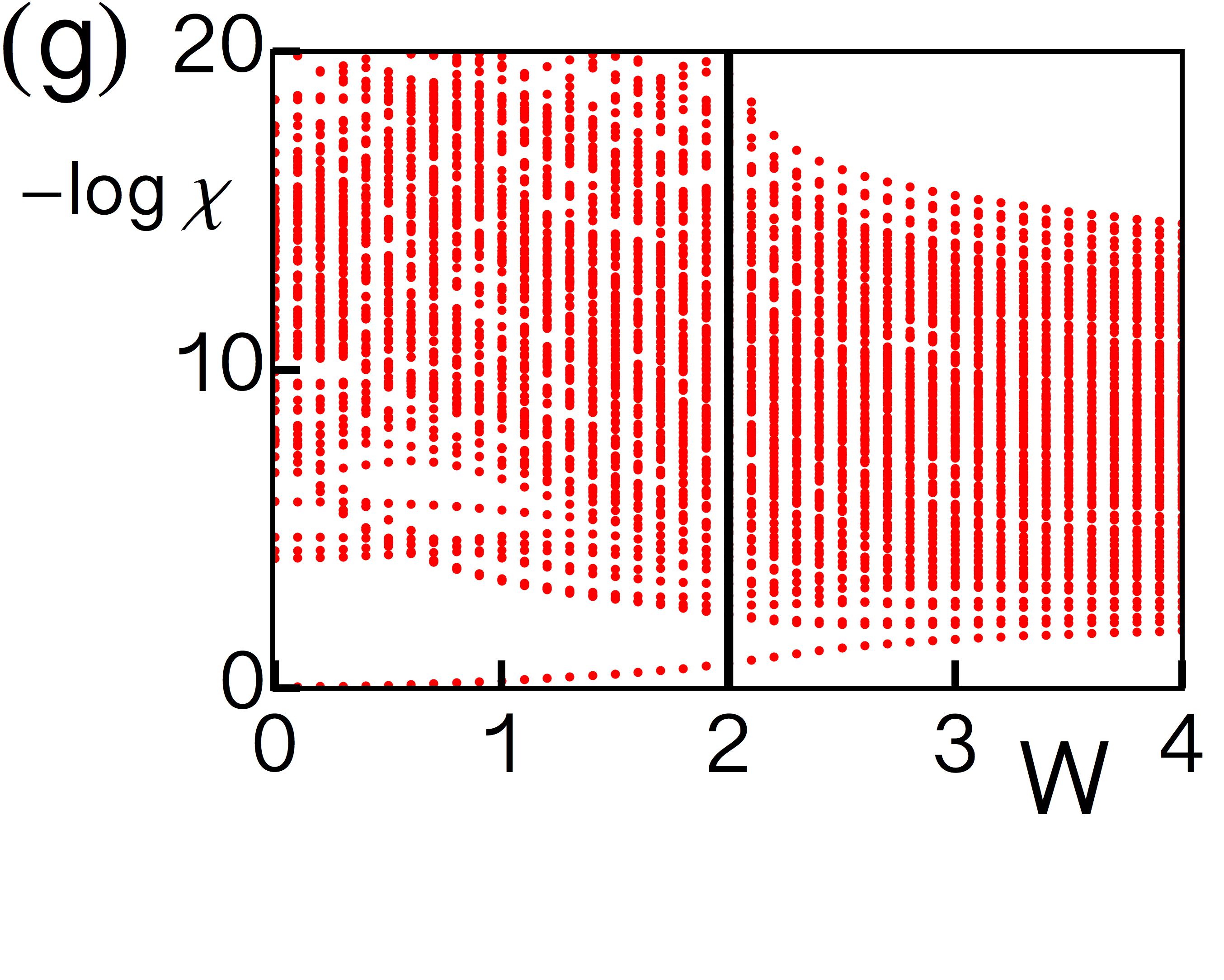}
\includegraphics[trim = 0mm 2cm 0cm 0cm, clip, scale=0.19]{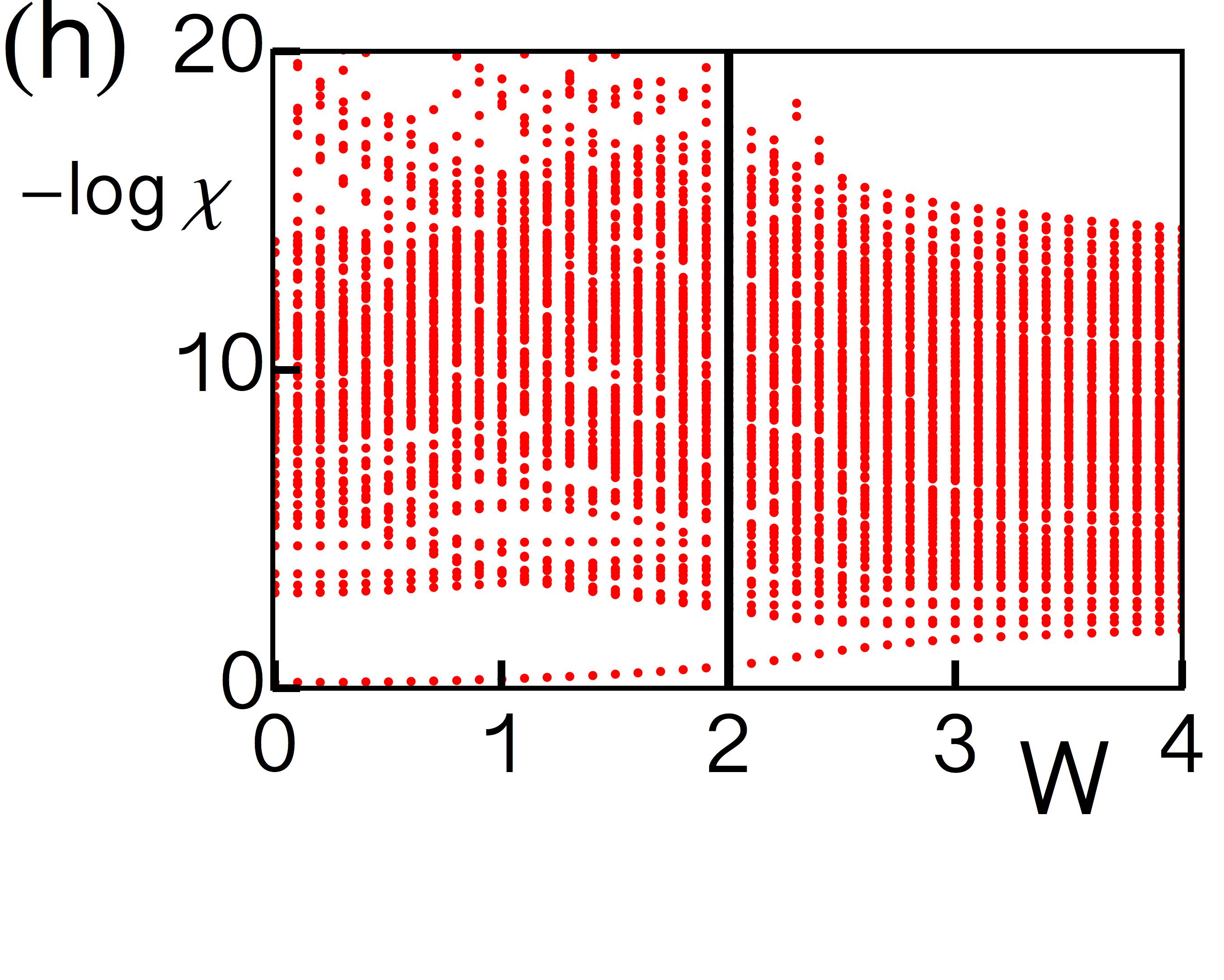}

\caption{Many-body entanglement spectrum of the interacting Aubry-Andr\'e model for: (top row) a spatial entanglement cut and (bottom row) a momentum entanglement cut. Each column corresponds to a Hubbard interaction strength $U=0.0,0.5,1.0,2.0$ respectively.}\label{Fig_manybody}
\end{figure*}

\subsubsection{Entanglement properties of the interacting model}
After clearly illustrating the behavior of the non-interacting AAM we now move on to the case with repulsive interactions. 
The simplest type of interaction that we can consider for spinless electrons is a nearest-neighbor Hubbard interaction of the form 
\begin{equation}
H_{int}= \sum_i U \hat{n}_n \hat{n}_{i+1}
\end{equation}\noindent where $n_i$ is the particle density on site $i.$
As we mentioned previously, we numerically diagonalize a system of size $N=14$ at half-filling, and choose $\alpha=24/14\approx1.64$. There have been several studies on the AAM with interactions including Ref. [\onlinecite{Eilmes1999}] where a two-particle  AAM with interactions was studied, and Ref. [\onlinecite{Lyer2013}] where the interacting AAM was studied at infinite temperature in the context of many-body localization.

In order to correlate the behavior of the spatial and momentum entanglement with whether there is a tendency for localization in the system, we compute two additional independent quantities that are not directly related to entanglement. One if these is the so-called configuration-basis inverse participation ratio computed in Ref. \onlinecite{Lyer2013}. If we express the ground state in terms of Slater determinants in the spatial basis, we can schematically write this as $\ket{\Omega}=\sum_{c} \psi(c)\ket{c}$, where $c$ corresponds to a specific spatial configuration of the fermions in the lattice. The normalized configuration-basis participation ratio is then given by $\eta=\frac{1}{\mathcal{N} P}$, where $P=\sum_{c}\vert \psi(c) \vert^4$ and $\mathcal{N}$ is the number of basis states. This quantity measures the weight distribution of the ground state in the basis of spatial Slater determinants. In the delocalized state, $\eta$ is expected to be of order $1$, whereas in the localized state $\eta$ should become vanishingly small. The second quantity we compute to probe localization is the change in ground state energy $\vert \Delta E\vert = \vert E(\pi)-E(0) \vert$ after twisting of the boundary conditions from $0$ to $\pi$. In this case, $\vert \Delta E\vert$ should be nonzero in the delocalized state and it should become vanishingly small in the localized state.

In Fig. \ref{Fig_set28FDEC}a,b we show the results of calculating $\log \eta$ and $\vert \Delta E \vert$ as a function of disorder strength for four values of the Hubbard interaction $U=0.0,0.5,1.0,2.0$. For the non-interacting case $U=0.0$, we obtain the behavior we expect as the system crosses the localization transition: there is a significant decrease in both quantities at the phase transition. As the interaction strength is increased, the curves maintain their overall shape, except that they appear to shift towards higher values of the disorder. The two lower plots Figs. \ref{Fig_set28FDEC}c,d, show the derivatives of $\log \eta$ and $\vert \Delta E \vert$ with respect to the disorder strength. These plots exhibit a dip, which approximately corresponds to the point at which localization transition occurs. The dip moves toward higher values of the disorder strength with increasing $U$, which indicates that as the interaction is increased, the parameter space over which delocalization is obtained is enlarged by interactions. We should emphasize that this conclusion might not extrapolate to the thermodynamic limit as there has been some evidence that the enhancement of delocalization in two-particle interacting AAM models vanishes in this limit \cite{Eilmes1999}. Notwithstanding this, it will be useful to consider these finite-size results, and to try to correlate the behavior of $\log \eta$ and $\vert \Delta E \vert$ with both the spatial and momentum entanglement.

Now we are will calculate the entanglement properties of the interacting AAM. As we mentioned previously for interacting systems, the simple relation between the eigenvalues of the correlation matrix and the entanglement entropy does not hold because the ground state will not, in general, be a single Slater determinant.  Instead, it is necessary to compute the eigenvalues of the reduced density matrix of the many-body system in order to obtain the entanglement entropy. We will briefly present here the method for calculating the many-body momentum entanglement.

To calculate the many-body momentum entanglement spectrum we start by writing the ground state as
\begin{equation}
\ket{\Omega}=\sum_{K_L, K_R} \psi(K_L,K_R)\ket{K_L,K_R}
\end{equation}
where $\ket{K_L,K_R}=\ket{k_{L1}\ldots k_{Ln_L}, k_{R1}\ldots k_{R n_R}}=\left[\prod^{n_L}_{k\in K_L}c^{\dagger}_{k}\right]\left[\prod^{n_R}_{k'\in K_R}c^{\dagger}_{k'}\right]\ket{0}$, and $n_L$ ($n_R$) is the number of left (right) movers in a given basis state $\ket{K_L,K_R}$. The matrix elements of the density matrix $\rho=\ket{\Omega}\bra{\Omega}$ are then given by
\begin{equation}
\left[\rho\right]^{K'_R,K'_L}_{K_R,K_L}= \psi^*(K_R,K_L)\psi(K'_R,K'_L)
\end{equation}
and so the reduced density matrix for the left movers is
\begin{equation}
\left[\rho_L\right]^{K'_L}_{K_L}= \sum_{K_R}\psi^*(K_R,K_L)\psi(K_R,K'_L)
\end{equation}
which describes the left-moving part of the interacting ground state $\ket{\Omega}$. The set of eigenvalues of $\rho_{L}$ which we previously called $\left\{\lambda_i^{2}\right\}$ from the Schmidt-decomposition, are then used to compute the entanglement entropy via $S=-\sum_i \lambda_i^2 \log \lambda_i^2$. An analogous calculation can be performed for the spatial entanglement entropy.

Let us now compute the entanglement for the same values of the Hubbard interaction we have used in Fig. \ref{Fig_set28FDEC}. We start by analyzing the spatial and momentum many-body entanglement spectra $\{\chi_i\}$ as a function of disorder potential strength. In Fig. \ref{Fig_manybody}, we show the set $\{-\log \chi_i\}$ as is customary in the literature. The closer the $(-\log \chi_i)$ for each mode gets to $1$, the higher the entanglement corresponding to that mode. 

In the spatial entanglement modes, where we have cut the system in half, the overall structure appears qualitatively the same for the four values of the interaction strength: in the region $W<W_c ( = 2 )$, the modes are spread out, with some modes near $-\log \chi_i=1$, whereas for $W>W_c$ the modes appear to move to higher (less entangled) values. The main effect of the interactions is to lower the entanglement modes toward $-\log \chi_i=1$ for all potential strengths. It is not, however, clear from these figures whether there is a definite behavior as the interaction strength is increased. The entanglement entropy computed using these modes will provide a clearer picture as we see below in Fig. \ref{IntEntropy}. 

In the momentum entanglement spectrum, on the other hand, there appears to be a gap around $-\log \chi_i=1$ when the system is in the delocalized state. In the localized state there is a large shift of many-body entanglement modes toward $-\log \chi_i=1$, which  signals a significant increase in momentum entanglement since the system is becoming localized. The main effect of the Hubbard interactions appears to be to systematically increase the momentum entanglement for all potential strengths. In particular, note that the momentum entanglement gap around $-\log \chi_i=1$ gets smaller as the potential strength is increased. In parallel to this, however, one can see that the gap actually increases slightly in the region $W>W_c$ as the interaction becomes larger. Since we know from the non-entanglement quantities $\log \eta$ and $\vert \Delta E \vert$ that interactions shift the point at which localization occurs toward higher values of $W$, the behavior of the momentum entanglement spectrum suggests that this entanglement gap is tracking the delocalized phase. This would constitute a generalization to interacting systems of the connection between suppressed momentum entanglement and the metallic state. 

\begin{figure}
\includegraphics[trim = 0cm 4cm 3cm 0cm, clip, scale=0.205]{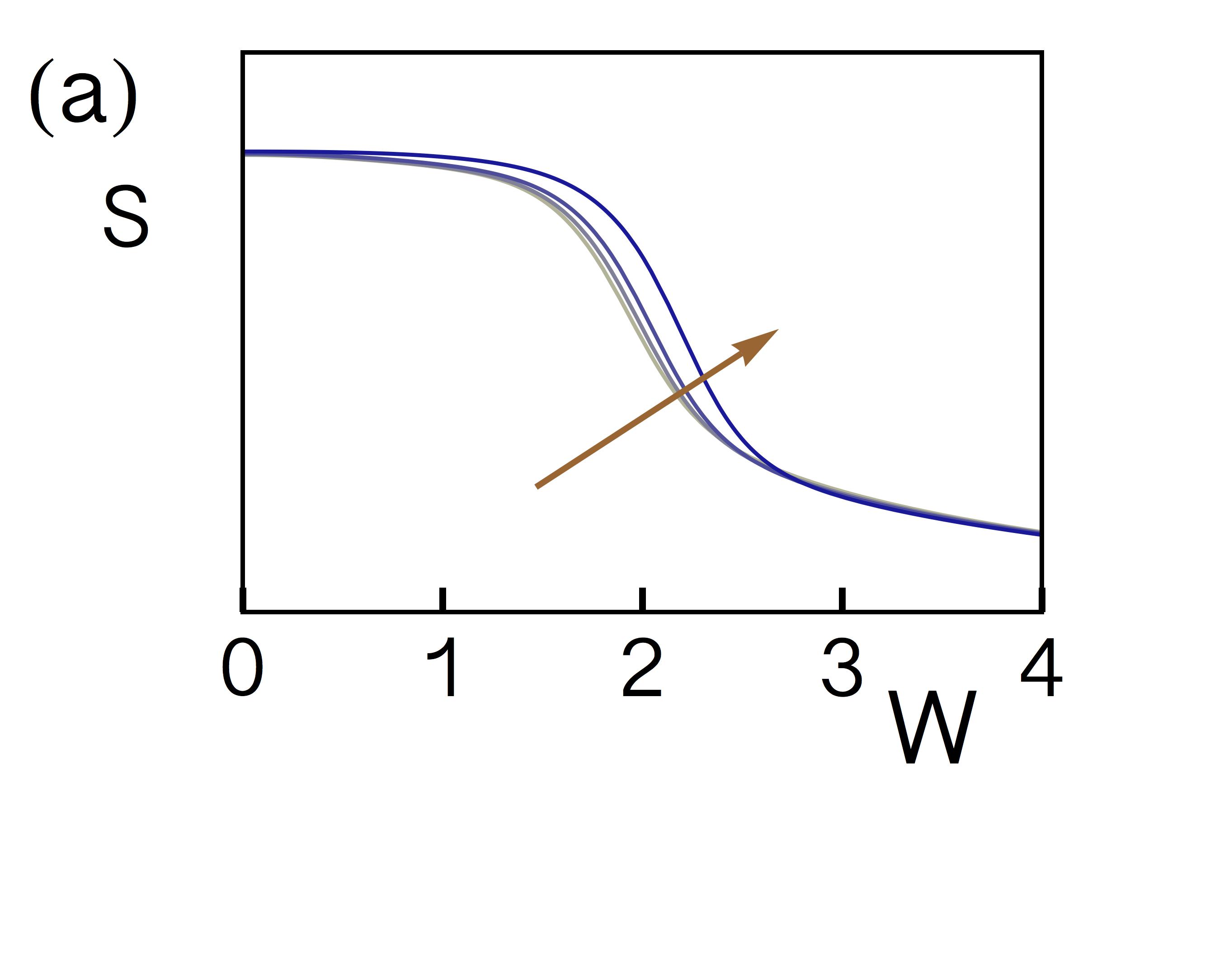}
\includegraphics[trim = 0mm 4cm 3cm 0cm, clip, scale=0.205]{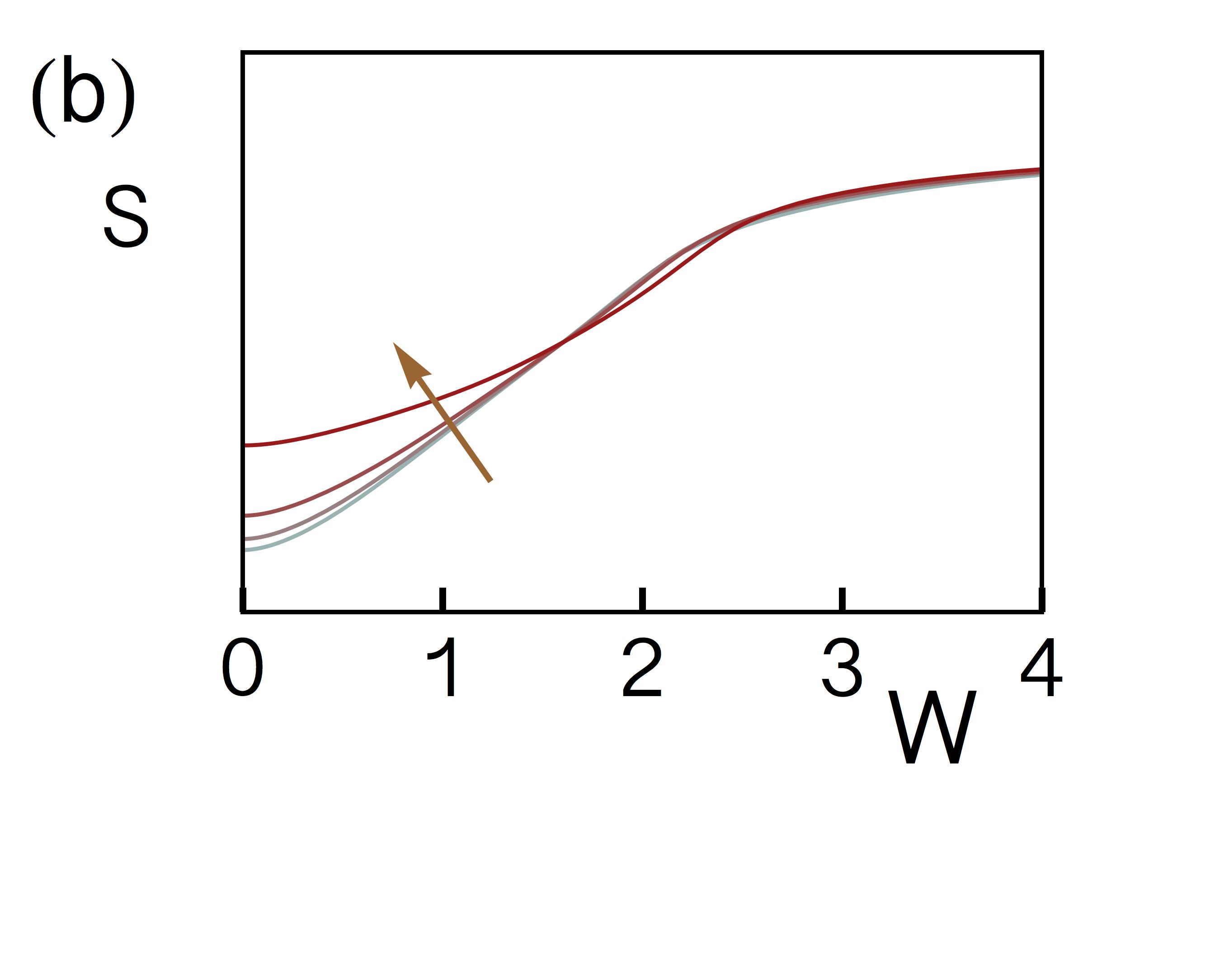}
\includegraphics[trim = 0mm 4cm 3cm 0cm, clip, scale=0.205]{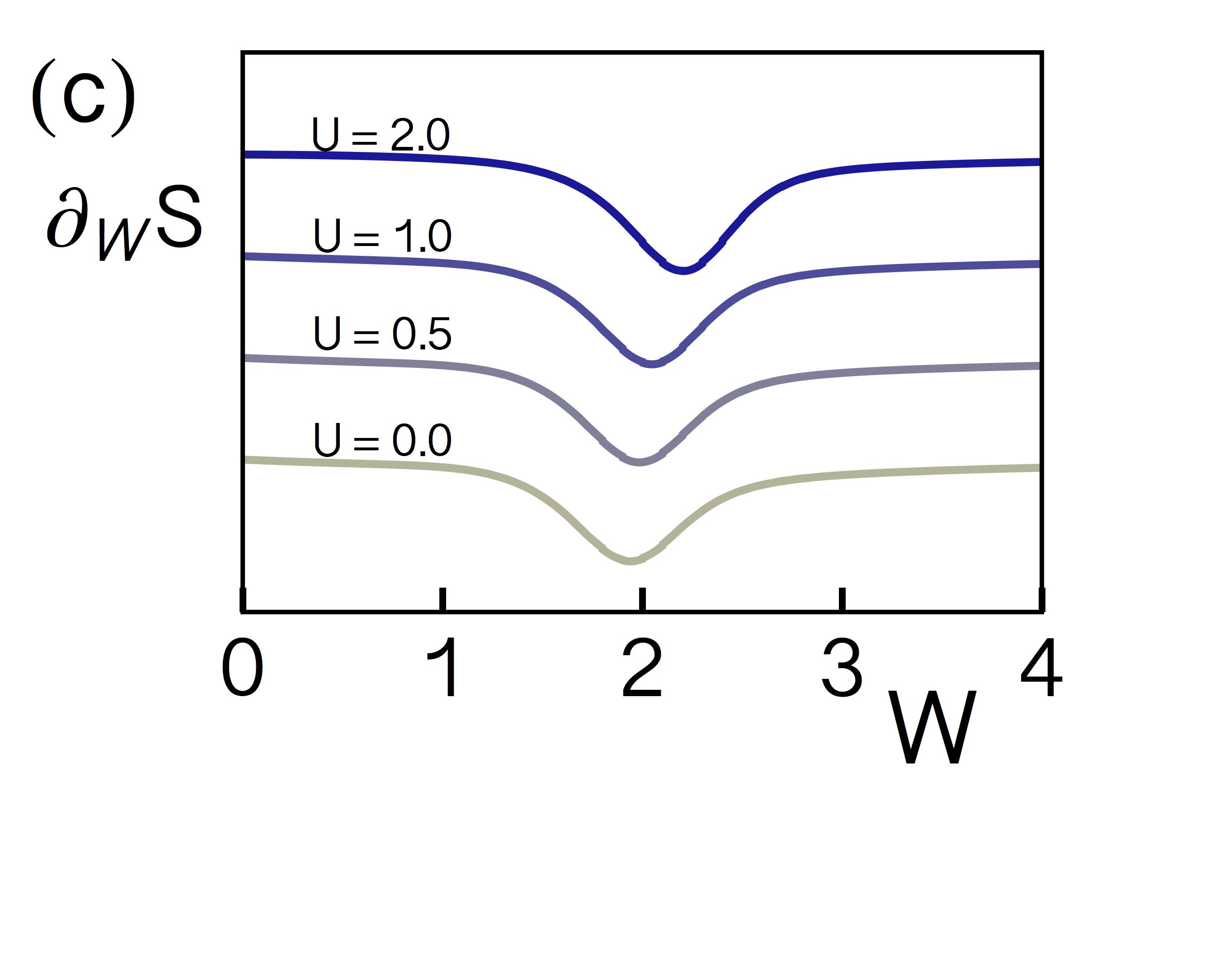}
\includegraphics[trim = 0mm 4cm 3cm 0cm, clip, scale=0.205]{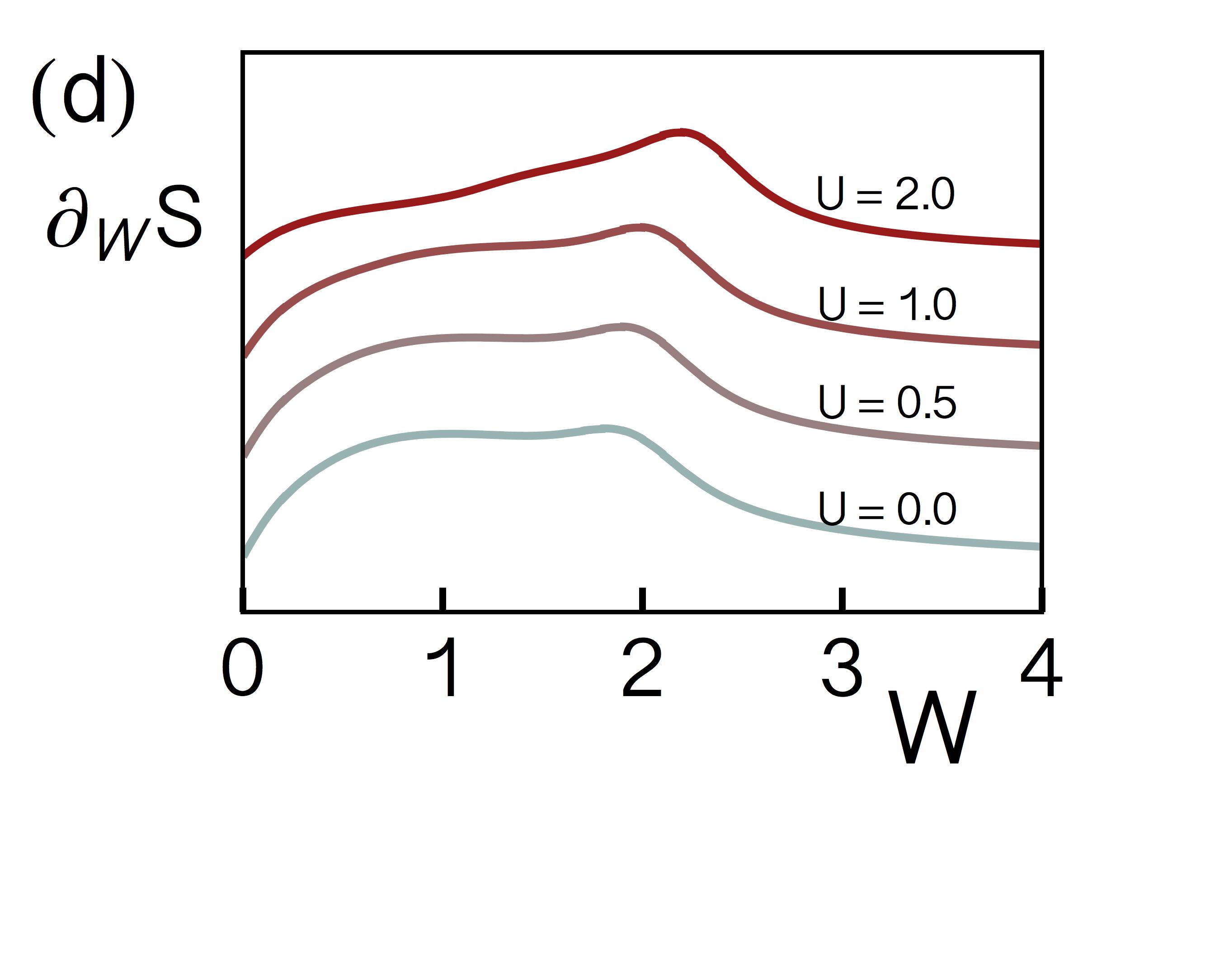}
\caption{Entanglement entropies for the interacting Aubry-Andr\'e model: (a) and (b) show the spatial and momentum entanglement entropy, respectively, as a function of disorder strength for $U=0.0,0.5,1.0,2.0$. The arrow shows the direction of increase of the interaction strength. (c) and (d) show the derivative of both types of entanglement entropy, which illustrate how the transition point shifts as a function of the interaction strength.}\label{IntEntropy}
\end{figure}

Having made these observations about the many-body entanglement spectra, let us now take a look at the corresponding entanglement entropies. This is shown in Fig. \ref{IntEntropy}a,b. It is clear that there is a drop (rise) in spatial (momentum) entanglement for all values of the interaction strength in the neighborhood of the non-interacting $W_c$, which we expect should be connected with the localization transition. By taking the derivative of these curves, which we show in Fig. \ref{IntEntropy}c,d, we can track the disorder strength for which both types of entanglement change their trend. In the derivative of the spatial entanglement we observe a dip  and in that of the momentum entanglement, we observe a peak. Both the dip and the peak shift toward larger values of the disorder strength as a function of the Hubbard interaction strength. 

The behavior of the entanglement entropies is consistent with that of $\log \eta$ and $\vert \Delta E \vert$. This instills confidence that both the spatial and the momentum entanglement are able to track the tendency to localize in the interacting system. In particular, the many-body momentum entanglement spectrum and the corresponding entanglement entropy appears to yield clear signatures of localization in such an interacting system even though the interactions will generate scattering between all momentum states. It is probable that there are other systems for which this behavior will also hold, especially in other interacting one-dimensional models.

\section{Effect of disorder on the internal configuration of disordered fermions}\label{sec:entanglement2}

We are now going to switch gears to study a slightly different problem. Whereas so far we have been considering the dual relationships between the spatial and momentum entanglement in simple models, we will now study a model that has internal degrees of freedom which are correlated with the single-particle momentum. 
In these systems we will explore how disorder affects the internal configurations of one-dimensional fermions. To study this, we will focus on disorder-induced phase transitions between two distinct insulating phases. A particular example of this type of transition is that which occurs between a topological insulator and trivial Anderson insulator. Topological states of non-interacting fermions in translationally invariant systems typically arise because of a coupling between the momentum and the internal spin/orbital space. The mapping between the Brillouin zone momentum and the orbitals of the particular occupied states via the Bloch functions can endow the electronic structure with robust topological properties. A natural question is then,  upon disordering the system, how robust is the entanglement between the momenta and the internal degrees of freedom? We will see that entanglement, and in particular a hybrid momentum/orbital partition, even without translation invariance, will be useful. Let us now move on to discuss a specific example.

\subsection{Topological insulator model with entangled internal degrees of freedom}

If we restrict ourselves to non-spatial discrete symmetries (namely particle-hole, time-reversal or chiral symmetry), there exist five classes that realize nontrivial ground states in one dimension and they are either classified by a $Z$ or a $Z_2$ topological invariant. The two simplest classes are the ones that only realize one of these symmetries: class D which satisfies exclusively particle-hole symmetry and class AIII which satisfies exclusively chiral symmetry (Hamiltonians with only time-reversal symmetry are all trivial in 1D). To simplify our discussion, we will focus on the latter.

Systems belonging to class AIII are those described by Hamiltonians that anti-commute with an operator $S$. The simplest such model  in this class  can be represented by a two-band model. Systems with more bands can be adiabatically deformed so that the topological properties are determined by copies of such a two-band model. We thus consider the following Hamiltonian
\begin{eqnarray}
H=\sum_{k} c^{\dagger}_k\left[h(k) \right] c_k=\sum_{k} c^{\dagger}_k\left[\sum_a d_a(k)\sigma^a \right] c_k\label{AIII}
\end{eqnarray}
where $\sigma^a$ ($a= 0, 1,2, 3$) are the identity and the Pauli matrices, and $d_a(k)$ is a momentum-dependent, real, four-component vector. This Hamiltonian has two energy bands given by $\epsilon_{\pm}(k)=d_0(k)\pm \sqrt{\sum_{i=1}^3d^2_i(k)}$. Since the Pauli matrices anti-commute,  chiral symmetry can be realized by setting $d_{0}(k)=0$ and any one of the remaining $d_{i}$ components to zero. Although this choice is arbitrary, we choose to set $d_{2}(k)=0$ for convenience. For this choice, the chiral operator is $S=\sigma^2$. A simple model in class AIII that has a non-trivial topological insulator phase is given by the Bloch Hamiltonian
\begin{equation}\label{Model}
h(k)=\cos k \sigma^1 +(m+\sin k)\sigma^3.
\end{equation}
Depending on the value of the parameter $m$, the ground state will either be a trivial band insulator ($m<-1$ and $1<m$) or a topological insulator ($-1<m<1$) protected by chiral symmetry. 

\begin{figure*}
\includegraphics[trim = 0cm 0cm 2.1cm 0cm, clip, scale=0.22]{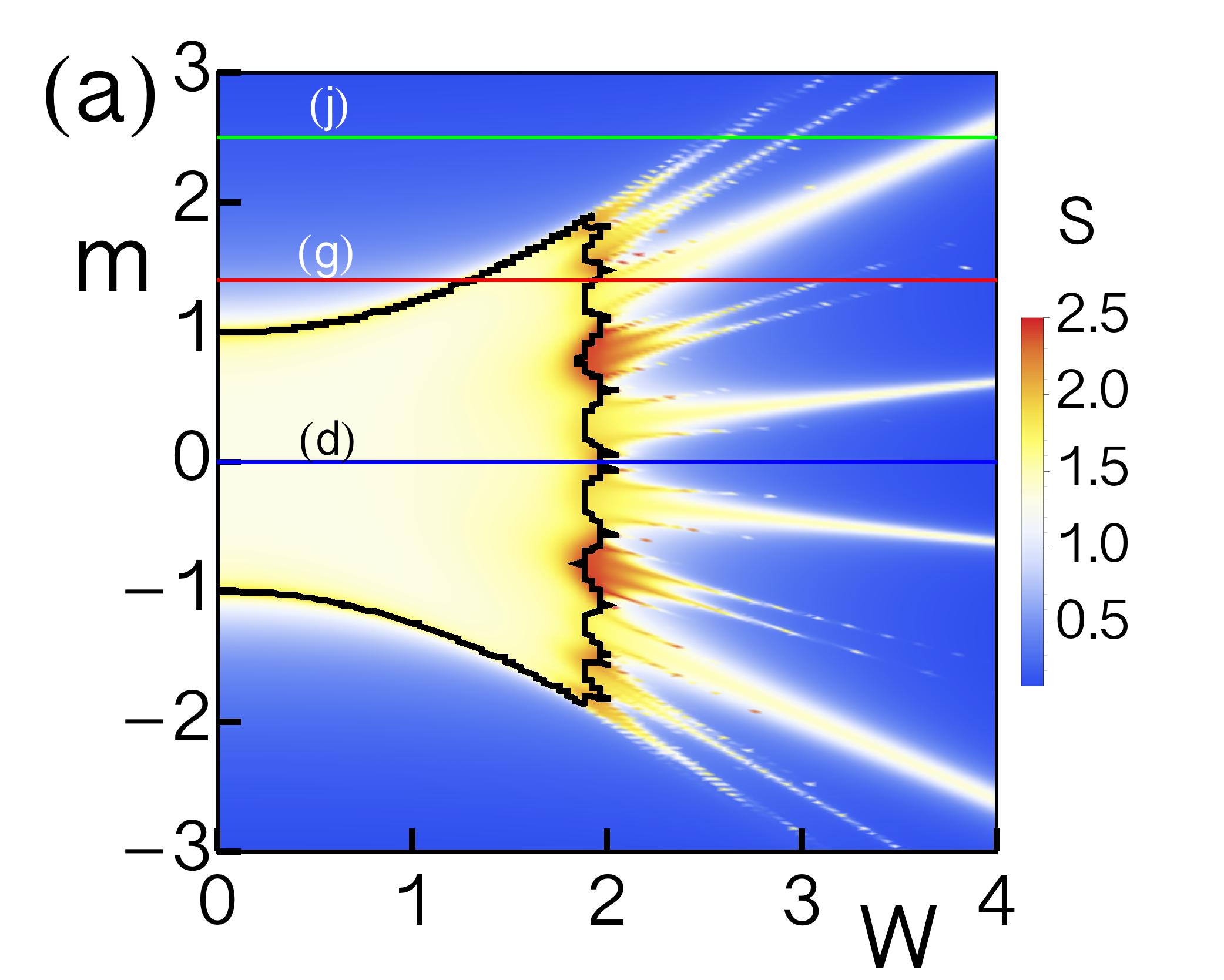}
\includegraphics[trim = 0cm 0cm 2.1cm 0cm, clip, scale=0.22]{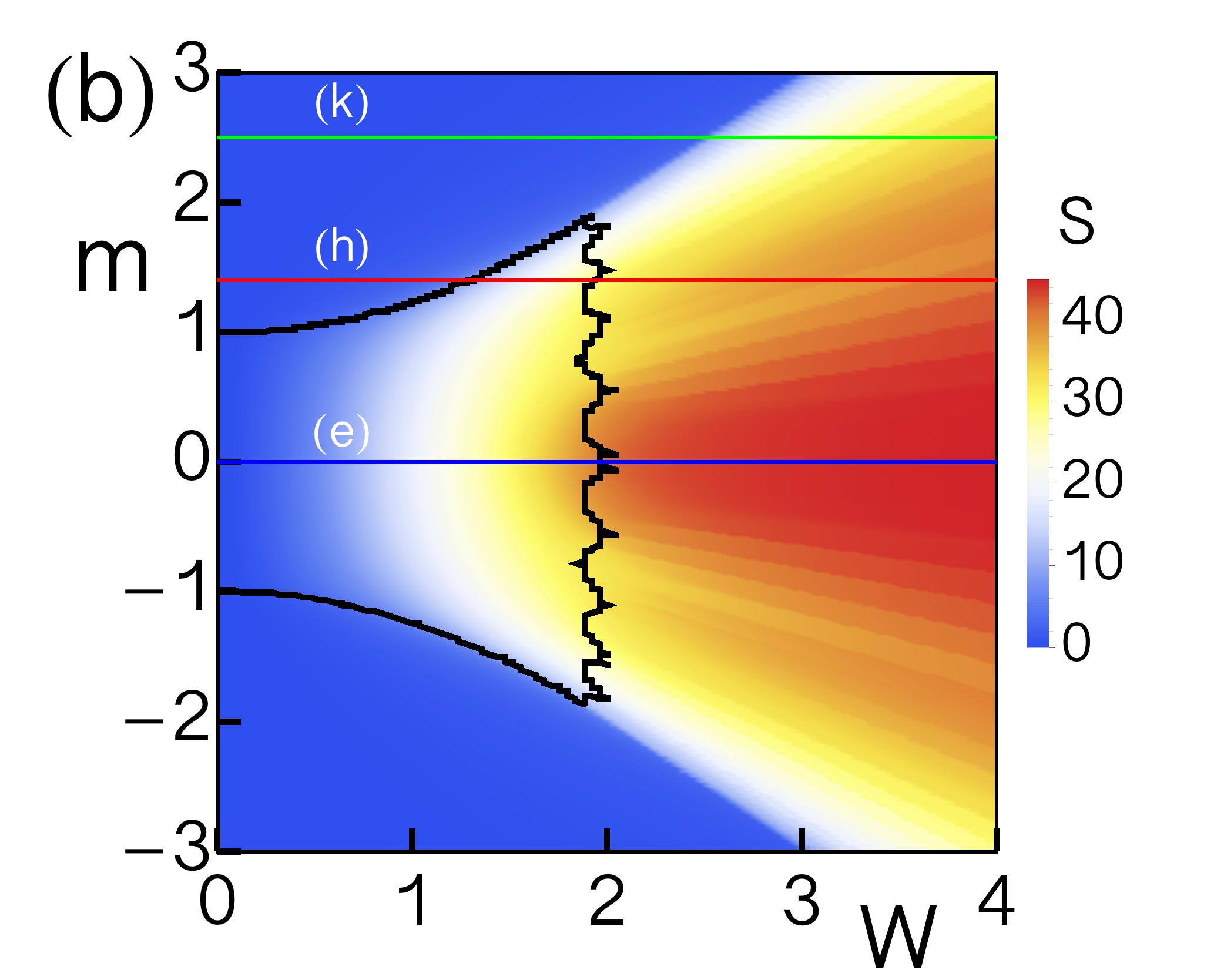}
\includegraphics[trim = 0cm 0cm 2.1cm 0cm, clip, scale=0.22]{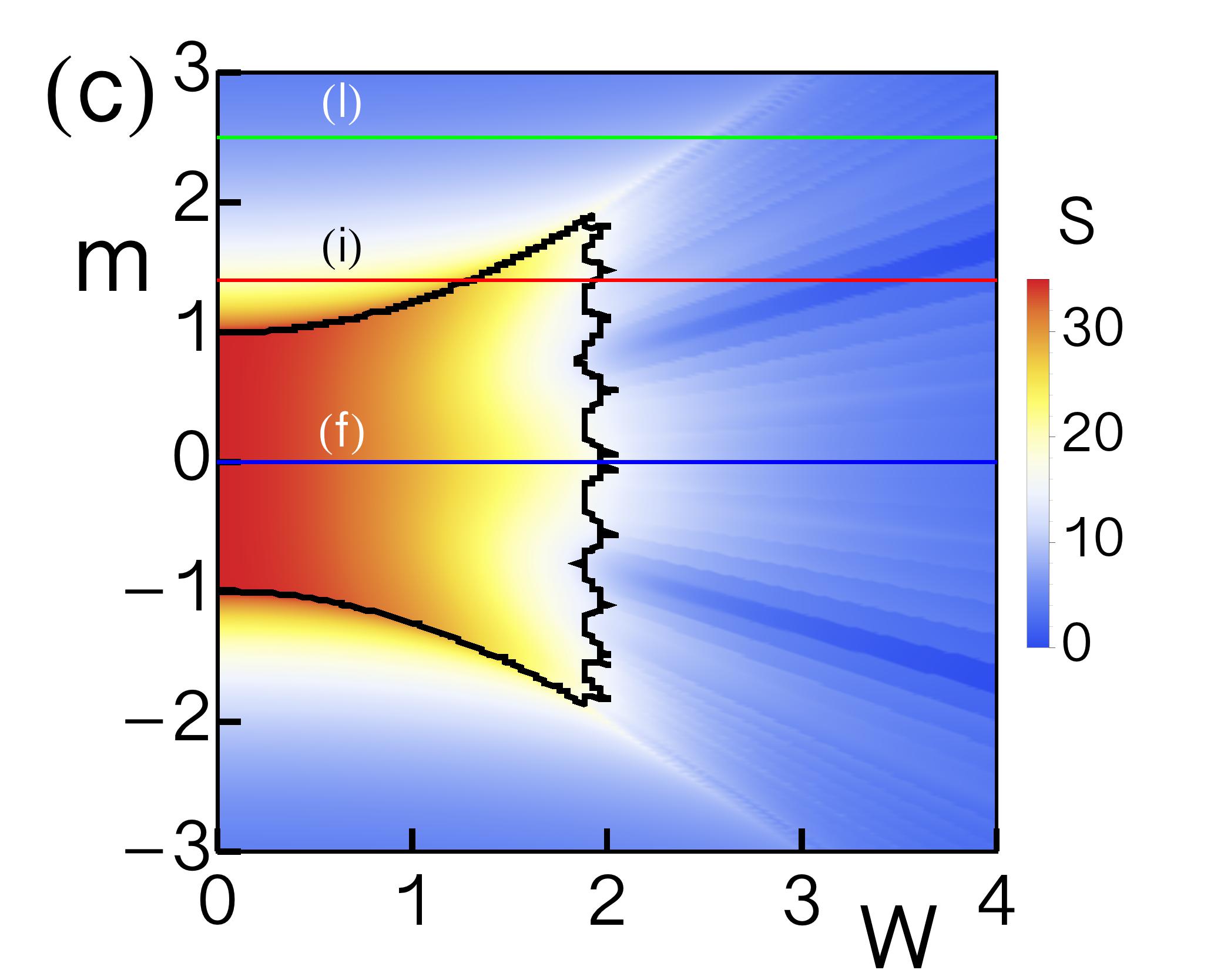}

\includegraphics[trim = 0cm 1cm 0cm 0mm, clip, scale=0.22]{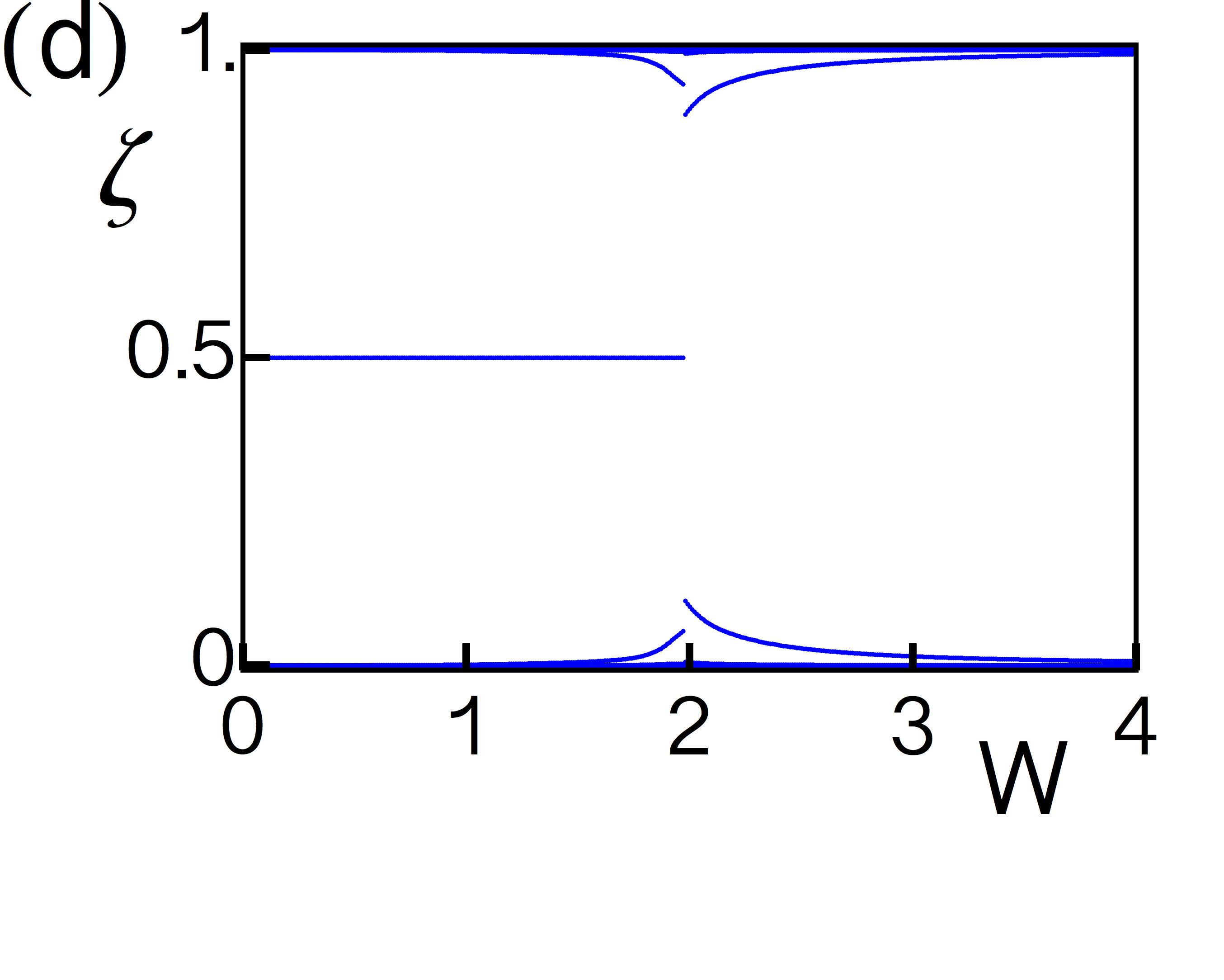}\hspace{0.22cm}
\includegraphics[trim = 0cm 1cm 0cm 0mm, clip, scale=0.22]{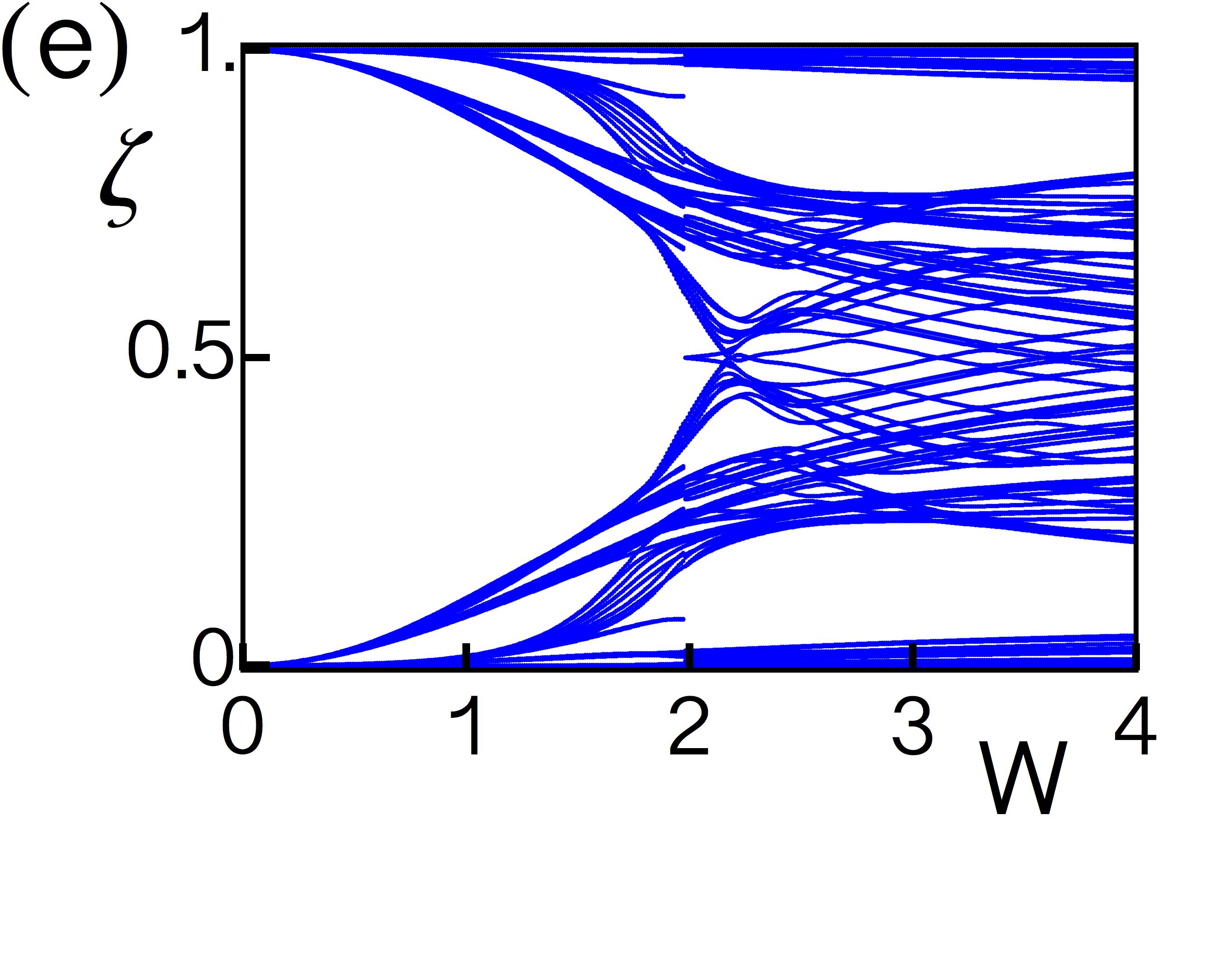}\hspace{0.22cm}
\includegraphics[trim = 0cm 1cm 0cm 0mm, clip, scale=0.22]{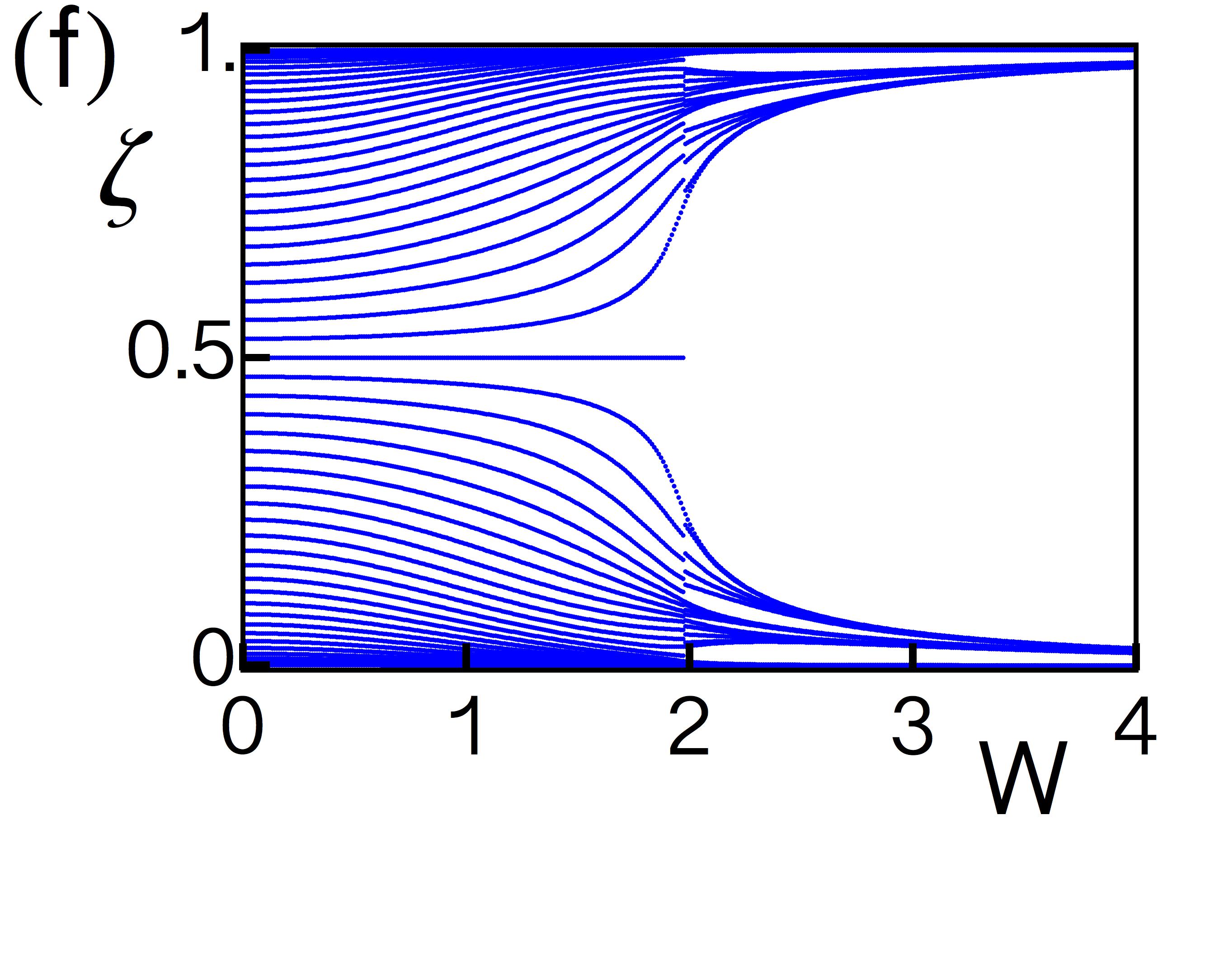}

\includegraphics[trim = 0cm 1cm 0cm 0mm, clip, scale=0.22]{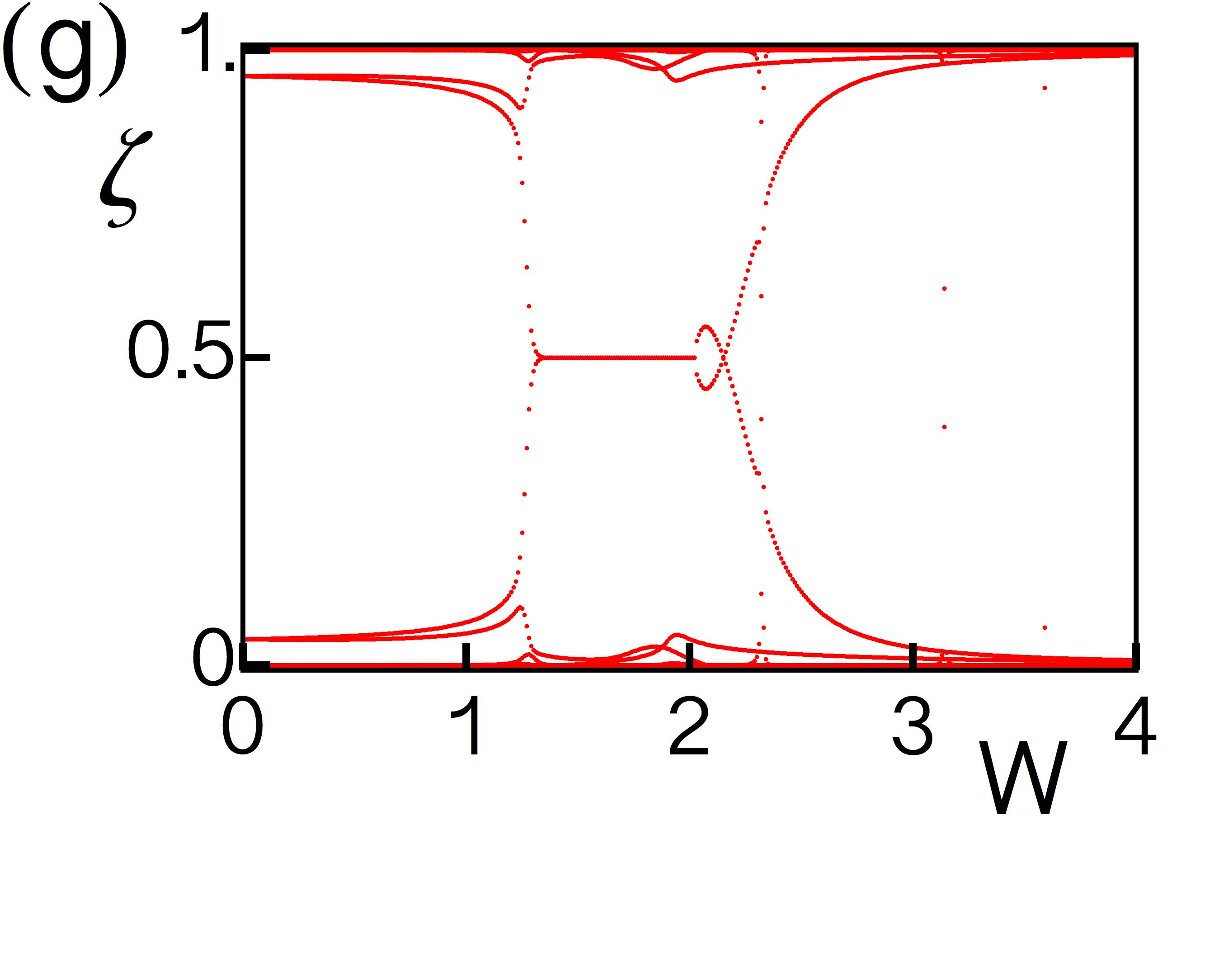}\hspace{0.22cm}
\includegraphics[trim = 0cm 1cm 0cm 0mm, clip, scale=0.22]{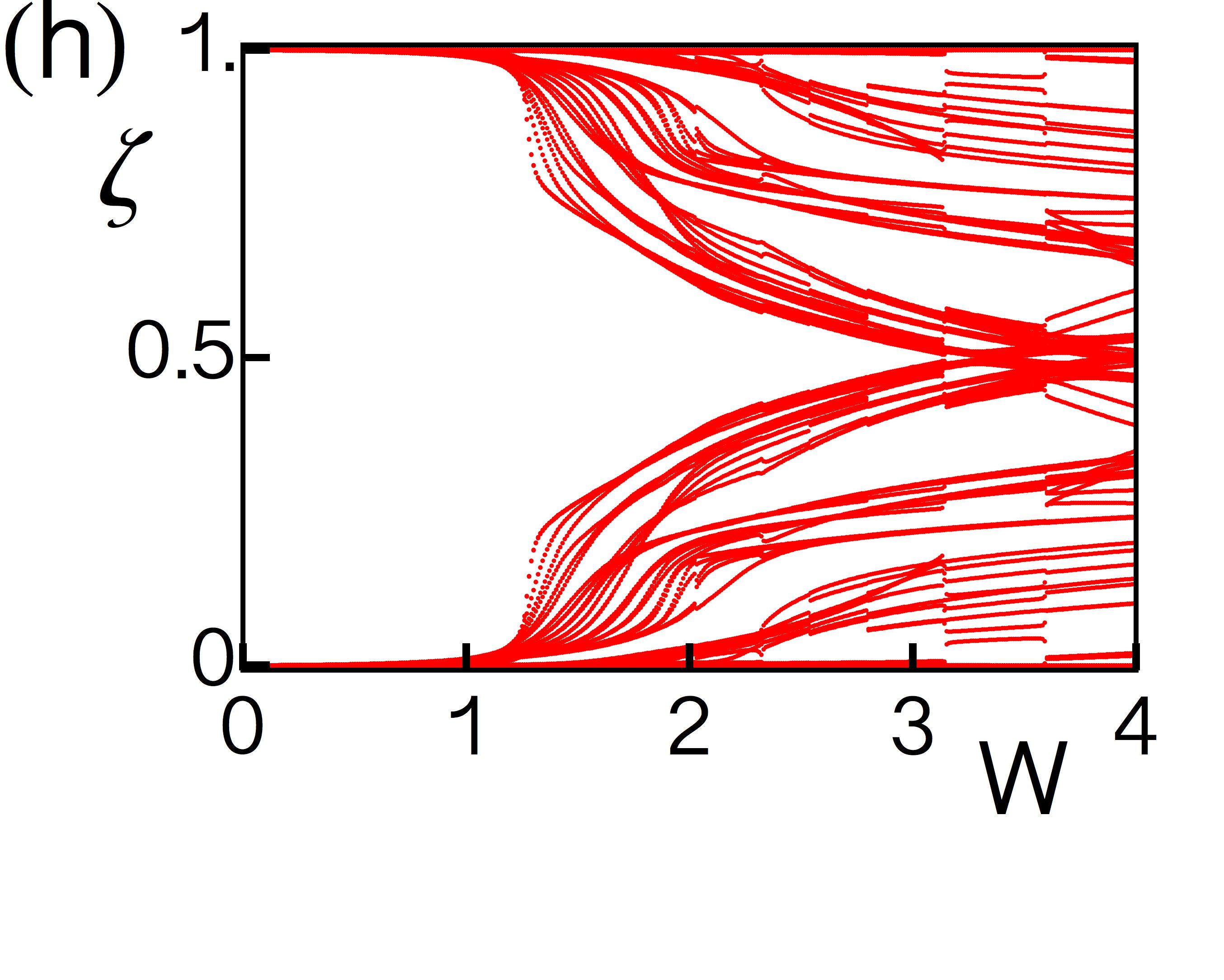}\hspace{0.22cm}
\includegraphics[trim = 0cm 1cm 0cm 0mm, clip, scale=0.22]{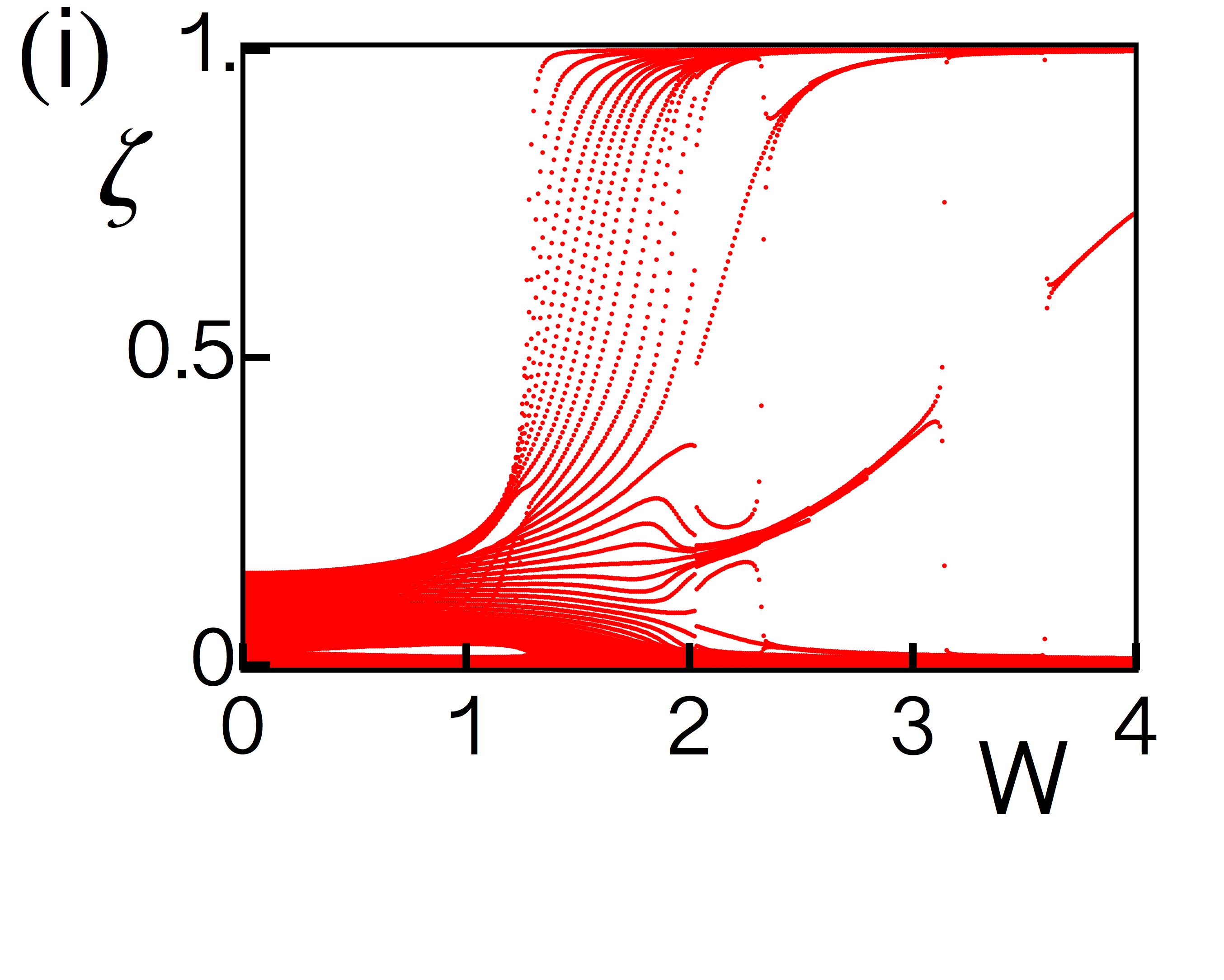}

\includegraphics[trim = 0cm 1cm 0cm 0mm, clip, scale=0.22]{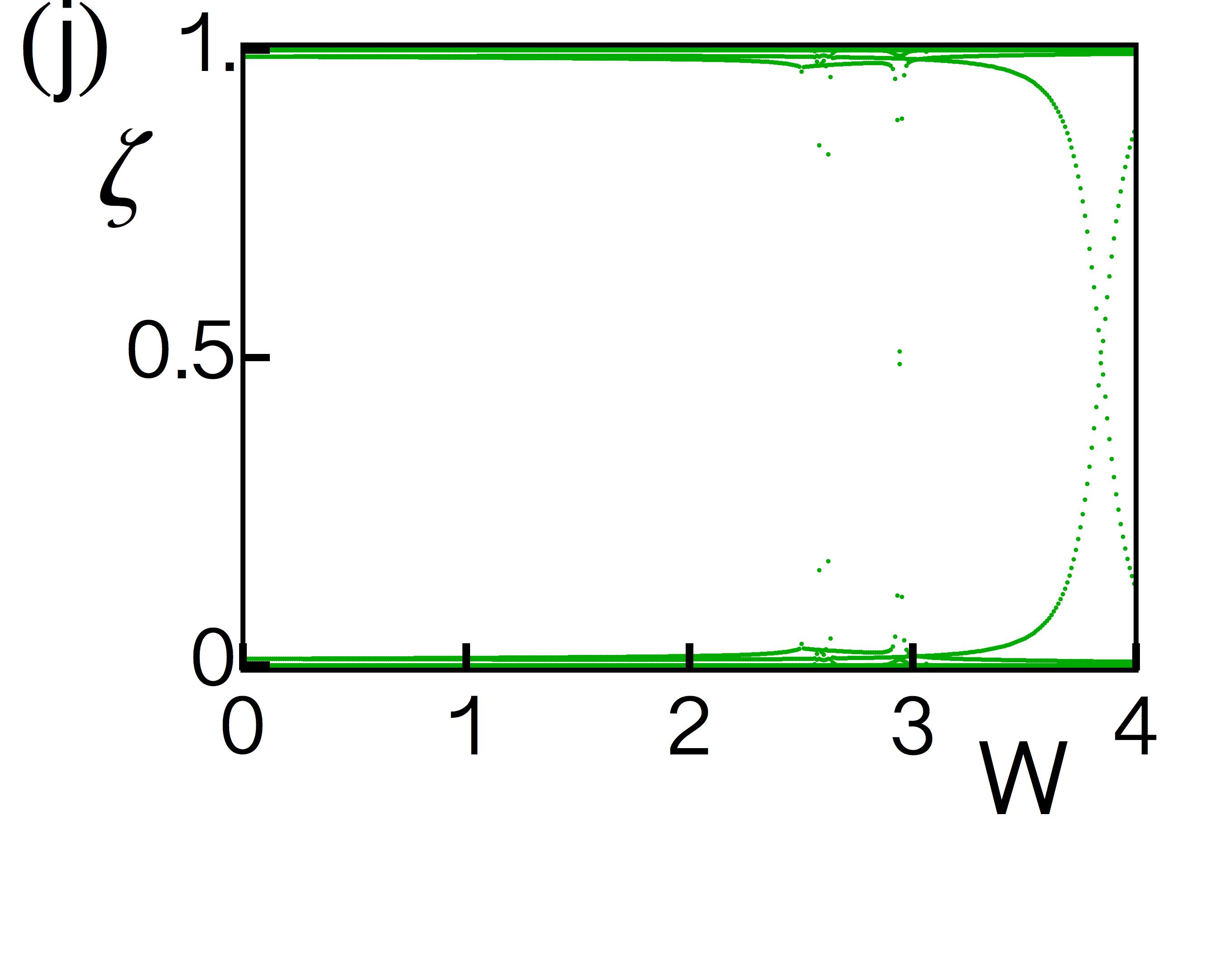}\hspace{0.22cm}
\includegraphics[trim = 0cm 1cm 0cm 0mm, clip, scale=0.22]{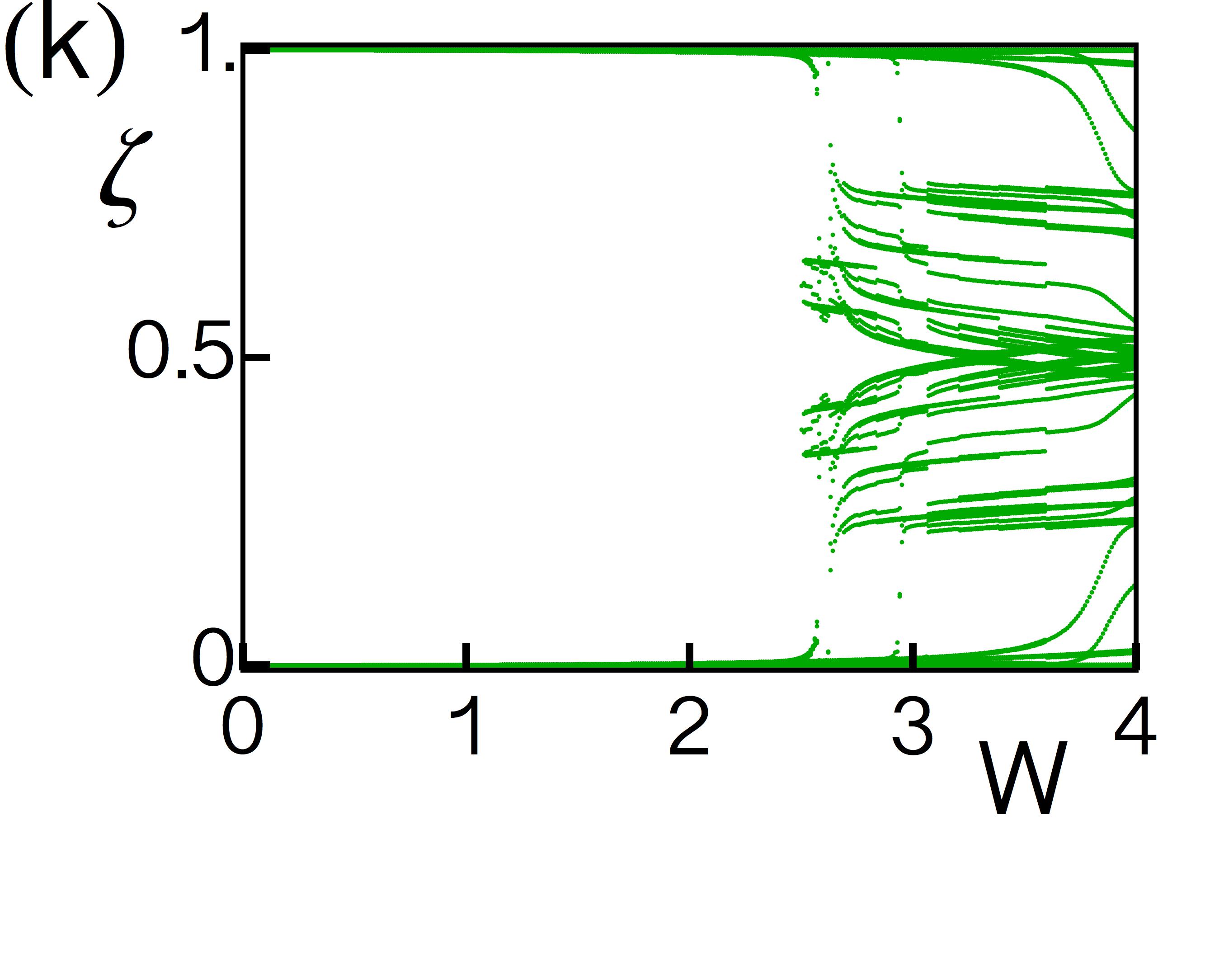}\hspace{0.22cm}
\includegraphics[trim = 0cm 1cm 0cm 0mm, clip, scale=0.22]{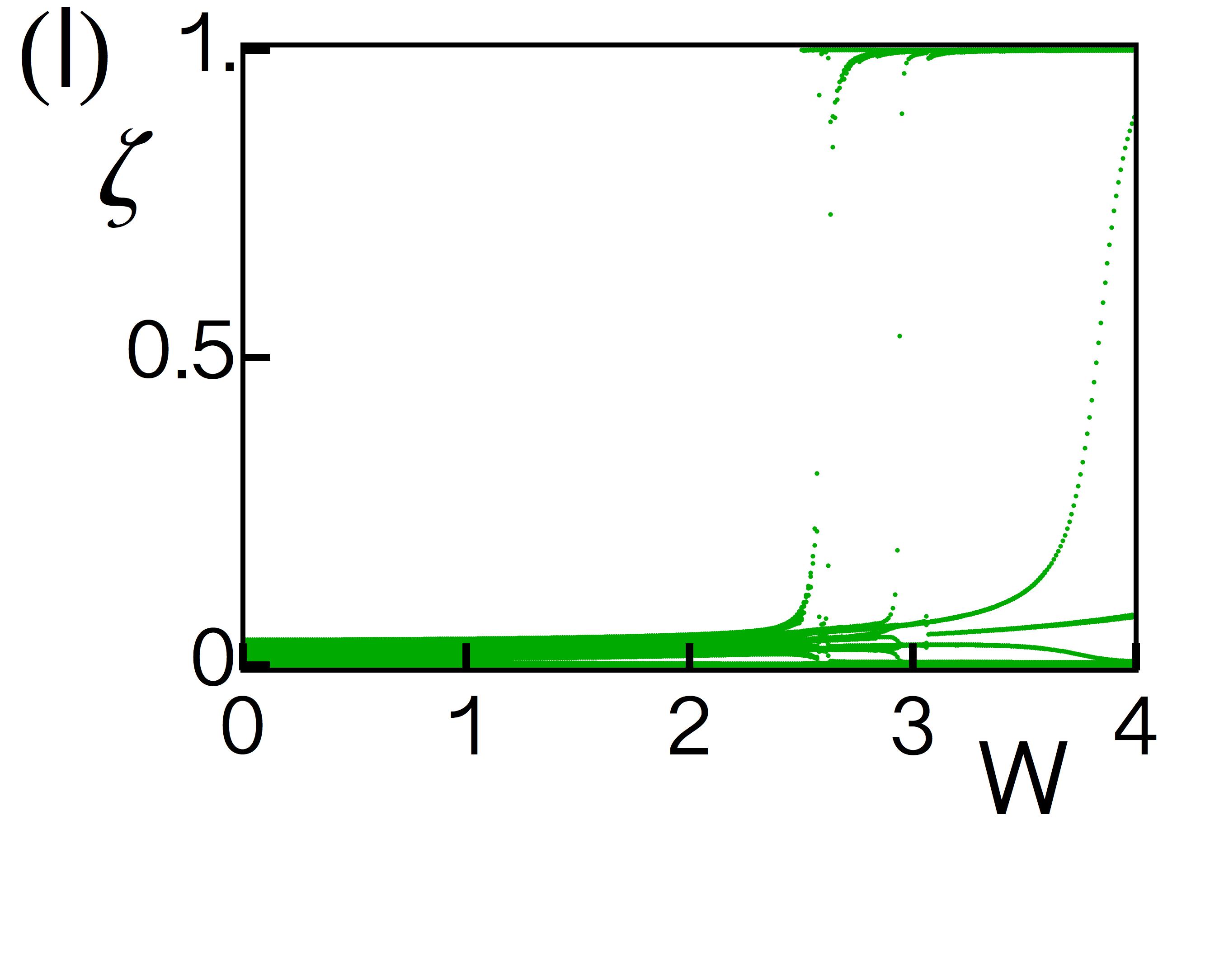}

\caption{Entanglement of a chiral topological insulator in class AIII subjected to an Aubry-Andr\'e potential. We show the spatial (left column), momentum (center column) and orbital (right column) entanglement plots. The top row corresponds to each type of entanglement in the $m \,\text{vs.} W$ parameter space. The overlayed black curves denote the boundary of the region where the system is topological according to the nontrivial polarization of the system. The second, third and fourth rows show the entanglement spectra as a function of $W$, for three values of the mass parameter which are shown in the top row by the colored lines.}\label{Fig_set15}
\end{figure*} 

This symmetry class has an integer topological invariant $\nu,$ and in the two-band model we will discuss, the invariant is given by the expression
\begin{equation}
\nu=\frac{1}{2\pi}\int^{2\pi}_0 dk \,  \epsilon_{ij} \hat{d}_i \partial_k \hat{d}_j,
\end{equation}
which counts the number of times the $\hat{d}$ vector winds around the unit circle in $\hat{d}$-space as $k$ traverses the Brillouin zone. This is the topological invariant that characterizes the nontrivial nature of the ground state. In the topological phase $(-1<m <1)$ our model has $\nu=1.$ 
One physical manifestation of a nonzero $\nu$ is the formation of protected boundary states when the system has open boundaries. These boundary modes indicate a non-zero charge polarization equal to $P=\tfrac{1}{2}(\nu\; \text{mod} 1)$ as shown in Ref. [\onlinecite{MS2013B}]. Since we will focus on the case $\nu=1$ case, the system will have $P=1/2$. In Appendix \ref{AppB}, we explain how we can compute $P$ when the system is disordered, which we will do to determine the topological phase boundary for strong disorder. More generally one can use the real-space winding number invariant introduced in Ref. [\onlinecite{MS2013B}].
When $\nu \neq 0$ the winding indicates that in the clean system the internal degree of freedom is entangled with the momentum, and we will use this idea to explore the entanglement properties of the disordered system.

We will add chiral-symmetry preserving disorder via the on-site potential given by
\begin{equation}
H_{\text{dis}}= \sum_{n}c^{\dagger}_{n\sigma} \left[ w_n \sigma^3_{\sigma \sigma'}\right] c_{n \sigma'}.
\end{equation}
For simplicity, we will primarily use the Aubry-Andr\'e quasicrystal potential for $w_n$ because there is no need to disorder-average. We will also show some results for the case in which $w_n$ represents uncorrelated disorder towards the end of the discussion as a comparison. We proceed by showing the results of spatial, momentum, and orbital entanglement cuts and use the results to generate a hybrid entanglement cut that combines a momentum and orbital cut.

\subsection{Conventional entanglement cuts}

In this section we perform three types of entanglement cuts, namely the same spatial and momentum entanglement cuts we made in previous sections and, in addition, we calculate the orbital entanglement by tracing out one of the components of the internal degree of freedom. As mentioned, the results in this section will further motivate a fourth type of entanglement cut that mixes the momentum and orbital cuts.

Before proceeding, we should emphasize that the spatial entanglement spectrum has been widely used to characterize topological insulator states \cite{Ryu2006,Fang2013,Hughes2011,Turner2010, Pollman2010, Turner2011,Fidkowski2010,Flammia2009,Prodan2010,Qi2012,Alexandradinata2011}. When the system is topological, the spatial entanglement spectrum exhibits single-particle mid-gap entanglement modes that are topologically protected. The fundamental reason for these modes is that the correlation matrix is equivalent to a spectrally flattened version of the Hamiltonian, so that upon an entanglement cut, boundary modes naturally arise\cite{Turner2010,Hughes2011,Fidkowski2010}. We will focus more on the momentum and orbital cuts since those have not been emphasized as much in the literature.

\begin{figure}
\includegraphics[trim = 0mm 0cm 5.5cm 0mm, clip, scale=0.4]{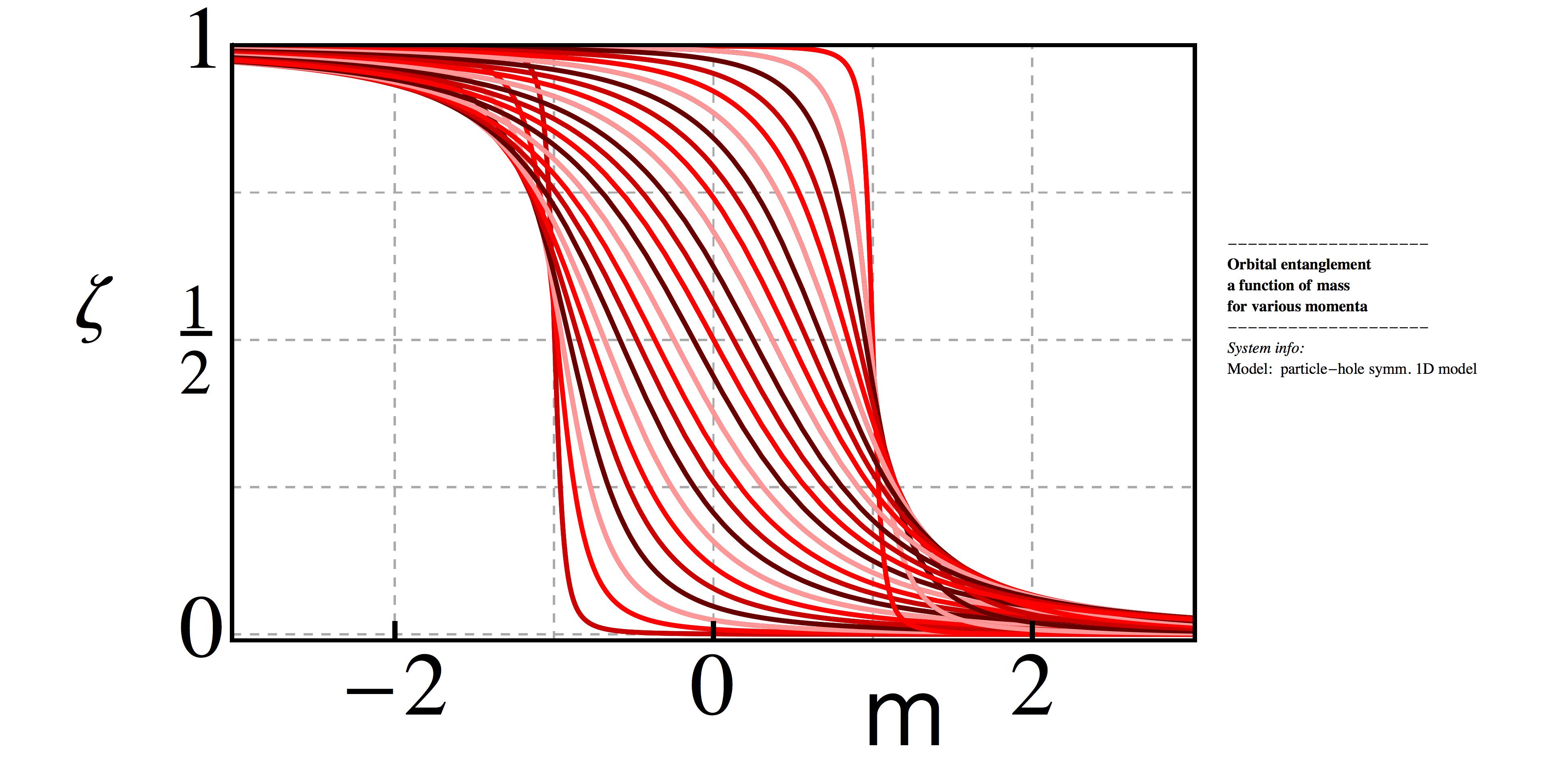}
\caption{Orbital entanglement spectrum of the lattice Dirac model in the clean limit as a function of the mass parameter. Each line corresponds to a fixed value of the momentum $k$ in the range $[0,  2\pi]$.}\label{Fig_cleanconf}
\end{figure}

\begin{figure*}
\includegraphics[trim = 0cm 0cm 0cm 0cm, clip, scale=0.35]{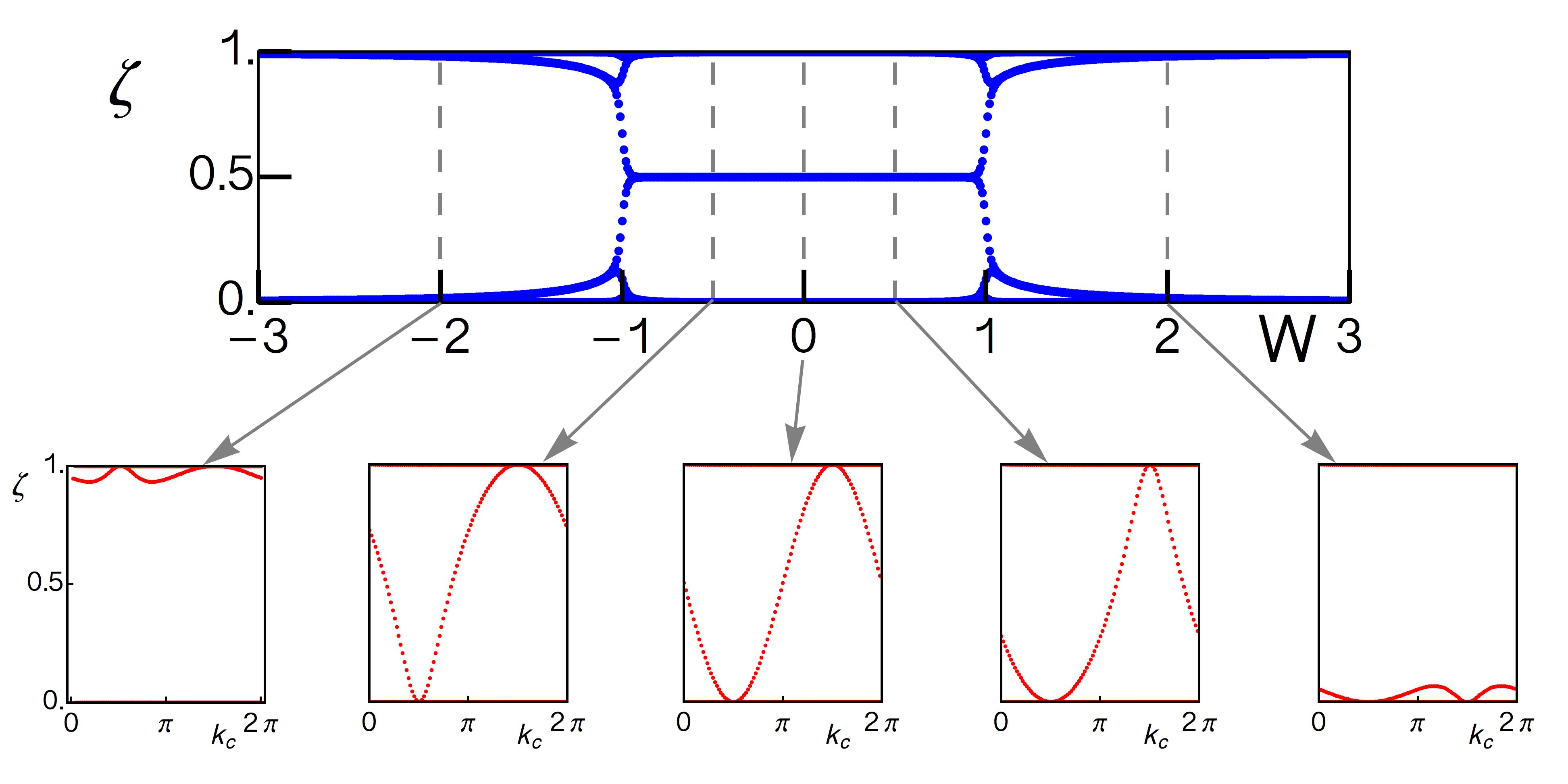}
\caption{The top plot shows the spatial entanglement spectrum as a function of the mass parameter when the system is clean, for a lattice size $N=102$. The lower red plots show the hybrid momentum-orbital entanglement spectrum for various mass values shown on the horizontal axis of top figure.}\label{Fig_CleanMomCut}
\end{figure*}

In Figs. \ref{Fig_set15}a,b,c we show density plots of the the spatial, momentum, and orbital entanglement entropy, respectively, over the $(W, m)$ parameter space. The phase boundary is given by the thick black line in each plot, which we obtained by calculating the polarization of the system numerically. The topological state is realized when the parameters are set within this black line where the polarization is exactly $P=1/2$, and the rest of parameter space corresponds to a trivial state where $P=0$.

From these plots, it seems clear that the phase diagram is most clearly identified in the spatial entanglement and orbital entanglement plots. When the system is topological there is a nearly constant, nonzero spatial entanglement entropy, as well as a significant degree of orbital entanglement. When the parameters cross the phase boundary, these two types of entanglement decrease significantly and become vanishingly small.  The match is not exact however, since there are some regions that appear to have spatial entanglement which lie outside of the topological phase boundary. It is unclear if these regions of nonzero spatial entanglement outside the topological regime are a peculiarity of the model, or are a signature of something more interesting. We leave this for future work. 

In contrast to these two types of entanglement, the momentum entanglement entropy does not appear to correlate clearly with the presence of the topological state. What appears to happen is that the momentum entanglement starts off from zero, as it must in the clean limit,  and increases as the disorder strength increases. There is a region in the parameter space at which the momentum entanglement exhibits a marked increase, which suggests that the Anderson insulator is realized beyond those points. However, this surge in momentum entanglement occurs regardless of whether the system begins in the  topological or trivial phase. For example, if one fixes $m$ to lie deep in the topological or trivial regimes the momentum entanglement entropy eventually surges as $W$ is increased for both cases. 

To better understand our observations about the entanglement entropy maps, let us take a look at the entanglement spectra for a few cases. We show examples of the spectra plotted vs. $W$ for three fixed values of the mass parameter in Figs. \ref{Fig_set15}d-l in the phase diagram. The particular choices of the mass are indicated with horizontal lines in the corresponding Figs. \ref{Fig_set15}a,b,c. Let us discuss the notable features observed in these figures.

The first line we consider is $m=0.0$ which is deep in the topological phase for weak disorder. The entanglement spectra for this line is given by the blue plots (second row): Fig.\ref{Fig_set15}d,e,f. Along this line, the topological phase transition occurs at around $W=2.0$. Up to this value of disorder, we observe entanglement properties which are reminiscent of the clean system: the spatial entanglement receives contributions almost exclusively from two $1/2$ modes; the momentum entanglement remains largely suppressed, albeit increasing gradually with disorder; and the orbital entanglement modes evenly occupy the whole region between $0$ and $1$ (compare the disordered orbital entanglement with the clean system shown in Fig. \ref{Fig_cleanconf}). As the disorder gets close to $W_c$, there are also clear signatures of when the topological phase transition occurs: at the critical point, the spatial and orbital entanglement modes begin to diverge toward $0$  and $1$, while the momentum entanglement modes collapse toward $1/2$. This is the type of behavior that is expected for the system as it transits toward the trivial Anderson insulator phase.

\begin{figure*}
\includegraphics[trim = 0cm 0cm 0cm 0cm, clip, scale=0.35]{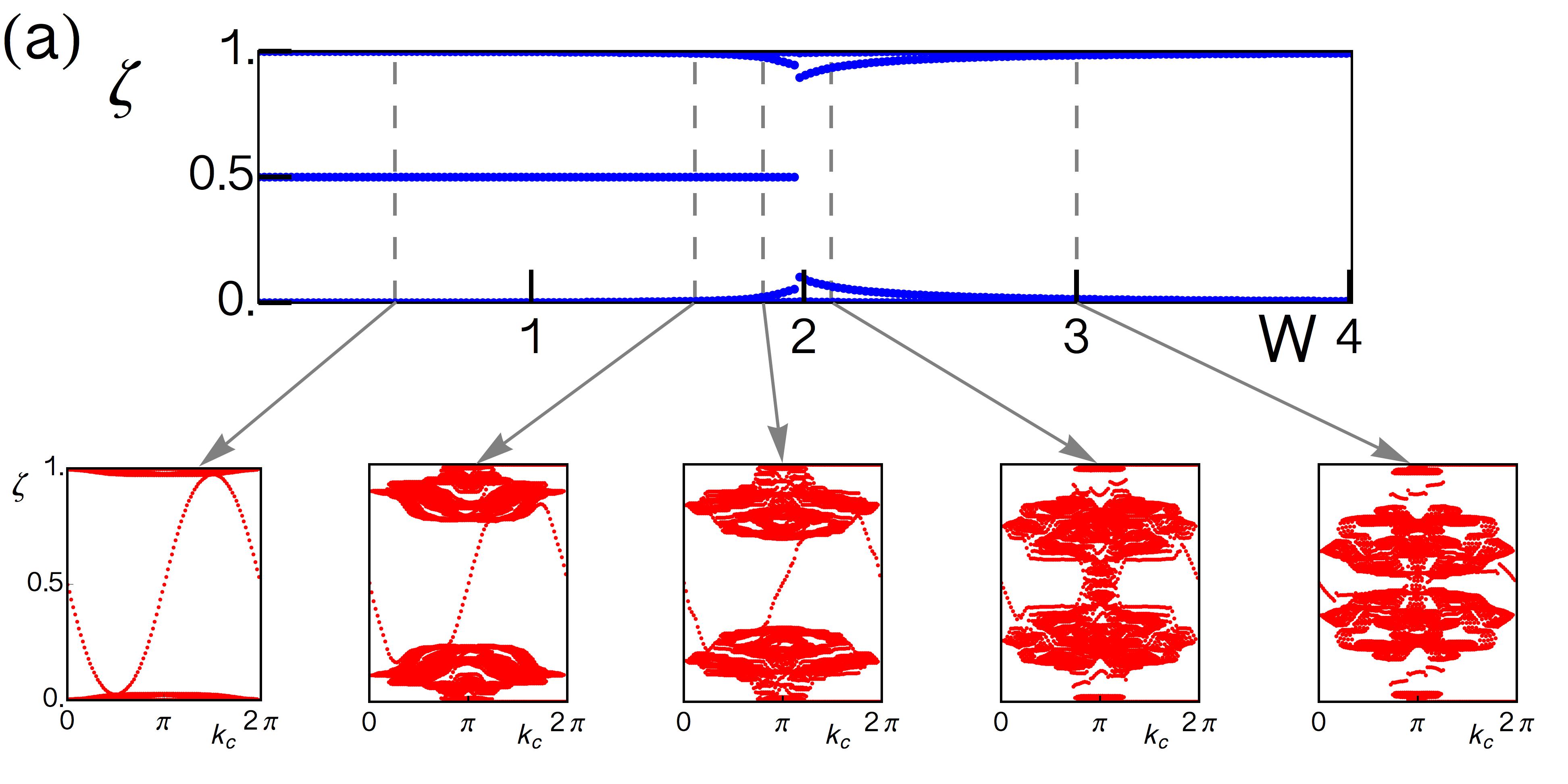}\vspace{0.5cm}
\includegraphics[trim = 0cm 0cm 0cm 0cm, clip, scale=0.35]{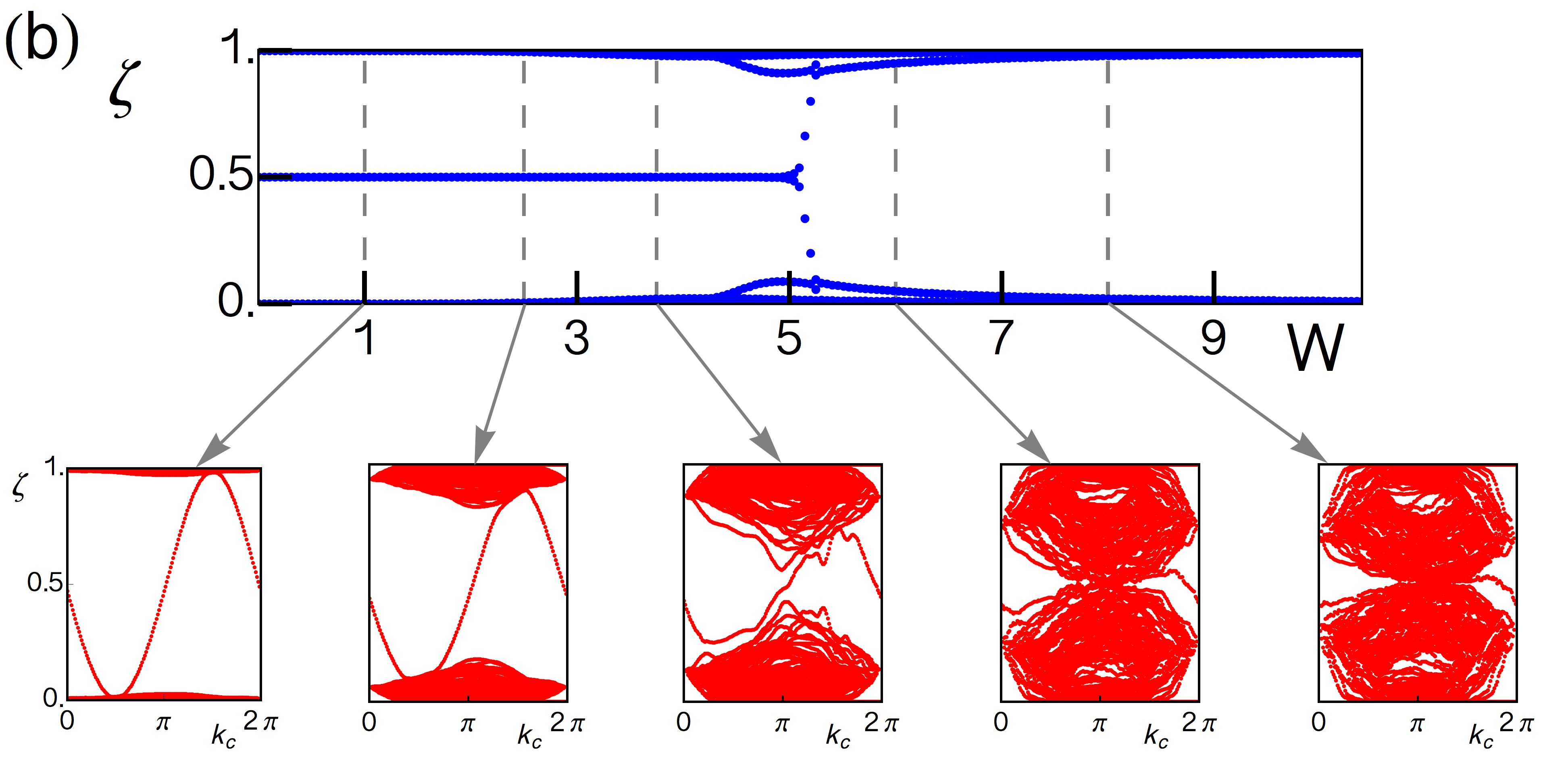}
\caption{(a) The large blue plot shows the spatial entanglement spectrum as a function of disorder when the mass parameter is set at $m=0$ for a lattice size $N=102$ and for the AA potential. The lower red plots show the hybrid momentum-orbital entanglement spectrum for various disorder strengths shown on the horizontal axis of top figure. (b) This set of subfigures correspond to the case when we use uncorrelated disorder with $N=150$ and $m=0$.}\label{Fig_wdMom}
\end{figure*}

The second mass line we consider is $m=1.4$ at which something interesting occurs. For this value of the mass parameter, the system is in the trivial phase at zero disorder. What is special about this value of the mass parameter is that, although it starts off from a trivial phase, the disorder is able to induce a topological state that occurs approximately between $W_1\approx 1.2$ and $W_2\approx 2.0$.  The entanglement spectra for this line are shown in red plots (third row) in Fig. \ref{Fig_set15}g,h,i. In this region of intermediate disorder, all three types of entanglement spectra exhibit the type of behavior we observed for the $m=0.0$ case in the topological regime. Beyond $W_2$, the trivial Anderson insulator is obtained, with the corresponding entanglement properties matching the expected behavior for such a phase. 

The final line we consider has $m=2.5$. For this value of the mass parameter, the system is in the trivial phase at zero disorder, and remains in that phase for all values of the disorder strength.  The entanglement spectra for this line is given by the green plots (fourth row) in Fig. \ref{Fig_set15}j,k,l. The behavior for all three types of entanglement is the same as what we would expect for a localized insulator. In the momentum entanglement spectrum in Fig. \ref{Fig_set15}k we see the feature on which we commented earlier, namely that even when starting in the trivial insulator phase there is a surge of momentum entanglement, presumably due to the transition from a band insulator to a localized Anderson insulator.  

From these results, we would like to bring attention to the behavior of both the momentum and orbital entanglement, as this will help motivate another type of entanglement cut. First, note that, even though the momentum entanglement does not reveal clearly the difference between the topological and trivial states clearly even at zero disorder, it does appear to reveal the point at which the trivial Anderson insulator is obtained when the disorder becomes strong enough. This is a nontrivial result, because even while the system is topological, the single-particle states are localized (they can be represented as exponentially localized Wannier functions) , which might naively have led us to assume that the momentum entanglement spectrum should saturate at $1/2$ even for weak disorder. However, from our results it is clear that the saturation near $1/2$ occurs only at a finite disorder strength. The connection of momentum entanglement saturation with the occurrence of the trivial Anderson insulator is consistent with what we observed for the metal-insulator insulator transitions of the previous sections. This thus suggests that this connection is generic, at least for one-dimensional systems. We will exploit this observation in what follows.

Regarding the orbital entanglement, we now argue that this type of entanglement is intimately related with the winding property of the ground state. To understand this, let us consider the orbital entanglement in the clean limit. Since the wave functions of the clean system can be computed analytically, the orbital entanglement can be obtained explicitly as
\begin{eqnarray}
\zeta(k,m)=\frac{1}{2}\left[1-\frac{m+\sin k}{\sqrt{\cos^2 k+\left(m+\sin k\right)^2}}\right]. \label{orbtent}
\end{eqnarray}
We show the orbital entanglement as a function of the mass parameter in Fig. \ref{Fig_cleanconf}. The curves followed by the orbital entanglement modes are very reminiscent of the behavior observed in Figs. \ref{Fig_set15} f,i,l, with the important difference that in the earlier set of figures it is the disorder which is being varied, whereas in the clean limit we only vary the mass parameter. 

We thus see that  the entanglement modes $\zeta(k,m)$ exhibit spectral flow from $0$ up until $1$ as $k$ goes from $0$ to $2\pi$ only when $m \in (-1,\,1),$ i.e., in the topological phase. Since the momentum part of the correlation matrix is diagonal in the clean limit, these entanglement modes originate exclusively from the entanglement of the internal degree of freedom at each value of $k$. Furthermore, the flow of entanglement modes when the system is topological occurs essentially because of the winding of the $\hat{d}$ vector as the Brillouin zone is traversed. This suggests that the flow of entanglement modes in Figs.\ref{Fig_set15} f,i,l are describing the unwinding of the internal degree of freedom as the system transits from a topological to a trivial state. Observations of this nature were also pointed out for two-dimensional topological models in Ref. [\onlinecite{Legner2013}], albeit exclusively for clean systems. This, together with the behavior of the momentum entanglement, suggests a useful quantity would be a mixed type entanglement which we discuss in the next section.

\subsection{Variation of the entanglement cut: winding in the momentum entanglement}

To define the hybrid momentum-orbital entanglement let us start by considering the clean system.  Let us allow for the variation of the momentum wave vector $k_c$ at which we perform the entanglement cut in the Brillouin zone, so that we keep $k\in[0,k_c].$ In our partition we keep all $k$-states between $[0,k_c]$ and we keep all the corresponding orbital degrees of freedom except for $k=k_c$ for which we also remove one of the orbitals.   Thus, in one subspace we have $N_A=2N_{k_c}-1$ states where $N_{k_c}$ is the number of allowed momenta in the set $[0,k_c]$, and in the other subspace we have $N_B=2(N-N_{k_c})+1$ remaining states where $N=\tfrac{1}{2}(N_A+N_B)$ is the total number of lattice sites. 

In the clean case, the correlation matrix is block-diagonal in the momentum $k$ so the the entanglement spectrum is composed of the set entanglement modes $\{1,0,\zeta(k_c,m)\}$ at each $k$ where $1(0)$ is $N_A-1 (N_B)$-fold degenerate and the non-trivial eigenvalue $\zeta(k_c,m)$ arises from the orbital cut at $k_c.$ In Fig. \ref{Fig_CleanMomCut}, we make a comparison between the spatial entanglement and the hybrid entanglement as a function of the mass parameter when the disorder is set to zero. In the hybrid entanglement there is a flow of entanglement modes between $0$ and $1$  when the system is topological as expected from Eq. \ref{orbtent}. In the trivial phase there is not a complete spectral flow as a function of $k_c.$  This cut is not so revealing in the clean system as it just re-expresses Eq. \ref{orbtent} if we plot the entanglement spectra for the whole range of possible $k_c.$

Let us now add disorder to the system. We consider the Aubry-Andr\'e potential first. In the sequence of plots shown in Fig. \ref{Fig_wdMom}a, we vary the disorder strength. For weak disorder, the entanglement spectrum largely retains the same structure we found in the clean system. As the disorder keeps increasing, and localization sets in, the momentum entanglement modes that were at $1$ and $0$ start to drift towards $1/2$, signaling an increase in momentum entanglement. Interestingly, the entanglement modes that traverse the entanglement gap are still present, and suggest that the system is still topological. This is corroborated with a calculation of the topological invariant, which is still quantized to $\nu=1$ at that disorder strength. Finally, when the disorder strength crosses the critical point, the momentum entanglement modes fill the gap. For comparison, and to illustrate the generality of our observations, we also show the the case of uncorrelated disorder, for one disorder configuration, in Fig. \ref{Fig_wdMom}b. It is nice to confirm that this type of entanglement can also be illustrative without disorder averaging. The overall behavior of the Anderson transition for uncorrelated disorder matches what is observed for the transition in the Aubry-Andr\'e potential for the topological insulator.

From these observations, we find that there is a real sense in which the topological winding of the ground state is still encoded in the disordered system. Even though the disorder induces scattering and tends to randomize the internal configuration of the single-particle states, the correlation between momentum and internal degree of freedom persists throughout the topological state. This manifests itself consistently  as a spectral flow of orbital entanglement modes as a function of the momentum entanglement cut. The topological phase transition toward the trivial state eventually occurs when the scattering is strong enough that entanglement modes saturate $1/2$, which we have identified throughout this work as leading to the trivial Anderson insulator. It would be interesting to see how these results can be applied to higher-dimensional topological systems. In higher dimensions the scattering kinematics are much less constrained and it is not clear how well the topological information would be encoded in the momentum entanglement. However, here we see that we can capture the correlations between the momentum and the internal degrees of freedom which is a fundamental feature of free-fermion topological phases, so perhaps this tool will still be useful. It may also be useful to consider this type of hybrid cut for numerically identifying 1D interacting topological phases of fermions or even in spin-systems.

			\section{Conclusions}

In this work we have discussed the signatures in the entanglement of disordered fermions when they undergo Anderson localization transitions. Disordered metallic states in 1D that have suppressed scattering exhibit marked signatures in the momentum entanglement spectrum. Delocalized states occur when momentum entanglement is suppressed and the Anderson insulator is obtained when there is a saturation of momentum entanglement modes near $1/2$. The momentum entanglement entropy is able to yield the disorder strength for which localization occurs, even in the presence of Hubbard-type interactions. Regarding disordered 1D topological states, we have found that the nontrivial nature of the ground state can be captured by performing an appropriate simultaneous entanglement cut in momentum and orbital space. This type of entanglement cut suggests that the ground state still has some correlation between the momentum and the orbital degree of freedom which manifests in the nontrivial topological boundary states of the system.

It would be interesting to further extend this study to higher-dimensional systems and with other symmetries. In 2D and in 3D there are interesting models that exhibit metal-insulator transitions as well as diverse nontrivial topological properties. Consequently, one might encounter further interesting and useful features in the momentum entanglement that probe the quantum properties of the ground state. However, because of the much wider range of scattering kinematics, we expect that the extension of the pure momentum entanglement will be challenging. Some work in this direction was carried out recently in Ref. [\onlinecite{Andrade2014}] where indeed they find the higher-dimensional extension to be challenging. 

			\section{Acknowledgements}
We would like to thank E. Prodan and J. Song for useful discussions, and the UIUC ICMT for support. This work was supported by ONR award N0014-12-1-0935.

\appendix

\section{Calculation of the localization length}\label{AppA}

In order to compute the localization length, we follow the calculation in terms of transfer matrices detailed in \cite{Sedrakyan2011}. This can be achieved by first computing numerically the matrix $\Lambda$ given by

\begin{equation}
\Lambda = \lim_{L\rightarrow \infty} \left[\prod_{n=1}^{L}T_n \prod_{n=L}^{1}T_n^{\dagger}\right]^{1/2L}
\end{equation}
where the transfer matrix $T_n$ at the position $n$ and energy $E$ is given by
\begin{eqnarray}
T_n=\left(\begin{array}{cc}
\frac{E-w_n}{t} & -1 \\
1& 0
\end{array}\right).
\end{eqnarray}

The eigenvalue of the matrix $\Lambda$ which is closest to unity can be written as $e^{\lambda(E)}$. Finally, the exponent $\lambda(E)$ is the so-called Lyapunov exponent which is related to the localization length by $\xi(E)=\lambda^{-1}(E)$. 

\section{Calculation of the polarization of 1D fermions}\label{AppB}

We can obtain the polarization of a system of 1D fermions by computing the eigenvalues of the projected position operator 
\begin{equation}
X_{\mathcal{P}}=\mathcal{P} X \mathcal{P}
\end{equation}
where $\mathcal{P}$ is the projection operator of the occupied states and $X$ is the position operator. In a finite system with periodic boundary conditions, the position operator is written in the exponential form
\begin{equation} 
X=\sum_{R} e^{\frac{2\pi R}{N} i} \ket{R}\bra{R}.
\end{equation}
where $\{\ket{R}\}$ are single-particle position basis states. Using the set of eigenvalues $\{ \xi_n\}$ of $X_{\mathcal{P}}$, the polarization is then given by
\begin{equation}
P=\sum_{n}\left(\frac{1}{2\pi}\text{Im} \log \xi_n-\frac{n}{N}\right)
\end{equation}
which just measures the relative shift of the Wannier centers with respect to the lattice sites.


\begin{thebibliography}{48}
\expandafter\ifx\csname natexlab\endcsname\relax\def\natexlab#1{#1}\fi
\expandafter\ifx\csname bibnamefont\endcsname\relax
  \def\bibnamefont#1{#1}\fi
\expandafter\ifx\csname bibfnamefont\endcsname\relax
  \def\bibfnamefont#1{#1}\fi
\expandafter\ifx\csname citenamefont\endcsname\relax
  \def\citenamefont#1{#1}\fi
\expandafter\ifx\csname url\endcsname\relax
  \def\url#1{\texttt{#1}}\fi
\expandafter\ifx\csname urlprefix\endcsname\relax\def\urlprefix{URL }\fi
\providecommand{\bibinfo}[2]{#2}
\providecommand{\eprint}[2][]{\url{#2}}

\bibitem[{\citenamefont{Evers and Mirlin}(2008)}]{Evers2008}
\bibinfo{author}{\bibfnamefont{F.}~\bibnamefont{Evers}} \bibnamefont{and}
  \bibinfo{author}{\bibfnamefont{A.~D.} \bibnamefont{Mirlin}},
  \bibinfo{journal}{Rev. Mod. Phys.} \textbf{\bibinfo{volume}{80}},
  \bibinfo{pages}{1355} (\bibinfo{year}{2008}),
  \urlprefix\url{http://link.aps.org/doi/10.1103/RevModPhys.80.1355}.

\bibitem[{\citenamefont{Schnyder et~al.}(2008)\citenamefont{Schnyder, Ryu,
  Furusaki, and Ludwig}}]{Schnyder2008}
\bibinfo{author}{\bibfnamefont{A.~P.} \bibnamefont{Schnyder}},
  \bibinfo{author}{\bibfnamefont{S.}~\bibnamefont{Ryu}},
  \bibinfo{author}{\bibfnamefont{A.}~\bibnamefont{Furusaki}}, \bibnamefont{and}
  \bibinfo{author}{\bibfnamefont{A.~W.~W.} \bibnamefont{Ludwig}},
  \bibinfo{journal}{Phys. Rev. B} \textbf{\bibinfo{volume}{78}},
  \bibinfo{pages}{195125} (\bibinfo{year}{2008}),
  \urlprefix\url{http://link.aps.org/doi/10.1103/PhysRevB.78.195125}.

\bibitem[{\citenamefont{Spiros~Evangelou}(2004)}]{Evangelou2004}
\bibinfo{author}{\bibfnamefont{B.~K.} \bibnamefont{Spiros~Evangelou},
  \bibfnamefont{Dimitris~Katsanos}}, \bibinfo{journal}{Lecture Notes in
  Physics} \textbf{\bibinfo{volume}{630}}, \bibinfo{pages}{203}
  (\bibinfo{year}{2004}).

\bibitem[{\citenamefont{Pal and Huse}(2010)}]{Pal2010}
\bibinfo{author}{\bibfnamefont{A.}~\bibnamefont{Pal}} \bibnamefont{and}
  \bibinfo{author}{\bibfnamefont{D.~A.} \bibnamefont{Huse}},
  \bibinfo{journal}{Phys. Rev. B} \textbf{\bibinfo{volume}{82}},
  \bibinfo{pages}{174411} (\bibinfo{year}{2010}),
  \urlprefix\url{http://link.aps.org/doi/10.1103/PhysRevB.82.174411}.

\bibitem[{\citenamefont{Lahini et~al.}(2008)\citenamefont{Lahini, Avidan,
  Pozzi, Sorel, Morandotti, Christodoulides, and Silberberg}}]{Lahini2008}
\bibinfo{author}{\bibfnamefont{Y.}~\bibnamefont{Lahini}},
  \bibinfo{author}{\bibfnamefont{A.}~\bibnamefont{Avidan}},
  \bibinfo{author}{\bibfnamefont{F.}~\bibnamefont{Pozzi}},
  \bibinfo{author}{\bibfnamefont{M.}~\bibnamefont{Sorel}},
  \bibinfo{author}{\bibfnamefont{R.}~\bibnamefont{Morandotti}},
  \bibinfo{author}{\bibfnamefont{D.~N.} \bibnamefont{Christodoulides}},
  \bibnamefont{and}
  \bibinfo{author}{\bibfnamefont{Y.}~\bibnamefont{Silberberg}},
  \bibinfo{journal}{Phys. Rev. Lett.} \textbf{\bibinfo{volume}{100}},
  \bibinfo{pages}{013906} (\bibinfo{year}{2008}),
  \urlprefix\url{http://link.aps.org/doi/10.1103/PhysRevLett.100.013906}.

\bibitem[{\citenamefont{Schulte et~al.}(2005)\citenamefont{Schulte,
  Drenkelforth, Kruse, Ertmer, Arlt, Sacha, Zakrzewski, and
  Lewenstein}}]{Schulte2005}
\bibinfo{author}{\bibfnamefont{T.}~\bibnamefont{Schulte}},
  \bibinfo{author}{\bibfnamefont{S.}~\bibnamefont{Drenkelforth}},
  \bibinfo{author}{\bibfnamefont{J.}~\bibnamefont{Kruse}},
  \bibinfo{author}{\bibfnamefont{W.}~\bibnamefont{Ertmer}},
  \bibinfo{author}{\bibfnamefont{J.}~\bibnamefont{Arlt}},
  \bibinfo{author}{\bibfnamefont{K.}~\bibnamefont{Sacha}},
  \bibinfo{author}{\bibfnamefont{J.}~\bibnamefont{Zakrzewski}},
  \bibnamefont{and}
  \bibinfo{author}{\bibfnamefont{M.}~\bibnamefont{Lewenstein}},
  \bibinfo{journal}{Phys. Rev. Lett.} \textbf{\bibinfo{volume}{95}},
  \bibinfo{pages}{170411} (\bibinfo{year}{2005}),
  \urlprefix\url{http://link.aps.org/doi/10.1103/PhysRevLett.95.170411}.

\bibitem[{\citenamefont{Kondov et~al.}(2011)\citenamefont{Kondov, McGehee,
  Zirbel, and DeMarco}}]{Kondov2011}
\bibinfo{author}{\bibfnamefont{S.~S.} \bibnamefont{Kondov}},
  \bibinfo{author}{\bibfnamefont{W.~R.} \bibnamefont{McGehee}},
  \bibinfo{author}{\bibfnamefont{J.~J.} \bibnamefont{Zirbel}},
  \bibnamefont{and} \bibinfo{author}{\bibfnamefont{B.}~\bibnamefont{DeMarco}},
  \bibinfo{journal}{Science} \textbf{\bibinfo{volume}{334}},
  \bibinfo{pages}{66} (\bibinfo{year}{2011}),
  \eprint{http://www.sciencemag.org/content/334/6052/66.full.pdf},
  \urlprefix\url{http://www.sciencemag.org/content/334/6052/66.abstract}.

\bibitem[{\citenamefont{Shi}(2004)}]{Shi2004}
\bibinfo{author}{\bibfnamefont{Y.}~\bibnamefont{Shi}}, \bibinfo{journal}{J.
  Phys. A: Math. Gen.} \textbf{\bibinfo{volume}{37}}, \bibinfo{pages}{6807}
  (\bibinfo{year}{2004}).

\bibitem[{\citenamefont{Amico et~al.}(2008)\citenamefont{Amico, Fazio,
  Osterloh, and Vedral}}]{Amico2008}
\bibinfo{author}{\bibfnamefont{L.}~\bibnamefont{Amico}},
  \bibinfo{author}{\bibfnamefont{R.}~\bibnamefont{Fazio}},
  \bibinfo{author}{\bibfnamefont{A.}~\bibnamefont{Osterloh}}, \bibnamefont{and}
  \bibinfo{author}{\bibfnamefont{V.}~\bibnamefont{Vedral}},
  \bibinfo{journal}{Rev. Mod. Phys.} \textbf{\bibinfo{volume}{80}},
  \bibinfo{pages}{517} (\bibinfo{year}{2008}),
  \urlprefix\url{http://link.aps.org/doi/10.1103/RevModPhys.80.517}.

\bibitem[{\citenamefont{Li and Haldane}(2008)}]{Li2008}
\bibinfo{author}{\bibfnamefont{H.}~\bibnamefont{Li}} \bibnamefont{and}
  \bibinfo{author}{\bibfnamefont{F.~D.~M.} \bibnamefont{Haldane}},
  \bibinfo{journal}{Phys. Rev. Lett.} \textbf{\bibinfo{volume}{101}},
  \bibinfo{pages}{010504} (\bibinfo{year}{2008}),
  \urlprefix\url{http://link.aps.org/doi/10.1103/PhysRevLett.101.010504}.

\bibitem[{\citenamefont{Fradkin}(2013)}]{Fradkin2013}
\bibinfo{author}{\bibfnamefont{E.}~\bibnamefont{Fradkin}},
  \emph{\bibinfo{title}{Field Theories of Condensed Matter Physics}}
  (\bibinfo{publisher}{Cambridge University Press}, \bibinfo{year}{2013}).

\bibitem[{\citenamefont{Pollmann and Moore}(2010)}]{Pollmann2010}
\bibinfo{author}{\bibfnamefont{F.}~\bibnamefont{Pollmann}} \bibnamefont{and}
  \bibinfo{author}{\bibfnamefont{J.}~\bibnamefont{Moore}},
  \bibinfo{journal}{New J. Phys.} \textbf{\bibinfo{volume}{12}},
  \bibinfo{pages}{025006} (\bibinfo{year}{2010}).

\bibitem[{\citenamefont{Varga and Méndez-Bermúdez}(2008)}]{Varga2008}
\bibinfo{author}{\bibfnamefont{I.}~\bibnamefont{Varga}} \bibnamefont{and}
  \bibinfo{author}{\bibfnamefont{J.~A.} \bibnamefont{Méndez-Bermúdez}},
  \bibinfo{journal}{physica status solidi (c)} \textbf{\bibinfo{volume}{5}},
  \bibinfo{pages}{867} (\bibinfo{year}{2008}), ISSN \bibinfo{issn}{1610-1642},
  \urlprefix\url{http://dx.doi.org/10.1002/pssc.200777589}.

\bibitem[{\citenamefont{Jia et~al.}(2008)\citenamefont{Jia, Subramaniam,
  Gruzberg, and Chakravarty}}]{Jia2008}
\bibinfo{author}{\bibfnamefont{X.}~\bibnamefont{Jia}},
  \bibinfo{author}{\bibfnamefont{A.~R.} \bibnamefont{Subramaniam}},
  \bibinfo{author}{\bibfnamefont{I.~A.} \bibnamefont{Gruzberg}},
  \bibnamefont{and}
  \bibinfo{author}{\bibfnamefont{S.}~\bibnamefont{Chakravarty}},
  \bibinfo{journal}{Phys. Rev. B} \textbf{\bibinfo{volume}{77}},
  \bibinfo{pages}{014208} (\bibinfo{year}{2008}),
  \urlprefix\url{http://link.aps.org/doi/10.1103/PhysRevB.77.014208}.

\bibitem[{\citenamefont{Chen et~al.}(2012)\citenamefont{Chen, Hsu, Hughes, and
  Fradkin}}]{Chen2012}
\bibinfo{author}{\bibfnamefont{X.}~\bibnamefont{Chen}},
  \bibinfo{author}{\bibfnamefont{B.}~\bibnamefont{Hsu}},
  \bibinfo{author}{\bibfnamefont{T.~L.} \bibnamefont{Hughes}},
  \bibnamefont{and} \bibinfo{author}{\bibfnamefont{E.}~\bibnamefont{Fradkin}},
  \bibinfo{journal}{Phys. Rev. B} \textbf{\bibinfo{volume}{86}},
  \bibinfo{pages}{134201} (\bibinfo{year}{2012}),
  \urlprefix\url{http://link.aps.org/doi/10.1103/PhysRevB.86.134201}.

\bibitem[{\citenamefont{Prodan et~al.}(2010)\citenamefont{Prodan, Hughes, and
  Bernevig}}]{Prodan2010}
\bibinfo{author}{\bibfnamefont{E.}~\bibnamefont{Prodan}},
  \bibinfo{author}{\bibfnamefont{T.~L.} \bibnamefont{Hughes}},
  \bibnamefont{and} \bibinfo{author}{\bibfnamefont{B.~A.}
  \bibnamefont{Bernevig}}, \bibinfo{journal}{Phys. Rev. Lett.}
  \textbf{\bibinfo{volume}{105}}, \bibinfo{pages}{115501}
  (\bibinfo{year}{2010}),
  \urlprefix\url{http://link.aps.org/doi/10.1103/PhysRevLett.105.115501}.

\bibitem[{\citenamefont{Gilbert et~al.}(2012)\citenamefont{Gilbert, Bernevig,
  and Hughes}}]{Gilbert2012}
\bibinfo{author}{\bibfnamefont{M.~J.} \bibnamefont{Gilbert}},
  \bibinfo{author}{\bibfnamefont{B.~A.} \bibnamefont{Bernevig}},
  \bibnamefont{and} \bibinfo{author}{\bibfnamefont{T.~L.}
  \bibnamefont{Hughes}}, \bibinfo{journal}{Phys. Rev. B}
  \textbf{\bibinfo{volume}{86}}, \bibinfo{pages}{041401}
  (\bibinfo{year}{2012}),
  \urlprefix\url{http://link.aps.org/doi/10.1103/PhysRevB.86.041401}.

\bibitem[{\citenamefont{Mondragon-Shem
  et~al.}(2013{\natexlab{a}})\citenamefont{Mondragon-Shem, Khan, and
  Hughes}}]{MS2013}
\bibinfo{author}{\bibfnamefont{I.}~\bibnamefont{Mondragon-Shem}},
  \bibinfo{author}{\bibfnamefont{M.}~\bibnamefont{Khan}}, \bibnamefont{and}
  \bibinfo{author}{\bibfnamefont{T.~L.} \bibnamefont{Hughes}},
  \bibinfo{journal}{Phys. Rev. Lett.} \textbf{\bibinfo{volume}{110}},
  \bibinfo{pages}{046806} (\bibinfo{year}{2013}{\natexlab{a}}),
  \urlprefix\url{http://link.aps.org/doi/10.1103/PhysRevLett.110.046806}.

\bibitem[{\citenamefont{Pouranvari and Yang}(2013)}]{Pouranvari2013}
\bibinfo{author}{\bibfnamefont{M.}~\bibnamefont{Pouranvari}} \bibnamefont{and}
  \bibinfo{author}{\bibfnamefont{K.}~\bibnamefont{Yang}}
  (\bibinfo{year}{2013}), \eprint{cond-mat/1311.4108}.

\bibitem[{\citenamefont{Andrade et~al.}(2013)\citenamefont{Andrade, Steudtner,
  and Vojta}}]{Andrade2014}
\bibinfo{author}{\bibfnamefont{E.~C.} \bibnamefont{Andrade}},
  \bibinfo{author}{\bibfnamefont{M.}~\bibnamefont{Steudtner}},
  \bibnamefont{and} \bibinfo{author}{\bibfnamefont{M.}~\bibnamefont{Vojta}}
  (\bibinfo{year}{2013}), \eprint{cond-mat/1403.2599}.

\bibitem[{\citenamefont{Peschel and Eisler}(2009)}]{peschel2009}
\bibinfo{author}{\bibfnamefont{I.}~\bibnamefont{Peschel}} \bibnamefont{and}
  \bibinfo{author}{\bibfnamefont{V.}~\bibnamefont{Eisler}},
  \bibinfo{journal}{Journal of Physics A: Mathematical and Theoretical}
  \textbf{\bibinfo{volume}{42}}, \bibinfo{pages}{504003}
  (\bibinfo{year}{2009}).

\bibitem[{\citenamefont{Peschel}(2003)}]{peschel2003}
\bibinfo{author}{\bibfnamefont{I.}~\bibnamefont{Peschel}},
  \bibinfo{journal}{Journal of Physics A: Mathematical and General}
  \textbf{\bibinfo{volume}{36}}, \bibinfo{pages}{L205} (\bibinfo{year}{2003}).

\bibitem[{\citenamefont{Haque et~al.}(2007)\citenamefont{Haque, Zozulya, and
  Schoutens}}]{Haque2007}
\bibinfo{author}{\bibfnamefont{M.}~\bibnamefont{Haque}},
  \bibinfo{author}{\bibfnamefont{O.}~\bibnamefont{Zozulya}}, \bibnamefont{and}
  \bibinfo{author}{\bibfnamefont{K.}~\bibnamefont{Schoutens}},
  \bibinfo{journal}{Phys. Rev. Lett.} \textbf{\bibinfo{volume}{98}},
  \bibinfo{pages}{060401} (\bibinfo{year}{2007}),
  \urlprefix\url{http://link.aps.org/doi/10.1103/PhysRevLett.98.060401}.

\bibitem[{\citenamefont{Legner and Neupert}(2013)}]{Legner2013}
\bibinfo{author}{\bibfnamefont{M.}~\bibnamefont{Legner}} \bibnamefont{and}
  \bibinfo{author}{\bibfnamefont{T.}~\bibnamefont{Neupert}},
  \bibinfo{journal}{Phys. Rev. B} \textbf{\bibinfo{volume}{88}},
  \bibinfo{pages}{115114} (\bibinfo{year}{2013}),
  \urlprefix\url{http://link.aps.org/doi/10.1103/PhysRevB.88.115114}.

\bibitem[{\citenamefont{Zozulya et~al.}(2007)\citenamefont{Zozulya, Haque,
  Schoutens, and Rezayi}}]{Zozulya2007}
\bibinfo{author}{\bibfnamefont{O.~S.} \bibnamefont{Zozulya}},
  \bibinfo{author}{\bibfnamefont{M.}~\bibnamefont{Haque}},
  \bibinfo{author}{\bibfnamefont{K.}~\bibnamefont{Schoutens}},
  \bibnamefont{and} \bibinfo{author}{\bibfnamefont{E.~H.}
  \bibnamefont{Rezayi}}, \bibinfo{journal}{Phys. Rev. B}
  \textbf{\bibinfo{volume}{76}}, \bibinfo{pages}{125310}
  (\bibinfo{year}{2007}),
  \urlprefix\url{http://link.aps.org/doi/10.1103/PhysRevB.76.125310}.

\bibitem[{\citenamefont{Arnesen et~al.}(2001)\citenamefont{Arnesen, Bose, and
  Vedral}}]{Arnesen2001}
\bibinfo{author}{\bibfnamefont{M.~C.} \bibnamefont{Arnesen}},
  \bibinfo{author}{\bibfnamefont{S.}~\bibnamefont{Bose}}, \bibnamefont{and}
  \bibinfo{author}{\bibfnamefont{V.}~\bibnamefont{Vedral}},
  \bibinfo{journal}{Phys. Rev. Lett.} \textbf{\bibinfo{volume}{87}},
  \bibinfo{pages}{017901} (\bibinfo{year}{2001}),
  \urlprefix\url{http://link.aps.org/doi/10.1103/PhysRevLett.87.017901}.

\bibitem[{\citenamefont{Thomale et~al.}(2010)\citenamefont{Thomale, Arovas, and
  Bernevig}}]{Thomale2010}
\bibinfo{author}{\bibfnamefont{R.}~\bibnamefont{Thomale}},
  \bibinfo{author}{\bibfnamefont{D.~P.} \bibnamefont{Arovas}},
  \bibnamefont{and} \bibinfo{author}{\bibfnamefont{B.~A.}
  \bibnamefont{Bernevig}}, \bibinfo{journal}{Phys. Rev. Lett.}
  \textbf{\bibinfo{volume}{105}}, \bibinfo{pages}{116805}
  (\bibinfo{year}{2010}),
  \urlprefix\url{http://link.aps.org/doi/10.1103/PhysRevLett.105.116805}.

\bibitem[{\citenamefont{Anderson}(1958)}]{Andreson1958}
\bibinfo{author}{\bibfnamefont{P.~W.} \bibnamefont{Anderson}},
  \bibinfo{journal}{Phys. Rev.} \textbf{\bibinfo{volume}{109}},
  \bibinfo{pages}{1492} (\bibinfo{year}{1958}),
  \urlprefix\url{http://link.aps.org/doi/10.1103/PhysRev.109.1492}.

\bibitem[{\citenamefont{Wu et~al.}(1992)\citenamefont{Wu, Goff, and
  Phillips}}]{Wu1992}
\bibinfo{author}{\bibfnamefont{H.-L.} \bibnamefont{Wu}},
  \bibinfo{author}{\bibfnamefont{W.}~\bibnamefont{Goff}}, \bibnamefont{and}
  \bibinfo{author}{\bibfnamefont{P.}~\bibnamefont{Phillips}},
  \bibinfo{journal}{Phys. Rev. B} \textbf{\bibinfo{volume}{45}},
  \bibinfo{pages}{1623} (\bibinfo{year}{1992}),
  \urlprefix\url{http://link.aps.org/doi/10.1103/PhysRevB.45.1623}.

\bibitem[{\citenamefont{Phillips and Wu}(1991)}]{Phillips1991}
\bibinfo{author}{\bibfnamefont{P.}~\bibnamefont{Phillips}} \bibnamefont{and}
  \bibinfo{author}{\bibfnamefont{H.-L.} \bibnamefont{Wu}},
  \bibinfo{journal}{Science} \textbf{\bibinfo{volume}{252}},
  \bibinfo{pages}{1805} (\bibinfo{year}{1991}),
  \eprint{http://www.sciencemag.org/content/252/5014/1805.full.pdf},
  \urlprefix\url{http://www.sciencemag.org/content/252/5014/1805.abstract}.

\bibitem[{\citenamefont{Aubry and Andr\'e}(1980)}]{AA1980}
\bibinfo{author}{\bibfnamefont{S.}~\bibnamefont{Aubry}} \bibnamefont{and}
  \bibinfo{author}{\bibfnamefont{G.}~\bibnamefont{Andr\'e}},
  \bibinfo{journal}{Ann. Isr. Phys. Soc.} \textbf{\bibinfo{volume}{3}},
  \bibinfo{pages}{33} (\bibinfo{year}{1980}).

\bibitem[{\citenamefont{Refael and Moore}(2004)}]{Refael2004}
\bibinfo{author}{\bibfnamefont{G.}~\bibnamefont{Refael}} \bibnamefont{and}
  \bibinfo{author}{\bibfnamefont{J.~E.} \bibnamefont{Moore}},
  \bibinfo{journal}{Phys. Rev. Lett.} \textbf{\bibinfo{volume}{93}},
  \bibinfo{pages}{260602} (\bibinfo{year}{2004}),
  \urlprefix\url{http://link.aps.org/doi/10.1103/PhysRevLett.93.260602}.

\bibitem[{\citenamefont{Phillips}(2012)}]{Phillips2012}
\bibinfo{author}{\bibfnamefont{P.}~\bibnamefont{Phillips}},
  \emph{\bibinfo{title}{Advanced Solid State Physics}}
  (\bibinfo{publisher}{Cambridge University Press}, \bibinfo{year}{2012}).

\bibitem[{\citenamefont{Iyer et~al.}(2013{\natexlab{a}})\citenamefont{Iyer,
  Oganesyan, Refael, and Huse}}]{Shankar2013}
\bibinfo{author}{\bibfnamefont{S.}~\bibnamefont{Iyer}},
  \bibinfo{author}{\bibfnamefont{V.}~\bibnamefont{Oganesyan}},
  \bibinfo{author}{\bibfnamefont{G.}~\bibnamefont{Refael}}, \bibnamefont{and}
  \bibinfo{author}{\bibfnamefont{D.~A.} \bibnamefont{Huse}},
  \bibinfo{journal}{Phys. Rev. B} \textbf{\bibinfo{volume}{87}},
  \bibinfo{pages}{134202} (\bibinfo{year}{2013}{\natexlab{a}}),
  \urlprefix\url{http://link.aps.org/doi/10.1103/PhysRevB.87.134202}.

\bibitem[{\citenamefont{Eilmes et~al.}(1999)\citenamefont{Eilmes, Grimm,
  Römer, and Schreiber}}]{Eilmes1999}
\bibinfo{author}{\bibfnamefont{A.}~\bibnamefont{Eilmes}},
  \bibinfo{author}{\bibfnamefont{U.}~\bibnamefont{Grimm}},
  \bibinfo{author}{\bibfnamefont{R.}~\bibnamefont{Römer}}, \bibnamefont{and}
  \bibinfo{author}{\bibfnamefont{M.}~\bibnamefont{Schreiber}},
  \bibinfo{journal}{The European Physical Journal B - Condensed Matter and
  Complex Systems} \textbf{\bibinfo{volume}{8}}, \bibinfo{pages}{547}
  (\bibinfo{year}{1999}), ISSN \bibinfo{issn}{1434-6028},
  \urlprefix\url{http://dx.doi.org/10.1007/s100510050721}.

\bibitem[{\citenamefont{Iyer et~al.}(2013{\natexlab{b}})\citenamefont{Iyer,
  Oganesyan, Refael, and Huse}}]{Lyer2013}
\bibinfo{author}{\bibfnamefont{S.}~\bibnamefont{Iyer}},
  \bibinfo{author}{\bibfnamefont{V.}~\bibnamefont{Oganesyan}},
  \bibinfo{author}{\bibfnamefont{G.}~\bibnamefont{Refael}}, \bibnamefont{and}
  \bibinfo{author}{\bibfnamefont{D.~A.} \bibnamefont{Huse}},
  \bibinfo{journal}{Phys. Rev. B} \textbf{\bibinfo{volume}{87}},
  \bibinfo{pages}{134202} (\bibinfo{year}{2013}{\natexlab{b}}),
  \urlprefix\url{http://link.aps.org/doi/10.1103/PhysRevB.87.134202}.

\bibitem[{\citenamefont{Mondragon-Shem
  et~al.}(2013{\natexlab{b}})\citenamefont{Mondragon-Shem, Song, Hughes, and
  Prodan}}]{MS2013B}
\bibinfo{author}{\bibfnamefont{I.}~\bibnamefont{Mondragon-Shem}},
  \bibinfo{author}{\bibfnamefont{J.}~\bibnamefont{Song}},
  \bibinfo{author}{\bibfnamefont{T.}~\bibnamefont{Hughes}}, \bibnamefont{and}
  \bibinfo{author}{\bibfnamefont{E.}~\bibnamefont{Prodan}}
  (\bibinfo{year}{2013}{\natexlab{b}}), \eprint{cond-mat/1311.5233}.

\bibitem[{\citenamefont{Ryu and Hatsugai}(2006)}]{Ryu2006}
\bibinfo{author}{\bibfnamefont{S.}~\bibnamefont{Ryu}} \bibnamefont{and}
  \bibinfo{author}{\bibfnamefont{Y.}~\bibnamefont{Hatsugai}},
  \bibinfo{journal}{Phys. Rev. B} \textbf{\bibinfo{volume}{73}},
  \bibinfo{pages}{245115} (\bibinfo{year}{2006}),
  \urlprefix\url{http://link.aps.org/doi/10.1103/PhysRevB.73.245115}.

\bibitem[{\citenamefont{Fang et~al.}(2013)\citenamefont{Fang, Gilbert, and
  Bernevig}}]{Fang2013}
\bibinfo{author}{\bibfnamefont{C.}~\bibnamefont{Fang}},
  \bibinfo{author}{\bibfnamefont{M.~J.} \bibnamefont{Gilbert}},
  \bibnamefont{and} \bibinfo{author}{\bibfnamefont{B.~A.}
  \bibnamefont{Bernevig}}, \bibinfo{journal}{Phys. Rev. B}
  \textbf{\bibinfo{volume}{87}}, \bibinfo{pages}{035119}
  (\bibinfo{year}{2013}),
  \urlprefix\url{http://link.aps.org/doi/10.1103/PhysRevB.87.035119}.

\bibitem[{\citenamefont{Hughes et~al.}(2011)\citenamefont{Hughes, Prodan, and
  Bernevig}}]{Hughes2011}
\bibinfo{author}{\bibfnamefont{T.~L.} \bibnamefont{Hughes}},
  \bibinfo{author}{\bibfnamefont{E.}~\bibnamefont{Prodan}}, \bibnamefont{and}
  \bibinfo{author}{\bibfnamefont{B.~A.} \bibnamefont{Bernevig}},
  \bibinfo{journal}{Phys. Rev. B} \textbf{\bibinfo{volume}{83}},
  \bibinfo{pages}{245132} (\bibinfo{year}{2011}),
  \urlprefix\url{http://link.aps.org/doi/10.1103/PhysRevB.83.245132}.

\bibitem[{\citenamefont{Turner et~al.}(2010)\citenamefont{Turner, Zhang, and
  Vishwanath}}]{Turner2010}
\bibinfo{author}{\bibfnamefont{A.~M.} \bibnamefont{Turner}},
  \bibinfo{author}{\bibfnamefont{Y.}~\bibnamefont{Zhang}}, \bibnamefont{and}
  \bibinfo{author}{\bibfnamefont{A.}~\bibnamefont{Vishwanath}},
  \bibinfo{journal}{Phys. Rev. B} \textbf{\bibinfo{volume}{82}},
  \bibinfo{pages}{241102} (\bibinfo{year}{2010}),
  \urlprefix\url{http://link.aps.org/doi/10.1103/PhysRevB.82.241102}.

\bibitem[{\citenamefont{Pollmann et~al.}(2010)\citenamefont{Pollmann, Turner,
  Berg, and Oshikawa}}]{Pollman2010}
\bibinfo{author}{\bibfnamefont{F.}~\bibnamefont{Pollmann}},
  \bibinfo{author}{\bibfnamefont{A.~M.} \bibnamefont{Turner}},
  \bibinfo{author}{\bibfnamefont{E.}~\bibnamefont{Berg}}, \bibnamefont{and}
  \bibinfo{author}{\bibfnamefont{M.}~\bibnamefont{Oshikawa}},
  \bibinfo{journal}{Phys. Rev. B} \textbf{\bibinfo{volume}{81}},
  \bibinfo{pages}{064439} (\bibinfo{year}{2010}),
  \urlprefix\url{http://link.aps.org/doi/10.1103/PhysRevB.81.064439}.

\bibitem[{\citenamefont{Turner et~al.}(2011)\citenamefont{Turner, Pollmann, and
  Berg}}]{Turner2011}
\bibinfo{author}{\bibfnamefont{A.~M.} \bibnamefont{Turner}},
  \bibinfo{author}{\bibfnamefont{F.}~\bibnamefont{Pollmann}}, \bibnamefont{and}
  \bibinfo{author}{\bibfnamefont{E.}~\bibnamefont{Berg}},
  \bibinfo{journal}{Phys. Rev. B} \textbf{\bibinfo{volume}{83}},
  \bibinfo{pages}{075102} (\bibinfo{year}{2011}),
  \urlprefix\url{http://link.aps.org/doi/10.1103/PhysRevB.83.075102}.

\bibitem[{\citenamefont{Fidkowski}(2010)}]{Fidkowski2010}
\bibinfo{author}{\bibfnamefont{L.}~\bibnamefont{Fidkowski}},
  \bibinfo{journal}{Phys. Rev. Lett.} \textbf{\bibinfo{volume}{104}},
  \bibinfo{pages}{130502} (\bibinfo{year}{2010}),
  \urlprefix\url{http://link.aps.org/doi/10.1103/PhysRevLett.104.130502}.

\bibitem[{\citenamefont{Flammia et~al.}(2009)\citenamefont{Flammia, Hamma,
  Hughes, and Wen}}]{Flammia2009}
\bibinfo{author}{\bibfnamefont{S.~T.} \bibnamefont{Flammia}},
  \bibinfo{author}{\bibfnamefont{A.}~\bibnamefont{Hamma}},
  \bibinfo{author}{\bibfnamefont{T.~L.} \bibnamefont{Hughes}},
  \bibnamefont{and} \bibinfo{author}{\bibfnamefont{X.-G.} \bibnamefont{Wen}},
  \bibinfo{journal}{Phys. Rev. Lett.} \textbf{\bibinfo{volume}{103}},
  \bibinfo{pages}{261601} (\bibinfo{year}{2009}),
  \urlprefix\url{http://link.aps.org/doi/10.1103/PhysRevLett.103.261601}.

\bibitem[{\citenamefont{Qi et~al.}(2012)\citenamefont{Qi, Katsura, and
  Ludwig}}]{Qi2012}
\bibinfo{author}{\bibfnamefont{X.-L.} \bibnamefont{Qi}},
  \bibinfo{author}{\bibfnamefont{H.}~\bibnamefont{Katsura}}, \bibnamefont{and}
  \bibinfo{author}{\bibfnamefont{A.~W.~W.} \bibnamefont{Ludwig}},
  \bibinfo{journal}{Phys. Rev. Lett.} \textbf{\bibinfo{volume}{108}},
  \bibinfo{pages}{196402} (\bibinfo{year}{2012}),
  \urlprefix\url{http://link.aps.org/doi/10.1103/PhysRevLett.108.196402}.

\bibitem[{\citenamefont{Alexandradinata
  et~al.}(2011)\citenamefont{Alexandradinata, Hughes, and
  Bernevig}}]{Alexandradinata2011}
\bibinfo{author}{\bibfnamefont{A.}~\bibnamefont{Alexandradinata}},
  \bibinfo{author}{\bibfnamefont{T.~L.} \bibnamefont{Hughes}},
  \bibnamefont{and} \bibinfo{author}{\bibfnamefont{B.~A.}
  \bibnamefont{Bernevig}}, \bibinfo{journal}{Phys. Rev. B}
  \textbf{\bibinfo{volume}{84}}, \bibinfo{pages}{195103}
  (\bibinfo{year}{2011}),
  \urlprefix\url{http://link.aps.org/doi/10.1103/PhysRevB.84.195103}.

\bibitem[{\citenamefont{Sedrakyan et~al.}(2011)\citenamefont{Sedrakyan,
  Kestner, and Das~Sarma}}]{Sedrakyan2011}
\bibinfo{author}{\bibfnamefont{T.~A.} \bibnamefont{Sedrakyan}},
  \bibinfo{author}{\bibfnamefont{J.~P.} \bibnamefont{Kestner}},
  \bibnamefont{and}
  \bibinfo{author}{\bibfnamefont{S.}~\bibnamefont{Das~Sarma}},
  \bibinfo{journal}{Phys. Rev. A} \textbf{\bibinfo{volume}{84}},
  \bibinfo{pages}{053621} (\bibinfo{year}{2011}),
  \urlprefix\url{http://link.aps.org/doi/10.1103/PhysRevA.84.053621}.

\end{thebibliography}

\end{document}